\documentclass[journal,onecolumn,draftclsnofoot,10pt]{IEEEtran}
\usepackage{multirow}
\usepackage{graphicx}
\usepackage{amsmath}
\usepackage{amsthm}
\usepackage{mathtools}
\usepackage[psamsfonts]{amssymb}
\usepackage{amsxtra}
\usepackage{threeparttable}
\usepackage{cleveref}
\usepackage{color}
\usepackage{url}
\usepackage{cite}
\usepackage[justification=centering]{caption}

\newtheorem{Theorem}{Theorem}[section]
\newtheorem{Lemma}[Theorem]{Lemma}

\newtheorem{Remark}[Theorem]{Remark}
\theoremstyle{definition}
\newtheorem{assump}{Assumption}

\newenvironment{myassump}[2][]
{\renewcommand\theassump{#2}\begin{assump}[#1]}
	{\end{assump}}
\def \tn {\mathbf{\theta}_{N}^{*}}
\def \gg {\mathbf{\gamma}_{G}}

\def \btheta{\boldsymbol{\theta}}
\begin{document}
	
	\title{Recursive Distributed Detection for Composite Hypothesis Testing: Nonlinear Observation Models in Additive Gaussian Noise}
	\author{Anit~Kumar~Sahu,~\IEEEmembership{Student Member,~IEEE} and Soummya~Kar,~\IEEEmembership{Member,~IEEE}
		\thanks{Copyright (c) 2017 IEEE. Personal use of this material is permitted.  However, permission to use this material for any other purposes must be obtained from the IEEE by sending a request to pubs-permissions@ieee.org.
			
			The authors are with the Department of Electrical and Computer Engineering, Carnegie Mellon University, Pittsburgh, PA 15213, USA (email: anits@andrew.cmu.edu, soummyak@andrew.cmu.edu). This work was supported in part by NSF under grants CCF-1513936 and ECCS-1408222.}}
	\maketitle
	\begin{abstract}
		This paper studies recursive composite hypothesis testing in a network of sparsely connected agents. The network objective is to test a simple null hypothesis against a composite alternative concerning the state of the field, modeled as a vector of (continuous) unknown parameters determining the parametric family of probability measures induced on the agents' observation spaces under the hypotheses. Specifically, under the alternative hypothesis, each agent sequentially observes an independent and identically distributed time-series consisting of a (nonlinear) function of the true but unknown parameter corrupted by Gaussian noise, whereas, under the null, they obtain noise only. Two distributed recursive generalized likelihood ratio test type algorithms of the \emph{consensus+innovations} form are proposed, namely $\mathcal{CIGLRT-L}$ and $\mathcal{CIGLRT-NL}$, in which the agents estimate the underlying parameter and in parallel also update their test decision statistics by simultaneously processing the latest local sensed information and information obtained from neighboring agents. For $\mathcal{CIGLRT-NL}$, for a broad class of nonlinear observation models and under a global observability condition, algorithm parameters which ensure asymptotically decaying probabilities of errors~(probability of miss and probability of false detection) are characterized. For $\mathcal{CIGLRT-L}$, a linear observation model is considered and upper bounds on large deviations decay exponent for the error probabilities are obtained.
	\end{abstract}
	
	
	\section{Introduction}

	\subsection{Background and Motivation}
	The focus of this paper is on distributed composite hypothesis testing in multi-agent networks in which the goal is not only to estimate the state (possibly high dimensional) of the environment but also detect as to which hypothesis is in force based on the sensed information across all the agents at all times. To be specific, we are interested in the design of recursive detection algorithms to decide between a simple null hypothesis and a composite alternative parameterized by a continuous vector parameter, which exploit available sensing resources to the maximum and obtain reasonable detection performance, i.e., have asymptotically (in the large sample limit) decaying probabilities of errors. Technically speaking, we are interested in the study of algorithms which can process sensed information as and when they are sensed and not wait till the end until all the sensed data has been collected. To be specific, the sensed data refers to the observations made across all the agents at all times. The problem of composite hypothesis testing is relevant to many practical applications, including cooperative spectrum sensing \cite{zarrin2009composite,font2010glrt,zou2010cooperative} and MIMO radars \cite{tajer2010optimal}, where the onus is also on achieving reasonable detection performance by utilizing as fewer resources as possible, which includes data samples, communication and sensing energy. In classical composite hypothesis testing procedures such as the Generalized Likelihood Ratio Test (GLRT) \cite{zeitouni1992generalized}, the detection procedure which uses the underlying parameter estimate based on all the collected samples as a plug-in estimate may not be initiated until a reasonably accurate parameter estimate, typically the maximum likelihood estimate of the underlying parameter (state) is obtained. Usually in setups which employ the classical (centralized) generalized likelihood ratio tests, the data collection phase precedes the parameter estimation and detection statistic update phase which makes the procedure essentially an \emph{offline} batch procedure. By \emph{offline} batch procedures, we mean algorithms where the sensing phase precedes any kind of information processing and the entire data is processed in batches.\footnote{We emphasize that, by offline, we strictly refer to the classical implementation of the GLRT. Recursive variants of GLRT type approaches have been developed for a variety of testing problems including sequential composite hypothesis testing and change detection (see, for example,~\cite{siegmund1995using,willsky1976generalized,chang1979recursive}), although in centralized processing scenarios.} The motivation behind studying recursive online detection algorithms in contrast to offline batch processing based detection algorithms is that in most multi-agent networked scenarios, which are typically energy constrained, the priority is to obtain reasonable inference performance by expending fewer amount of resources. Moreover, in centralized scenarios, where the communication graph is all-to-all, the implementation suffers from high communication overheads, synchronization issues and high energy requirements. Motivated by requirements such as the latter, we propose distributed recursive composite hypothesis testing algorithms, where the inter-agent collaboration is restricted to a pre-assigned, possibly sparse communication graph and the detection and estimation schemes run in a parallel fashion with a view to reduce energy and resource consumption while achieving reasonable detection performance.
	
	In the domain of hypothesis testing, when one of the hypotheses is composite, i.e., the hypothesis is parameterized by a continuous vector parameter and the underlying parameter is unknown apriori, one of the most well-known algorithms is the Generalized Likelihood Ratio Testing (GLRT). The GLRT has an estimation procedure built into it, where the underlying parameter estimate is used as a plug-in estimate for the decision statistic. In a centralized setting or in a scenario where the inter-agent communication graph is all-to-all, the fusion center has access to all the sensed information and the parameter estimates across all the agents at all times. The procedure of obtaining the underlying parameter estimate, which in turn employs a maximization, achieves reasonable performance in general, but has a huge communication overhead which makes it infeasible to be implemented in practice, especially in networked environments. In contrast to the fully centralized setup, we focus on a fully distributed setup where the communication between the agents is restricted to a pre-assigned possibly sparse communication graph. In this paper, we propose two algorithms namely,  $\emph{consensus}+\emph{innovations}$ GLRT Non-Linear ~($\mathcal{CIGLRT-NL}$) and $\emph{consensus}+\emph{innovations}$ GLRT Linear~($\mathcal{CIGLRT-L}$), which are of the $\emph{consensus}+\emph{innovations}$ form and are based on fully distributed setups. We specifically focus on a setting in which the agents obtain conditionally Gaussian and independent and identically distributed observations and update their parameter estimates and decision statistics by simultaneous assimilation of the information obtained from the neighboring agents (\emph{consensus}) and the latest locally sensed information (\emph{innovation}). Also similar, to the classical GLRT, both of our algorithms involve a parameter estimation scheme and a detection algorithm. This justifies the names $\mathcal{CIGLRT-L}$ and $\mathcal{CIGLRT-NL}$ which are distributed GLRT type algorithms of the $\emph{consensus}+\emph{innovations}$ form. In this paper, so as to replicate typical practical sensing environments accurately, we model the underlying vector parameter as a static parameter, whose dimension is $M$~(possibly large) and every agent's observations, say for agent $n$, is $M_{n}$ dimensional, where $M_{n} << M$, thus rendering the parameter locally unobservable at each agent. We show that, under a minimal global observability condition imposed on the collective observation model and connectedness of the communication graph, the parameter estimate sequences are consistent and the detection schemes achieve asymptotically decaying probabilities of errors in the large sample limit. The main contributions of the paper are as follows:\\
	
	\textbf{Main Contribution 1: Distributed Composite Hypothesis Testing Algorithms.} We propose two distributed recursive composite hypothesis testing algorithms, where the composite alternative concerning the state of the field is modeled as a vector of (continuous) unknown parameters determining the parametric family of probability measures induced on the agents' observation spaces under the hypotheses. Moreover, we focus on fully distributed setups where the underlying parameter may not be locally observable at any of the agents and hence no agent can conduct hypothesis testing to achieve reasonable decision performance by itself. The agents only collaborate locally in their neighborhood, where the collaboration dynamics are specified by a possibly sparse pre-assigned communication graph.\\
	
	\textbf{Main Contribution 2: Recursive Detection Algorithm with decaying probabilities of errors.} We show that in spite of being a recursive algorithm (hence suboptimal\footnote{The sub-optimality with respect to GLRT is due to inaccurate parameter estimates being incorporated into the decision statistic in the proposed algorithm in contrast to the \emph{optimal} parameter estimate incorporated into the decision statistic in case of the classical GLRT.}), the proposed algorithm $\mathcal{CIGLRT-NL}$, which is based on general non-linear observation models, guarantees asymptotically decaying probabilities of false alarm and miss under minimal conditions of global observability and connectivity of the inter-agent communication graph. We also characterize the feasible choice of thresholds and other algorithm design parameters for which such an asymptotic decay of probabilities of errors in the large sample~(time) limit can be guaranteed. \\
	
	\textbf{Main Contribution 3: Recursive Detection Algorithm with exponentially decaying errors.} Through algorithm $\mathcal{CIGLRT-L}$, we focus on a linear observation setup, where we not only characterize thresholds and other algorithm parameters which ensure exponentially decaying probabilities of error, but also analyze the upper bounds of the associated large deviations exponent of the probabilities of error under global observability as functions of the network and model parameters.

	\textbf{Related Work:}
	Existing work in the literature on distributed detectors can be broadly classified into three classes. The first class includes architectures which are characterized by presence of a fusion center and all the agents transmit their decision or local measurements or test statistics or its quantized version to the fusion center (see, for example \cite{blum1997distributed,tsitsiklis1993decentralized}) and subsequently the estimation and detection schemes are conducted by the fusion center. The second class consists of consensus schemes (see, for example \cite{kar2007consensus,olfati2006belief}) with no fusion center and in which in the first phase the agents collect information over a long period of time from the environment followed by the second phase, in which agents exchange information (through consensus or gossip type procedures ~\cite{kar2007consensus,olfatisaberfaxmurray07,jadbabailinmorse03})~in their respective neighborhoods which are in turn specified by a pre-assigned communication graph or a sequence of possibly sparse time-varying communication graphs satisfying appropriate connectivity conditions. The third class consists of schemes which perform simultaneous assimilation of information obtained from sensing and communication (see, for example \cite{bajovic2011distributed,kar2011distributed,cattivelli2011distributed}).
	Distributed inference has been studied extensively in the literature. In particular, the diffusion and \emph{consensus+innovations} schemes have been extensively used for various distributed inference problems, which include distributed parameter estimation (see, for example \cite{KarMoura-LinEst-JSTSP-2011,cattivelli2010diffusion,bajovic2015distributed}), distributed detection (see, for example \cite{kar2011distributed,jakovetic2012distributed,sahu2015distributed,cattivelli2011distributed}), distributed reinforcement learning (see, for example \cite{kar2013learning}), distributed information processing in presence of faulty agents or imperfect model information~(see, for example \cite{Zhou-ICASSP-2012}) and multi-task learning (see, for example \cite{chen2014multitask}) to name a few. 	More relevant to the proposed algorithms in this paper, are distributed detection algorithms from the third class of detectors described above, which can be further sub-categorized to three classes, namely the running consensus approach \cite{braca2008enforcing,braca2010asymptotic}, the diffusion approach \cite{cattivelli2009diffusion,cattivelli2009distributed,cattivelli2011distributed,zou2010cooperative,matta2014diffusion,matta2015exact,braca2014large,zou2010cooperative} and the \emph{consensus+innovations} approach \cite{bajovic2011distributed,jakovetic2012distributed,kar2011distributed}. These works address important questions pertaining to binary simple hypothesis and also characterize the fundamental limits of the detection scheme through large deviations analysis. Other relevant recent work include \cite{lalitha2014social, lalitha2015social,nedic2014nonasymptotic}. However, there is a fundamental difference between the algorithms proposed in this work and the algorithms discussed above. To be specific, the objective of the detection scheme in this paper is to decide between a simple null hypothesis and a composite alternative which is parameterized by a vector parameter which can take values in a continuous space, in contrast with other distributed detection schemes, where the hypotheses involved are either binary simple hypothesis (see, for example \cite{kar2011distributed,bajovic2011distributed,jakovetic2012distributed}) or multiple simple hypothesis or composite testing scenarios involving finite parametric alternatives~(see, for example \cite{lalitha2014social,lalitha2015social,nedic2014nonasymptotic,jadbabaie2012non}). In a similar vein, we note that consensus or gossip type strategies have been developed for other distributed testing tasks such as in change detection~\cite{ilic2012consensus} and quickest detection~\cite{li2015distributed} problems in multi-agent networks.
	In the context of distributed composite hypothesis testing, the proposed algorithm involves a recursive parameter estimation update and a decision statistic update running in parallel. The proposed algorithms, $\mathcal{CIGLRT-NL}$ and $\mathcal{CIGLRT-L}$ involve parameter estimate updates in a non-linear observation model and a linear observation model setting respectively. The linear parameter estimation scheme of the $\emph{consensus}+\emph{innovations}$ form which is employed in the algorithm $\mathcal{CIGLRT-L}$ has been studied before in \cite{kar2011convergence}. Non-linear parameter estimation scheme of the \emph{consensus}+\emph{innovations} form has been studied in \cite{kar2014asymptotically}. However, for the non-linear estimation parameter estimation scheme, employed in in the algorithm $\mathcal{CIGLRT-NL}$ there is a subtle difference between the one considered in \cite{kar2014asymptotically} in terms of innovation gains. The innovation gains in the distributed estimation algorithm proposed in \cite{kar2014asymptotically} is generated from an auxiliary estimate sequence and hence the convergence of the innovation gain processes is not tied to the convergence of the parameter estimate sequence. However, the innovation gain sequence in this paper is tied to the parameter estimate sequence itself which needed refined proof techniques which we address in this paper.\\
	In the context of characterization of the probability of errors in terms of the exponent, for recursive distributed detection schemes, the works most related to the proposed algorithms are \cite{bajovic2011distributed,jakovetic2012distributed}. In these works, in addition to the hypotheses under consideration being binary simple hypotheses, the decision statistic update involves an innovation sequence which is independent and identically distributed (i.i.d.). However, the detection scheme in this paper decides between a simple null hypothesis and a composite alternative which is parameterized by a vector parameter that takes values in a continuous space. Moreover, the decision statistic update in this paper involves an innovation term which is neither independent nor identically distributed, due to the coupling between the estimation and detection update rules. It is to be noted that in the context of centralized detection literature, the weak convergence of the decision statistics under the null and the alternate hypothesis~(see, \cite{wilks1938large} for example)~ is usually not enough to establish the decay rates of the probability of errors. To be specific, in this paper we extend Wilks' theorem to the distributed recursive setup and characterize the asymptotic normality of the decision statistic sequence. However, the statistical dependencies exhibited in the decision static update due to the parameter estimation scheme and the decision statistic update running in a parallel fashion warrants the development of technical machinery so as to address concentration of measure for sums of non i.i.d random variables which in turn helps characterize the decay exponent of the probability of errors which we develop in this paper. \\
	
	\noindent\textbf{Paper Organization :} The rest of the paper is organized as follows. Section \ref{subsec:not} presents the notation to be used throughout the paper. The sensing model is discussed is Section \ref{subsec:sys_model}, whereas the preliminaries pertaining to classical GLRT are summarized in Section \ref{subsec:prel_glrt}, which in turn motivates distributed online detection algorithms proposed in this paper. Section \ref{subsec:cilrt} presents the $\mathcal{CIGLRT-NL}$ algorithm whereas the $\mathcal{CIGLRT-L}$ algorithm is discussed in Section \ref{subsec:lo-cilrt}. The main results of the algorithm $\mathcal{CIGLRT-NL}$ concerning consistency of the parameter estimate and asymptotically decaying probabilities of errors are provided in Section \ref{subsec:main_res_ciglrt}, while the main results concerning the algorithm $\mathcal{CIGLRT-L}$, which include consistency of the parameter estimate sequence and the characterization of the upper bound of the large deviations exponent of the probabilities of errors are presented in Section \ref{subsec:main_res_clirt}. Section \ref{sec:illust_eg} contains an illustrative example which provides more intuition on the upper bounds of the large deviations exponent obtained for the $\mathcal{CIGLRT-L}$ algorithm and presents the simulation results. The proof of main results appear in Sections \ref{sec:proof_res_ciglrt} and \ref{sec:proof_res_cilrt}, while Section \ref{sec:conc} concludes the paper.
	
	\subsection{Notation}
	\label{subsec:not}
	\noindent We denote by~$\mathbb{R}$ the set of reals, $\mathbb{R}_{+}$ the set of non-negative reals, and by~$\mathbb{R}^{k}$ the $k$-dimensional Euclidean space.
	\noindent The set of $k\times k$ real matrices is denoted by $\mathbb{R}^{k\times k}$. The set of integers is denoted by $\mathbb{Z}$, whereas, $\mathbb{Z}_{+}$ denotes the subset of non-negative integers. We denote vectors and matrices by bold faced characters. We denote by $A_{ij}$ or $[A]_{ij}$ the $(i,j)$-th entry of a matrix $\mathbf{A}$; $a_{i}$ or $[a]_{i}$ the $i$-th entry of a vector $\mathbf{a}$. The symbols $\mathbf{I}$ and $\mathbf{0}$ are used to denote the $k\times k$ identity matrix and the $k\times k$ zero matrix respectively, the dimensions being clear from the context. We denote by $\mathbf{e_{i}}$ the $i$-th column of $\mathbf{I}$. The symbol $\top$ denotes matrix transpose. We denote the determinant and trace of a matrix by $\det(.)$ and $\operatorname{tr}(.)$ respectively. The $k\times k$ matrix $\mathbf{J}=\frac{1}{k}\mathbf{1}\mathbf{1^{\top}}$ where $\mathbf{1}$ denotes the $k\times 1$ vector of ones. The operator $|| . ||$ applied to a vector denotes the standard Euclidean $\mathcal{L}_{2}$ norm, while applied to matrices it denotes the induced $\mathcal{L}_{2}$ norm, which is equivalent to the spectral radius for symmetric matrices. For a matrix $A$ with real eigenvalues, the notation $\lambda_{\mbox{\scriptsize{min}}}(A)$ and $\lambda_{\mbox{\scriptsize{max}}}(A)$ will be used to denote its smallest and largest eigenvalues respectively.
	Throughout the paper, the true (but unknown) value of the parameter is denoted by $\mathbf{\theta}^{*}$. The estimate of $\mathbf{\theta}^{*}$ at time $t$ at agent $n$ is denoted by $\mathbf{\theta}_{n}(t) \in  \mathbb{R}^{M\times 1}$. All the logarithms in the paper are with respect to base $e$ and represented as $\log(\cdot)$. The operators $\mathbb{E}_{0}[\cdot]$ and $\mathbb{E}_{\theta}[\cdot]$ denote expectation conditioned on hypothesis $\mathcal{H}_{0}$ and $\mathcal{H}_{\theta}$, where $\theta$ in the parametric alternative respectively. $\mathbb{P}(\cdot)$ denotes the probability of an event and $\mathbb{P}_{0}( . )$ and $\mathbb{P}_{\theta}( . )$ denote the probability of the event conditioned on the null hypothesis $\mathcal{H}_{0}$ and $\mathcal{H}_{\theta}$, where $\theta$ is the parametric alternative. For deterministic $\mathbb{R}_{+}$-valued sequences $\{a_{t}\}$ and $\{b_{t}\}$, the notation $a_{t}=O(b_{t})$ denotes the existence of a constant $c>0$ such that $a_{t}\leq cb_{t}$ for all $t$ sufficiently large; the notation $a_{t}=o(b_{t})$ denotes $a_{t}/b_{t}\rightarrow 0$ as $t\rightarrow\infty$. The order notations $O(\cdot)$ and $o(\cdot)$ will be used in the context of stochastic processes as well in which case they are to be interpreted almost surely or path-wise.
	
	\noindent{\bf Spectral Graph Theory} For an undirected graph $G=(V, E)$, $V$ denotes the set of agents or vertices with cardinality $|V|=N$, and $E$ the set of edges with $|E|=M$. The unordered pair $(i,j) \in E$ if there exists an edge between agents $i$ and $j$. We only consider simple graphs, i.e., graphs devoid of self loops and multiple edges. A path between agents $i$ and $j$ of length $m$ is a sequence ($i=p_{0},p_{1},\cdots,p_{m}=j)$ of vertices, such that $(p_{t}, p_{t+1})\in E$, $0\le t \le m-1$. A graph is connected if there exists a path between all the possible agent pairings.
	The neighborhood of an agent $n$ is given by $\Omega_{n}=\{j \in V|(n,j) \in E\}$. The degree of agent $n$ is given by $d_{n}=|\Omega_{n}|$. The structure of the graph may be equivalently represented by the symmetric $N\times N$ adjacency matrix $\mathbf{A}=[A_{ij}]$, where $A_{ij}=1$ if $(i,j) \in E$, and $0$ otherwise. The degree matrix is represented by the diagonal matrix $\mathbf{D}=diag(d_{1}\cdots d_{N})$. The graph Laplacian matrix is represented by
	\begin{align}
		\label{eq:0.2}
		\mathbf{L}=\mathbf{D}-\mathbf{A}.
	\end{align}
	\noindent The Laplacian is a positive semidefinite matrix, hence its eigenvalues can be sorted and represented in the following manner
	\begin{align}
		\label{eq:0.3}
		0=\lambda_{1}(\mathbf{L})\le\lambda_{2}(\mathbf{L})\le \cdots \le\lambda_{N}(\mathbf{L}).
	\end{align}
\noindent Furthermore, a graph is connected if and only if $\lambda_{2}(\mathbf{L})>0$ (see~\cite{chung1997spectral}~for instance).
	
\section{Problem Formulation}
\label{sec:prob_form}
\subsection{System Model and Preliminaries}
\label{subsec:sys_model}
\noindent  There are $N$ agents deployed in the network. Every agent $n$  at time index $t$ makes a noisy observation $\mathbf{y}_{n}(t)$. The observation $\mathbf{y}_{n}(t)$ is a $M_{n}$-dimensional vector, a noisy nonlinear function of $\mathbf{\theta}^{*}$ which is a $M$-dimensional parameter, i.e., $\mathbf{\theta}^{*}\in\mathbb{R}^{M}$. The observation $\mathbf{y}_{n}(t)$ comes from a probability distribution $\mathbb{P}_{0}$ under the hypothesis $\mathcal{H}_{0}$, whereas, under the composite alternative $\mathcal{H}_{1}$, the observation is sampled from a probability distribution which is a member of a parametric family $\{\mathbb{P}_{\mathbf{\theta}^{*}}\}$. We emphasize here that the parameter $\mathbf{\theta}^{*}$ is deterministic but unknown.
Formally,
\begin{align}
\label{eq:1}
&\mathcal{H}_{1} : \mathbf{y}_{n}(t)=\mathbf{h}_{n}(\theta^{*}) + \mathbf{\gamma}_{n}(t) \nonumber\\
&\mathcal{H}_{0} : \mathbf{y}_{n}(t)=\mathbf{\gamma}_{n}(t),
\end{align}

\noindent where $\mathbf{h}_{n}(.)$ is, in general, non-linear function, $\{\mathbf{y}_{n}(t)\}$ is a $\mathbb{R}^{M_{n}}$-valued observation sequence for the $n$-th agent, where typically $M_{n}<<M$ and $\{\mathbf{\gamma}_{n}(t)\}$ is a zero-mean temporally i.i.d Gaussian noise sequence at the $n$-th agent with nonsingular covariance matrix $\mathbf{\Sigma}_{n}$, where $\mathbf{\Sigma}_{n}\in\mathbb{R}^{M_{n}\times M_{n}}$. Moreover, the noise sequences at two agents $n,l$ with $n\neq l$ are independent.

\noindent By taking $\mathbf{h}_{n}(\mathbf{0})=\mathbf{0},~\forall n$ and certain other identifiability and regularity conditions outlined below, in the above formulation the null hypothesis corresponds to $\mathbf{\theta}^{*}=0$ and the
composite alternative to the case $\mathbf{\theta}^{*}\neq\mathbf{0}$.

\noindent Since, the sources of randomness in our formulation are the observations $\mathbf{y}_{n}(t)$'s made by the agents in the network, we define the natural filtration $\{\mathcal{F}_{t}\}$ generated by the random observations,~i.e.,
\begin{align}
\label{eq:filt_1}
\mathcal{F}_{t}=\mathbf{\sigma}\left(\{\{\mathbf{y}_{n}(s)\}_{n=1}^{N}\}_{s=0}^{t-1}\right),
\end{align}

\noindent which is the sequence of $\sigma$-algebras induced by the observation processes, in order to model the overall available network information at all times. Finally, a stochastic process $\{\mathbf{x}(t)\}$ is said to be $\{\mathcal{F}_{t}\}$-adapted if the $\mathbf{\sigma}$-algebra $\mathbf{\sigma}\left(\mathbf{x}(t)\right)$ is a subset of $\mathcal{F}_{t}$ at each $t$.
\subsection{Preliminaries : Generalized Likelihood Ratio Tests}
\label{subsec:prel_glrt}
\noindent We start by reviewing some concepts from the classical theory of Generalized Likelihood Ratio Tests~(GLRT). Consider, for instance, a generalized target detection problem in which the absence of target is  modeled by a simple hypothesis $\mathcal{H}_{0}$, whereas, its presence corresponds to a composite alternative $\mathcal{H}_{1}$, as it is parametrized by a continuous vector parameter (perhaps modeling its location and other attributes) which is unknown apriori. Let $\mathbf{y}(t)$ denote the collection of the data from the agents, i.e., $\mathbf{y}(t)=\left[\mathbf{y}_{1}^{\top}(t)\cdots \mathbf{y}_{N}^{\top}(t)\right]^{\top}$, at time $t$, which is $\sum_{n=1}^{N}M_{n}$ dimensional. In a centralized setup, where there is a fusion center having access to the entire $\mathbf{y}(t)$ at all times $t$, a classical testing approach is the generalized likelihood ratio test (GLRT) (see, for example \cite{zeitouni1992generalized}). Specifically, the GLRT decision procedure decides on the hypothesis\footnote{It is important to note that the considered setup does not admit uniquely most powerful tests~\cite{scharf1991statistical}.} as follows:
\begin{align}
\label{eq:prel_glrt_1}
\mathcal{H}=
\begin{cases}
\mathcal{H}_{1}, & \mbox{if}~~\max_{\mathbf{\theta}}\sum_{t=0}^{T}\log\frac{f_{\mathbf{\theta}}(\mathbf{y}(t))}{f_{0}(\mathbf{y}(t))} > \eta,\\
\mathcal{H}_{0}, & \mbox{otherwise},
\end{cases}
\end{align}

\noindent where $\eta$ is a predefined threshold, $T$ denotes the number of sensed observations and assuming that the data from the agents are conditionally independent $f_{\mathbf{\theta}}(\mathbf{y}(t))=f_{\mathbf{\theta}}^{1}(\mathbf{y}_{1}(t))\cdots f_{\mathbf{\theta}}^{N}(\mathbf{y}_{N}(t))$ denotes the likelihood of observing $\mathbf{y}(t)$ under $\mathcal{H}_{1}$ and realization $\mathbf{\theta}$ of the parameter and $f_{\mathbf{\theta}}^{n}(\mathbf{y}_{n}(t))$ denotes the likelihood of observing $\mathbf{y}_{n}(t)$ at the $n$-th agent under $\mathcal{H}_{1}$ and realization $\mathbf{\theta}$ of the parameter ; similarly, $f_{0}(\mathbf{y}(t))=f_{0}^{1}(\mathbf{y}_{1}(t))\cdots f_{0}^{N}(\mathbf{y}_{N}(t))$ denotes the likelihood of observing $\mathbf{y}(t)$ under $\mathcal{H}_{0}$ and $f_{0}^{N}(\mathbf{y}_{N}(t))$ denotes the likelihood of observing $\mathbf{y}_{n}(t)$ at the $n$-th agent under $\mathcal{H}_{0}$. The key bottleneck in the implementation of the classical GLRT as formulated in \eqref{eq:prel_glrt_1} is the maximization
\begin{align}
\label{eq:prel_glrt_2}
\max_{\mathbf{\theta}}\sum_{t=0}^{T}\log\frac{f_{\mathbf{\theta}}(\mathbf{y}(t))}{f_{0}(\mathbf{y}(t))}=\max_{\mathbf{\theta}}\sum_{t=0}^{T}\sum_{n=1}^{N}\log\frac{f_{\mathbf{\theta}}^{n}(\mathbf{y}_{n}(t))}{f_{0}^{n}(\mathbf{y}_{n}(t))}
\end{align}
\noindent which involves the computation of the generalized log-likelihood ratio, i.e., the decision statistic. In general, a maximizer of \eqref{eq:prel_glrt_2} is not known beforehand as it depends on the entire sensed data collected across all the agents at all times, and hence as far as communication complexity in the GLRT implementation is concerned, the maximization step incurs the major overhead -- in fact, a direct implementation of the maximization \eqref{eq:prel_glrt_2} requires access to the entire raw data $\mathbf{y}(t)$ at all times $t$ at the fusion center.

\section{Distributed Generalized Likelihood Ratio Testing}
\label{sec:DLGRT}
%\emph{Note : This is the case where every sensor starts estimating the location parameters after taking its observations. I will update the synchronized case where the sensors take observations and estimate the location parameters simultaneously in the next draft.}\\
\noindent To mitigate the communication overhead, we present distributed message passing schemes in which agents, instead of forwarding raw data to a fusion center, participate in a collaborative iterative process to obtain a maximizing $\theta$. The agents also maintain a copy of their local decision statistic, where the decision statistic is updated by assimilating local decision statistics from the neighborhood and the latest sensed information. In order to obtain reasonable decision performance with such localized communication, we propose a distributed detector of the $\emph{consensus}+\emph{innovations}$ type. To this end, we propose two algorithms, namely\\
1)~$\mathcal{CIGLRT-NL}$, which is a general algorithm based on a non-linear observation model with additive Gaussian noise. We specifically show that the decision errors go to zero asymptotically as time $t\to\infty$ or equivalently, in the large sample limit, if the thresholds are chosen appropriately, and \\2)~$\mathcal{CIGLRT-L}$, where we specifically consider a linear observation model. In the case of $\mathcal{CIGLRT-L}$, we not only show that the probabilities of errors go to zero asymptotically, but also, we characterize the large deviations exponent upper bounds for the probabilities of errors arising from the decision scheme under minimal assumptions of global observability and connectedness of the communication graph.

\noindent The algorithms $\mathcal{CIGLRT-NL}$ and $\mathcal{CIGLRT-L}$ are motivated from the bottlenecks that one would encounter from centralized batch processing. To be specific, if a hypothetical fusion center which had access to all the agents' observations at all times were to conduct the parameter estimation in a recursive way, it would do so in the following way:
	\begin{align*}
	\mathbf{\theta}_{c}(t+1)=\mathbf{\theta}_{c}(t)+\underbrace{\alpha_{t}\sum_{n=1}^{N}\mathbf{\nabla}\mathbf{h}_{n}\left(\mathbf{\theta}_{c}(t)\right)\mathbf{\Sigma}_{n}^{-1}\left(\mathbf{y}_{n}(t)-\mathbf{h}_{n}\left(\mathbf{\theta}_{c}(t)\right)\right)}_{\text{Global Innovation}},
	\end{align*}
	where $\left\{\mathbf{\theta}_{c}(t)\right\}$ represents the centralized estimate sequence. Similarly, the centralized decision statistic update can be represented as follows:
	\begin{align*}
	z_{c}(t+1)=\frac{t}{t+1}z_{c}(t)+\underbrace{\frac{1}{t+1}\sum_{n=1}^{N}\log\frac{f_{\theta_{c}(t)}(y_{n}(t))}{f_{0}(y_{n}(t))}}_{\text{Global Innovation}},
	\end{align*}
	where $\left\{z_{c}(t)\right\}$ represents the centralized decision statistic sequence. It is to be noted that the centralized scheme may not be implementable in our distributed multi-agent setting with sparse inter-agent interaction primarily due to the fact that the desired global innovation computation requires instantaneous access to the entire set of network sensed data at all times at a central computing resource.
	If in the case of a distributed setup, an agent $n$ in the network were to replicate the centralized update by replacing the global innovation in accordance with its local innovation, the updates for the parameter estimate and the decision statistic would be as follows:
	\begin{align*}
	\widehat{\mathbf{\theta}}_{n}(t+1)=\widehat{\mathbf{\theta}}_{n}(t)+\underbrace{\alpha_{t}\mathbf{\nabla}\mathbf{h}_{n}\left(\widehat{\mathbf{\theta}}_{n}(t)\right)\mathbf{\Sigma}_{n}^{-1}\left(\mathbf{y}_{n}(t)-\mathbf{h}_{n}\left(\widehat{\mathbf{\theta}}_{n}(t)\right)\right)}_{\text{Local Innovation}},
	\end{align*}
	where $\left\{\widehat{\mathbf{\theta}}_{n}(t)\right\}$ represents the estimate sequence at agent $n$. Similarly, the decision statistic update at agent $n$ would have been:
	\begin{align*}
	\widehat{z}_{n}(t+1)=\frac{t}{t+1}\widehat{z}_{n}(t)+\underbrace{\frac{1}{t+1}\log\frac{f_{\widehat{\theta}_{n}(t)}(y_{n}(t))}{f_{0}(y_{n}(t))}}_{\text{Local Innovation}},
	\end{align*}
	where $\left\{\hat{z}_{n}(t)\right\}$ represents the decision statistic sequence at agent $n$. The above correspond to purely decentralized local processing with no inter-agent collaboration whatsoever. However, note that in absence of local observability both the parameter estimates and decision statistics would be erroneous and sub-optimal.
	Hence, as a surrogate to the global innovation in the centralized recursions, the local estimators compute a local innovation based on the locally sensed data as an agent has access to the information in its neighborhood. However, they intend to compensate for the resulting information loss by incorporating an agreement or consensus potential into their updates in which the individual estimators.

\noindent We, first present the algorithm $\mathcal{CIGLRT-NL}$.

\subsection{Non-linear Observation Models : Algorithm $\mathcal{CIGLRT-NL}$}
\noindent Consider the sensing model described in \eqref{eq:1}. It is to be noted that the formulation assumes no indifference zone\footnote{By the absence of an indifference zone, we mean that $\mathcal{H}_{1}\cup\mathcal{H}_{0}=\mathbb{R}^{M}$. To be specific, there is no zone which has been taken out from the parameter set. We direct the interested reader to \cite{lehmann2006testing} for a more general treatment of indifference zone.}, however, as expected\footnote{Even with an indifference zone, in general, there exists no uniformly most powerful test for the considered vector nonlinear scenario.}, the performance of the proposed distributed approach (i.e., the various error probabilities) under the composite alternative will depend on the specific instance of $\mathbf{\theta}^{*}$ in force.
\noindent We start by making some identifiability assumptions on our sensing model before stating the algorithm.
\begin{myassump}{A1}
	\label{as:1}
	\emph{The sensing model is globally observable, i.e., any two distinct values of $\mathbf{\theta}$ and $\mathbf{\theta}^{*}$ in the parameter space $\mathbb{R}^{M}$ satisfy
		\begin{align}
		\label{eq:as1}
		\sum_{n=1}^{N}\left\|\mathbf{h}_{n}(\mathbf{\theta})-\mathbf{h}_{n}(\mathbf{\theta^{*}})\right\|^{2}=0
		\end{align}
		if and only if $\mathbf{\theta}$ = $\mathbf{\theta}^{*}$.}
\end{myassump}

\label{subsec:cilrt}
\noindent We propose a distributed detector of the \emph{consensus}+\emph{innovations} form for the scenario outlined in \eqref{eq:1}. Before discussing the details of our algorithm, we state an assumption on the inter-agent communication graph.
\begin{myassump}{A2}
	\label{as:2}
	\emph{The inter-agent communication graph is connected, i.e., $\lambda_{2}(\mathbf{L}) > 0$, where $\mathbf{L}$ denotes the associated graph Laplacian matrix}.
\end{myassump}

We now present the distributed $\mathcal{CIGLRT-NL}$ algorithm. The sequential decision procedure consists of three interacting recursive processes operating in parallel, namely, a parameter estimate update process, a decision statistic update process, and a detection decision formation rule, as described below. We state an assumption on the sensing functions before stating the algorithm.

\begin{myassump}{A3}
	\label{as:3}
	\emph{For each agent $n$, $\forall \mathbf{\theta}\neq \mathbf{\theta}_{1}$, the sensing functions $\mathbf{h}_{n}$ are continuously differentiable on $\mathbb{R}^{M}$ and Lipschitz continuous with constants $k_{n} >0$, i.e.,
		\begin{align}
		\label{eq:as2_1}
		\left\|\mathbf{h}_{n}\left(\mathbf{\theta}\right)-\mathbf{h}_{n}\left(\mathbf{\theta}_{1}\right)\right\| \le k_{n}\left\|\mathbf{\theta}-\mathbf{\theta}_{1}\right\|.
		\end{align}}
\end{myassump}

\noindent\textbf{Parameter Estimate Update.}
\noindent The algorithm $\mathcal{CIGLRT-NL}$ generates a sequence $\{\mathbf{\theta}_{n}(t)\} \in \mathbb{R}^{M}$ of estimates of the parameter $\mathbf{\theta}^{*}$ at the $n$-th agent according to the distributed recursive scheme

\begin{align}
\label{eq:est_upt1}
\mathbf{\theta}_{n}(t+1)=\mathbf{\theta}_{n}(t)-\underbrace{\beta_{t}\sum_{l\in\Omega_{n}}\left(\mathbf{\theta}_{n}(t)-\mathbf{\theta}_{l}(t)\right)}_{\text{neighborhood consensus}}+\underbrace{\alpha_{t}\mathbf{\nabla}\mathbf{h}_{n}\left(\mathbf{\theta}_{n}(t)\right)\mathbf{\Sigma}_{n}^{-1}\left(\mathbf{y}_{n}(t)-\mathbf{h}_{n}\left(\mathbf{\theta}_{n}(t)\right)\right)}_{\text{local innovation}},
\end{align}

\noindent where $\Omega_{n}$ denotes the communication neighborhood of agent $n$ and $\nabla\mathbf{h}_{n}\left(.\right)$ denotes the gradient of $\mathbf{h}_{n}$, which is a matrix of dimension $\mathbf{M}\times\mathbf{M}_{n}$, with the $(i,j)$-th entry given by $\frac{\partial \left[\mathbf{h}_{n}\left(\mathbf{\theta}_{n}(t)\right)\right]_{j}}{\partial \left[\mathbf{\theta}_{n}(t)\right]_{i}}$. Finally, $\{\beta_{t}\}$ and $\{\alpha_{t}\}$ are consensus and innovation weight sequences respectively (to be specified shortly).
%where $k(.)=\frac{1}{N}h^{*}(.)$, with $h^{*}(\mathbf{\theta})=\left[h_{1}(\theta)^{\top}\cdots h_{N}(\theta)^{\top}\right]^{\top}$ and %$g_{n}(\mathbf{y})=\left[\mathbf{0}_{M_{1}}^{\top} \mathbf{0}_{M_{2}}^{\top}\cdots\mathbf{y}^{\top}\cdots\mathbf{0}_{M_{N}}^{\top}\right]^{\top}$.

\noindent The update in \eqref{eq:est_upt1} can be written in a compact manner as follows:
\begin{align}
\label{eq:est_upt2}
\theta(t+1)=\theta(t)-\beta_{t}\left(\mathbf{L}\otimes\mathbf{I}_{M}\right)\theta(t)+\alpha_{t}\mathbf{G}(\mathbf{\theta}(t))\mathbf{\Sigma}^{-1}\left(\mathbf{y}(t)-\mathbf{h}\left(\mathbf{\theta}(t)\right)\right),
\end{align}
\noindent where $\theta(t)=\left[\theta_{1}^{\top}(t)\cdots\theta_{N}^{\top}(t)\right]^{\top}$, $\mathbf{h}(\mathbf{\theta}(t))=\left[\mathbf{h}_{1}^{\top}(\mathbf{\theta}_{1}(t))\cdots \mathbf{h}_{N}^{\top}(\mathbf{\theta}_{N}(t))\right]^{\top}$, $\mathbf{\Sigma}^{-1}=\textit{diag}\left[\mathbf{\Sigma}_{1}^{-1},\cdots,\mathbf{\Sigma}_{N}^{-1}\right]$ and \\$\mathbf{G}\left(\mathbf{\theta}(t)\right)=\textit{diag}\left[\nabla\mathbf{h}_{1}\left(\mathbf{\theta}_{1}(t)\right), \cdots, \nabla\mathbf{h}_{N}\left(\mathbf{\theta}_{N}(t)\right)\right]$.

\begin{Remark}
\label{rm:1}
\noindent Note that the parameter estimate update has an innovation term, which has in turn a state dependent innovation gain. The key in analyzing the convergence of distributed stochastic algorithms of the form \eqref{eq:est_upt1}-\eqref{eq:est_upt2} is to obtain conditions that ensure the existence of appropriate stochastic Lyapunov functions. Hence, we propose two conditions on the sensing functions which also involve the state dependent innovation gains that enable the convergence of the distributed estimation procedure, by guaranteeing existence of Lyapunov functions.
\end{Remark}

\begin{myassump}{A4}
\label{as:4}
\emph{There exists a constant $c_{1}>0$ for each pair of $\mathbf{\theta}$ and $\acute{\mathbf{\theta}}$ with $\mathbf{\theta}\neq \acute{\mathbf{\theta}}$ such that the following aggregate strict monotonicity condition holds
\begin{align}
\label{eq:as5_1}
\sum_{n=1}^{N}\left(\mathbf{\theta}-\acute{\mathbf{\theta}}\right)^{\top}\left(\nabla\mathbf{h}_{n}\left(\mathbf{\theta}\right)\right)\mathbf{\Sigma}_{n}^{-1}\left(\mathbf{h}_{n}\left(\mathbf{\theta}\right)-\mathbf{h}_{n}\left(\acute{\mathbf{\theta}}\right)\right) \geq c_{1}\left\|\mathbf{\theta}-\acute{\mathbf{\theta}}\right\|^{2}.
\end{align}}
\end{myassump}
\noindent The distributed parameter estimation algorithm we employ for the algorithm $\mathcal{CIGLRT-NL}$ is recursive in nature and hence, to ensure convergence we need Lyapunov type conditions, which in turn is specified by Assumption \ref{as:4}. Moreover, \ref{as:4} is only a sufficient condition.
For example, in assumption \ref{as:4}, if $h_{n}(.)$'s are linear, i.e., $h_{n}(\theta^{*})=\mathbf{H}_{n}\theta^{*}$, where $\mathbf{H}_{n}\in\mathbb{R}^{M_{n}\times M}$ the monotonicity condition is trivially satisfied by the positive definiteness of the matrix $\sum_{n=1}^{N}\mathbf{H}_{n}^{\top}\mathbf{\Sigma}_{n}^{-1}\mathbf{H}_{n}$.

\noindent We make the following assumption on the weight sequences $\{\alpha_{t}\}$ and $\{\beta_{t}\}$:
\begin{myassump}{A5}
	\emph{\label{as:5}
	The weight sequences $\{\alpha_{t}\}_{t\geq0}$ and $\{\beta_{t}\}_{t\geq0}$ are given by
	\begin{align}
	\label{eq:as6_1}
	\alpha_{t}=\frac{a}{\left(t+1\right)}~\beta_{t}=\frac{b}{\left(t+1\right)^{\tau_{2}}},
	\end{align}
	where $ac_{1}\geq 1$ with $c_{1}$ is as defined in Assumption \ref{as:4}.}
\end{myassump}

\noindent where $0 <\tau_{2} < 1/2, b>0$.

\noindent\textbf{Decision Statistic Update.}
\noindent The algorithm $\mathcal{CIGLRT-L}$ generates a scalar-valued decision statistic sequence $\{z_{n}(t)\}$ at the $n$-th agent according to the distributed recursive scheme
\begin{align}
\label{eq:dec_upt1}
z_{n}(t+1)=\frac{t}{t+1}\left(z_{n}(t)-\underbrace{\delta\sum_{l\in\Omega_{n}}(z_{n}(t)-z_{l}(t))}_{\text{neighborhood consensus}}\right)+\underbrace{\frac{1}{t+1}\log\frac{f_{\theta_{n}(t)}(y_{n}(t))}{f_{0}(y_{n}(t))}}_{\text{local innovation}},
\end{align}

\noindent where $f_{\mathbf{\theta}}(.)$ and $f_{0}(.)$ represent the likelihoods under $\mathcal{H}_{1}$ and $\mathcal{H}_{0}$ respectively,
\begin{align}
\label{eq:dec_upt_l}
\delta\in \left(0, \frac{2}{\lambda_{N}\left(\mathbf{L}\right)}\right),
\end{align}
\noindent and
\begin{align}
\label{eq:dec_upt2}
\log\frac{f_{\theta_{n}(t)}(y_{n}(t))}{f_{0}(y_{n}(t))}=\mathbf{h}_{n}^{\top}(\theta_{n}(t))\mathbf{\Sigma}_{n}^{-1}y_{n}(t)-\frac{\mathbf{h}_{n}^{\top}(\theta_{n}(t))\mathbf{\Sigma}_{n}^{-1}\mathbf{h}_{n}(\theta_{n}(t))}{2},
\end{align}
which follows due to the Gaussian noise assumption in the observation model in \eqref{eq:1}. However, we specifically choose $\delta=\frac{2}{\lambda_{2}\left(\mathbf{L}\right)+\lambda_{N}\left(\mathbf{L}\right)}$ for subsequent analysis.

\noindent The decision statistic update in \eqref{eq:dec_upt1} can be written in a compact manner as follows:
\begin{align}
\label{eq:dec_upt3}
\mathbf{z}(t+1)=\frac{t}{t+1}(\mathbf{I}_{N}-\delta\mathbf{L})\mathbf{z}(t)+\frac{1}{\left(t+1\right)}\mathbf{h}^{*}(\mathbf{\theta}(t))\mathbf{\Sigma}^{-1}\left(\mathbf{y}(t)-\frac{\mathbf{h}(\mathbf{\theta}(t))}{2}\right),
\end{align}
where $\mathbf{z}(t)=\left[z_{1}(t)\cdots z_{N}(t)\right]$, $\mathbf{h}^{*}(\theta(t))=\textit{diag}[\mathbf{h}_{1}^{\top}(\theta_{1}(t)), \mathbf{h}_{2}^{\top}(\theta_{2}(t)),\cdots, \mathbf{h}_{N}^{\top}(\theta_{N}(t))]$, $\mathbf{\Sigma}=diag\left[\mathbf{\Sigma}_{1},\cdots, \mathbf{\Sigma}_{N}\right]$ and $\mathbf{h}(\mathbf{\theta}(t))=\left[\mathbf{h}_{1}^{\top}(\theta_{1}(t))\cdots \mathbf{h}_{N}^{\top}(\theta_{N}(t))\right]^{\top}$.\\
\noindent It is to be noted that $\delta$ is chosen in such a way that $\mathbf{W}=\mathbf{I}_{N}-\delta\mathbf{L}$ is non-negative, symmetric, irreducible and stochastic, i.e., each row of $\mathbf{W}$ sums to one.
\noindent Furthermore, the second largest eigenvalue in magnitude of $\mathbf{W}$, denoted by $r$, is strictly less than one~(see \cite{dimakis2010gossip}). Moreover, by the stochasticity of $\mathbf{W}$, the quantity $r$ satisfies $r=||\mathbf{W}-\mathbf{J}||$, where $\mathbf{J}=\frac{1}{N}\mathbf{1}_{N}\mathbf{1}_{N}^{\top}$.\\

\noindent\textbf{Decision Rule.}
\noindent The following decision rule is adopted at all times $t$ at all agents $n$ :
\begin{align}
\label{eq:dec_rule_1}
\mathcal{H}_{n}(t)=
\begin{cases}
\mathcal{H}_{0} & z_{n}(t) \le \eta\\
\mathcal{H}_{1} & z_{n}(t) > \eta,
\end{cases}
\end{align}
where $\mathcal{H}_{n}(t)$ denotes the local selection (decision) at agent $n$ at time $t$.
\noindent Under the aegis of such a decision rule, the associated probability of errors are as follows:
\begin{align}
\label{eq:dec_rule_2}
&\mathbb{P}_{M,\theta^{*}}(t)=\mathbb{P}_{1,\mathbf{\theta}^{*}}\left(z_{n}(t) \le \eta\right)\nonumber\\
&\mathbb{P}_{FA}(t)=\mathbb{P}_{0}\left(z_{n}(t) > \eta\right),
\end{align}
\noindent where $\mathbb{P}_{M,\theta^{*}}$ and $\mathbb{P}_{FA}$ refer to probability of miss and probability of false alarm respectively. One of the major aims of this paper is to characterize thresholds which ensure that $\mathbb{P}_{M,\theta^{*}}(t), \mathbb{P}_{FA}(t) \to 0$ as $t\to\infty$. We emphasize that, since the alternative $\mathcal{H}_{1}$ is composite, the associated probability of miss is a function of the parameter value $\mathbf{\theta}^{*}$ in force.
\noindent We refer to the parameter estimate update, the decision statistic update and the decision rule in \eqref{eq:est_upt2}, \eqref{eq:dec_upt3} and \eqref{eq:dec_rule_1} respectively, as the $\mathcal{CIGLRT-NL}$ algorithm.
\begin{Remark}
	\label{rm:2}
	It is to be noted that the decision statistic update is recursive and distributed and runs parallelly with the parameter estimate update. Hence, no additional sensing resources are required as in the case of the decision statistic update of the classical GLRT. Owing to the fact that the sensing resources utilized by the parameter estimate update and the decision statistic update are the same, the proposed $\mathcal{CIGLRT-NL}$ algorithm is recursive and online in contrast to the offline batch processing nature of the classical GLRT. However, with the initial parameter estimates being incorporated into the decision statistic makes it sub-optimal with respect to the classical GLRT decision statistic as the initial parameter estimates may be inaccurate. As, we will show later in spite of the sub-optimality with respect to the classical GLRT, the algorithm guarantees reasonable
	detection performance with the probabilities of errors decaying to 0 asymptotically in the large sample limit. Another useful distributed parameter estimation approach is the diffusion approach (see, for example \cite{cattivelli2010diffusion,cattivelli2011distributed})~in which constant weights are employed for incorporating the neighborhood information and the latest local sensed information. The constant weights in diffusion implementations enable potential adaptation and help in tracking adaptive systems. The price that one has to pay though for adaptivity is in terms of inconsistency of the statistics generated from the procedure thereby resulting in residual errors.  If constant weights are used for the consensus and innovation terms, the parameter estimation update scheme and the decision statistic update scheme will be affected by inconsistent estimates of the parameter of interest and the decision statistic respectively. As such, the log likelihood ratios incorporated into the decision statistic are sub-optimal as they are evaluated at the latest parameter estimate. The parameter estimate is in general possibly erroneous to a large extent in the initial time instants which in turn also potentially affects the $\mathcal{CIGLRT-NL}$ algorithm. On top of this, if the parameter estimate sequence is derived from an inconsistent procedure, the degree of sub-optimality could possibly be magnified. The further degree of sub-optimality would be due to estimate sequences generated from the estimate update with constant weights being inconsistent and having a steady state error. In particular, this might affect the time asymptotics of the probabilities of errors. In diffusion-based implementations, the residual error is a function of the constant step size and, as shown in \cite{matta2014diffusion,matta2015exact,braca2014large,zou2010cooperative,matta2016distributed}, goes to zero in the limit of small step size. Therefore, in order to characterize a GLRT-type diffusion implementation, it would be necessary to characterize the interplay between the estimation error and the detection error, as functions of the vanishing step-size. 
\end{Remark}
\subsection{Linear Observation Models : Algorithm $\mathcal{CIGLRT-L}$}
\label{subsec:lo-cilrt}
\noindent In this section, we develop the algorithm $\mathcal{CIGLRT-L}$ for linear observation models which lets us specifically characterize the large deviations exponent upper bounds for probability of miss and probability of false alarm.

\noindent There are $N$ agents deployed in the network. Every agent $n$ at time index $t$ makes a noisy observation $\mathbf{y}_{n}(t)$, a noisy function of $\theta^{*}$ which is a $M$-dimensional parameter. Formally the observation model for the $n$-th agent is given by,
\begin{align}
\label{eq:sens_ob}
\mathbf{y}_{n}(t)=\mathbf{H}_{n}\theta^{*}+\mathbf{\gamma}_{n}(t),
\end{align}
\noindent where $\{\mathbf{y}_{n}(t)\} \in \mathbb{R}^{M_{n}}$ is the observation sequence for the $n$-th agent and $\{\mathbf{\gamma}_{n}(t)\}$ is a zero mean temporally i.i.d Gaussian noise sequence at the $n$-th agent with nonsingular covariance $\mathbf{\Sigma}_{n}$, where $\mathbf{\Sigma}_{n}\in\mathbb{R}^{M_{n}\times M_{n}}$. The noise processes are independent across different agents.
\noindent If $M$ is large, in practical applications each agent's observations may only correspond to a subset of the components of $\theta^{*}$, with $M_{n} << M$, which basically renders the parameter of interest $\theta^{*}$ locally unobservable at each agent. Under local unobservability, in isolation, an agent cannot estimate the entire parameter. However under appropriate observability conditions, it may be possible for each agent to get a consistent estimate of $\theta^{*}$. Moreover, depending on as to which hypothesis is in force, the observation model is formalized as follows:
\begin{align}
\label{eq:sens_ob_hyp}
&\mathcal{H}_{1} : \mathbf{y}_{n}(t)=\mathbf{H}_{n}\theta^{*} + \mathbf{\gamma}_{n}(t) \nonumber\\
&\mathcal{H}_{0} : \mathbf{y}_{n}(t)=\mathbf{\gamma}_{n}(t).
\end{align}

\noindent We formalize the assumptions on the inter-agent communication graph and global observability.
\begin{myassump}{B1}
	\label{bs:1}
	\emph{We require the following global observability condition. The matrix $\mathbf{G}$}
	\begin{align}
	\label{eq:global_obs_1}
	\mathbf{G}=\sum_{n=1}^{N}\mathbf{H}_{n}^{\top}\mathbf{\Sigma}_{n}^{-1}\mathbf{H}_{n}
	\end{align}
	\emph{is full rank}.
\end{myassump}
\begin{Remark}
	\label{rm:1a}
	It is to be noted that Assumption~\ref{as:1} reduces to Assumption~\ref{bs:1} for linear models, i.e., by taking $\mathbf{h}_{n}\left(\mathbf{\theta}^{*}\right)=\mathbf{H}_{n}\mathbf{\theta}^{*}$.
\end{Remark}

\begin{myassump}{B2}
	\label{bs:2}
	\emph{The inter-agent communication graph is connected, i.e., $\lambda_{2}(\mathbf{L}) > 0$, where $\mathbf{L}$ denotes the associated graph Laplacian matrix}.
\end{myassump}

\noindent\textbf{Algorithm $\mathcal{CIGLRT-L}$}\\

\noindent The algorithm $\mathcal{CIGLRT-L}$ consists of three parts, namely, parameter estimate update, decision statistic update and the decision rule.\\
\textbf{Parameter Estimate Update.}
\noindent The algorithm $\mathcal{CIGLRT-L}$ generates a sequence $\{\theta_{n}(t)\}\in\mathbb{R}^{M}$ which are estimates of $\mathbf{\theta}^{*}$ at the $n$-th agent according to the following recursive scheme
\begin{align}
\label{eq:est_upt_lin_1}
\theta_{n}(t+1)=\theta_{n}(t)-\underbrace{\beta_{t}\sum_{l\in\Omega_{n}}(\theta_{n}(t)-\theta_{l}(t))}_{\text{neighborhood consensus}}+\underbrace{\alpha_{t}\nabla_{\theta}\log\frac{f_{\theta_{n}(t)}(\mathbf{y}_{n}(t))}{f_{0}(\mathbf{y}_{n}(t))}}_{\text{local innovation}},
\end{align}
\noindent where $\Omega_{n}$ denotes the communication neighborhood of agent $n$, $\nabla\left( . \right)$ denotes the gradient and $\{\beta_{t}\}$ and $\{\alpha_{t}\}$ are consensus and innovation weight sequences respectively (to be specified shortly) and
\begin{align}
\label{eq:est_upt_lin_2}
\log\frac{f_{\theta_{n}(t)}(\mathbf{y}_{n}(t))}{f_{0}(\mathbf{y}_{n}(t))}=\theta_{n}(t)^{\top}\mathbf{H}_{n}^{\top}\mathbf{\Sigma}_{n}^{-1}\mathbf{y}_{n}(t)-\frac{\theta_{n}(t)^{\top}\mathbf{H}_{n}^{\top}\mathbf{\Sigma}_{n}^{-1}\mathbf{H}_{n}\theta_{n}(t)}{2}.
\end{align}
It is to be noted that the parameter estimation update of the $\mathcal{CIGLRT-L}$ algorithm is a special case of the $\mathcal{CIGLRT-NL}$ algorithm with $\mathbf{h}_{n}(\theta^{*})=\mathbf{H}_{n}\theta^{*}$.\\
\noindent The update in \eqref{eq:est_upt_lin_1} can be written in a compact manner as follows:
\begin{align}
\label{eq:est_upt_lin3}
\mathbf{\theta}(t+1)=\mathbf{\theta}(t)-\beta_{t}(\mathbf{L}\otimes \mathbf{I}_{M})\mathbf{\theta}(t)+\alpha_{t}\mathbf{G}_{H}\mathbf{\Sigma}^{-1}\left(\mathbf{y}(t)-\mathbf{G}_{H}^{\top}\mathbf{\theta}(t)\right),
\end{align}
\noindent where $\mathbf{\theta}(t)=[\theta_{1}^{\top}(t)~\theta_{2}^{\top}(t)\cdots \theta_{N}^{\top}(t)]^{\top}$, $\mathbf{G}_{H}=diag[\mathbf{H}_{1}^{\top}, \mathbf{H}_{2}^{\top},\cdots, \mathbf{H}_{N}^{\top}]$, $\mathbf{y}(t)=[\mathbf{y}_{1}^{\top}(t) ~\mathbf{y}_{2}^{\top}(t) \cdots \mathbf{y}_{N}^{\top}(t)]^{\top}$ and $\mathbf{\Sigma}=diag\left[\mathbf{\Sigma}_{1},\cdots,\mathbf{\Sigma}_{N}\right]$. \\

\noindent We make the following assumptions on the weight sequences $\{\alpha_{t}\}$ and $\{\beta_{t}\}$.
\begin{myassump}{B3}
	\label{bs:3}
	\emph{The weight sequences $\{\alpha_{t}\}$ and $\{\beta_{t}\}$ are of the form
		\begin{align}
		\label{eq:ab_bs3}
		\alpha_{t}=\frac{a}{(t+1)}~\beta_{t}=\frac{a}{(t+1)^{\delta_{2}}},
		\end{align}
		where $a\geq1$ and $0 < \delta_{2}\le 1$.}
\end{myassump}
\noindent\textbf{Decision Statistic Update.} The algorithm $\mathcal{CIGLRT-L}$ generates a decision statistic sequence $\{z_{n}(t)\}$ at the $n$-th agent according to the distributed recursive scheme
\begin{align}
\label{eq:dec_stat_lin_1}
\hat{z}_{n}(kt-k+1)=\mathbf{\theta}_{n}(k(t-1))^{\top}\mathbf{H}_{n}^{\top}\mathbf{\Sigma}_{n}^{-1}\left(\mathbf{s}_{n}(k(t-1))-\frac{\mathbf{H}_{n}\mathbf{\theta}_{n}(k(t-1))}{2}\right),
\end{align}
\noindent where $\mathbf{s}_{n}(k(t-1))=\sum_{i=0}^{k(t-1)}\frac{\mathbf{y}_{n}(i)}{k(t-1)+1}$, i.e., the time averaged sum of local observations at agent $n$, and the underlying parameter estimate used in the test statistic is the estimate at time $k(t-1)$. In other words, at every time instant $k(t-1)+1$ (times which are one modulo $k$), where $k$ is a pre-determined positive integer~($k$ to be specified shortly), an agent $n$, incorporates its local observations made in the past $k$ time instants, in the above mentioned manner in \eqref{eq:dec_stat_lin_1}. It is to be noted that, independent of the decision statistic update, $\mathbf{s}_{n}(k(t-1))$ is updated as and when a new observation is made at agent $n$.
After incorporating the local observations, every agent $n$ undergoes $k-1$ rounds of consensus, which can be expressed in a compact form as follows:
\begin{align}
\label{eq:dec_stat_lin_2}
\hat{\mathbf{z}}(kt)=\mathbf{W}^{k-1}\mathbf{G}_{\theta}(k(t-1))\mathbf{\Sigma}^{-1}\left(\mathbf{s}(k(t-1))-\frac{\mathbf{G}_{H}^{\top}\mathbf{\theta}(k(t-1))}{2}\right),
\end{align}
\noindent where $\mathbf{z}(t)=\left[z_{1}(t)\cdots z_{N}(t)\right]$, $\mathbf{G}_{\theta}(t)=diag\left[\theta_{1}^{\top}(t)\mathbf{H}_{1}^{\top}, \theta_{2}^{\top}(t)\mathbf{H}_{2}^{\top}, \cdots, \theta_{N}^{\top}(t)\mathbf{H}_{N}^{\top}\right]$ and $\mathbf{s}(t)=\left[\mathbf{s}_{1}^{\top}(t)~\mathbf{s}_{2}^{\top}(t)\cdots \mathbf{s}_{n}^{\top}(t)\right]^{\top}$, where $\mathbf{W}$ is a $N\times N$ weight matrix, where we assign $w_{ij}=0$, if $(i,j)\notin E$.
The sequence $\left\{\hat{z}_{n}(t)\right\}$ is an auxiliary sequence and the decision statistic sequence $\left\{z_{n}(t)\right\}$ is generated from the auxiliary sequence in the following way:
\begin{align}
\label{eq:dec_stat_lin_12}
z_{n}(kt)=\hat{z}_{n}(kt), \forall  t,
\end{align}
where as in the interval $\left[k(t-1),kt-1\right]$, the value of the decision statistic stays constant corresponding to its value at $z_{n}(kt-k),~\forall t$.
\begin{Remark}
	\label{rm:2_1}
	The $\mathcal{CIGLRT-NL}$ algorithm is for general non-linear observations with a very intuitive decision statistic update which seeks to track the time average of the log-likelihood ratios over time. The $\mathcal{CIGLRT-NL}$ algorithm can be extended to general linear models but the linear model version of the $\mathcal{CIGLRT-NL}$ algorithm is different from the $\mathcal{CIGLRT-L}$. In particular, the results of this paper are inconclusive as to whether the linear model version of $\mathcal{CIGLRT-NL}$ algorithm is able to achieve exponential decay in terms of the probability of miss or not. This is why we introduce the $\mathcal{CIGLRT-L}$ algorithm for the specific linear case for which we are able to establish exponential decays for both the probabilities of miss and false alarm. The algorithms are different.
	Intuitively speaking, the performance of the $\mathcal{CIGLRT-NL}$ algorithm in terms of decay of the probability of miss is affected as there is no mechanism in the decision statistic update so as to get rid of the initial bad parameter estimates, which are weighed equally as the later more accurate parameter estimates. The decision statistics update for the $\mathcal{CIGLRT-L}$ algorithm however ensures that at any time after $kt$, the previous parameter estimates i.e., $\theta(ks), s=1,\cdots,t-1$ do not contribute to the decision statistic. The key observation which lets us prove the exponential decay for $\mathcal{CIGLRT-L}$ can be best explained by term $(t1)$ in the probability of miss characterization in \eqref{eq:g2}. Due to the nature of the decision statistic update, for the $\mathcal{CIGLRT-L}$ it can be shown that $t\left\|\mathbf{\theta}(t)-\tn\right\|^{2}=\frac{t\mathbf{P}_{t}}{2}\mathbf{\gamma}_{G,t}\mathbf{\gamma}_{G,t}^{\top}$, where
	\begin{align*}
	\mathbf{\gamma}_{G,t}=[\gg^{\top}(0)~\gg^{\top}(1)~\cdots~\gg^{\top}(t-1)]^{\top},
	\end{align*}
	\noindent where $\gg(t)=\mathbf{G}_{H}\mathbf{\Sigma}^{-1}\mathbf{\gamma}(t)$ and $\mathbf{P}_{t}$ is a block matrix of dimension $NMt\times NMt$, whose $(i,j)$-th block $i,j=0,\cdots,t-1$ is given as follows:
	\begin{align*}
	\left[\mathbf{P}_{t}\right]_{ij}=\alpha_{i}\alpha_{j}\prod_{u=0}^{t-2-i}\mathbf{A}(t-1-u)\prod_{v=j+1}^{t-1}\mathbf{A}(v),
	\end{align*}
	where $\mathbf{A}(t)=\mathbf{I}_{NM}-\beta_{t}\left(\mathbf{L}\otimes\mathbf{I}_{M}\right)-\alpha_{t}\mathbf{G}_{H}\mathbf{\Sigma}^{-1}\mathbf{G}_{H}^{\top}$.
	The matrix $t\left\|\mathbf{P}_{t}\right\|$ can be bounded as shown in Lemma \ref{wish}. For the $\mathcal{CIGLRT-NL}$ though, instead of $t\left\|\mathbf{\theta}(t)-\tn\right\|^{2}$, we have to deal with $\sum_{s=1}^{t}\left\|\mathbf{\theta}(s)-\tn\right\|^{2}$ which in turn does not stay bounded as $t\rightarrow\infty$. To sum it up, $\mathcal{CIGLRT-L}$ is indeed a different algorithm from $\mathcal{CIGLRT-NL}$ with a carefully designed decision statistic update so as to be able to characterize the exponential decay of both the probabilities of error.
\end{Remark}

\noindent Now we state some design assumptions on the weight matrix $\mathbf{W}$.
\begin{myassump}{B4}
	\label{bs:4}
	\emph{The entries in the weight matrix $\mathbf{W}$ are designed in such a way that $\mathbf{W}$ is non-negative, symmetric, irreducible and stochastic, i.e., each row of $\mathbf{W}$ sums to one}.
\end{myassump}

\noindent We remark that, if Assumption \ref{bs:4} is satisfied, then the second largest eigenvalue in magnitude of $\mathbf{W}$, denoted by $r$, turns out to be strictly less than one, see for example \cite{dimakis2010gossip}.  Note that, by the stochasticity of $\mathbf{W}$, the quantity $r$ satisfies
\begin{align}
\label{eq:dec_stat_lin_3}
r=||\mathbf{W}-\mathbf{J}||,
\end{align}
\noindent where $\mathbf{J}=\frac{1}{N}\mathbf{1}_{N}\mathbf{1}_{N}^{\top}$.

\noindent A intuitive way to design $\mathbf{W}$ is to assign equal combination weights, in which case we have,
\begin{align}
\label{eq:dec_stat_lin_4}
\mathbf{W}=\mathbf{I}_{N}-\delta\mathbf{L},
\end{align}
\noindent where $\delta\in\left(0,\frac{2}{\lambda_{N}(\mathbf{L})}\right)$. For subsequent analysis, we specifically choose $\delta=\frac{2}{\lambda_{2}(\mathbf{L})+\lambda_{N}(\mathbf{L})}$. \\
\textbf{Decision Rule.}
\noindent The following decision rule is adopted at all times $t$ :
\begin{align}
\label{eq:dec_rule_lin_1}
\mathcal{H}_{n}(t)=
\begin{cases}
\mathcal{H}_{0} & z_{n}(t) \le \eta\\
\mathcal{H}_{1} & z_{n}(t) > \eta,
\end{cases}
\end{align}
where $\mathcal{H}_{n}(t)$ is the local decision at time $t$ at agent $n$.
\begin{Remark}
	\label{rm:3.0}
For both the proposed algorithms the agents reach asymptotic agreement or consensus in terms of the parameter estimate and the decision statistic. The decisions in the initial few time steps might be different. But, with subsequent cooperation among the agents, each agent gets to the same local decision eventually with probability one. Hence, the overall decision then is the decision of any local agent. It is to be noted that, to reach decision consensus the agents need to reach consensus on the indicator function with respect to the threshold. However, the indicator function $\mathbb{I}_{\left\{z_{n}(t)>\eta\right\}}$ is discontinuous at the threshold. But, from Theorems 4.2 and 4.3, we have that the decision statistics not only reach consensus but converge to $\frac{\mathbf{h}^{\top}\left(\mathbf{\theta}_{N}^{*}\right)\mathbf{\Sigma}^{-1}\mathbf{h}\left(\mathbf{\theta}_{N}^{*}\right)}{2N}$ in expectation under $\mathcal{H}_{1}$ and $0$ under $\mathcal{H}_{0}$ and hence the threshold is chosen in such a way that the decision statistics of different agents reach consensus to a value strictly different from the threshold so that the indicator function $\mathbb{I}_{\left\{z_{n}(t)>\eta\right\}}$ is continuous at the limiting consensus value. This ensures in turn that the binary decisions at the agents also reach consensus.
\end{Remark}
\noindent Under the aegis of such a decision rule, the associated probability of errors are as follows:
\begin{align}
\label{eq:dec_rule_lin_2}
&\mathbb{P}_{M,\theta^{*}}(t)=\mathbb{P}_{1,\mathbf{\theta}^{*}}\left(z_{n}(t) \le \eta\right)\nonumber\\
&\mathbb{P}_{FA}(t)=\mathbb{P}_{0}\left(z_{n}(t) > \eta\right),
\end{align}
\noindent where $\mathbb{P}_{M,\theta^{*}}$ and $\mathbb{P}_{FA}$ refer to probability of miss and probability of false alarm respectively. \noindent In Section \ref{subsec:main_res_clirt}, we not only characterize thresholds which ensure that $\mathbb{P}_{M,\theta^{*}}(t), \mathbb{P}_{FA}(t) \to 0$ as $t\to\infty$ but also derive the large deviations exponent upper bounds for $\mathbb{P}_{M,\theta^{*}}(t)$ and  $\mathbb{P}_{FA}(t)$.
\begin{Remark}
	\label{rm:3}
Note that, the decision statistic update requires the agents to store a copy of the running time-average of their observations. The additional memory requirement to store the running average stays constant, as the average $\mathbf{s}_{n}(t)$, say for agent $n$, can  be updated recursively. It is to be noted that the decision statistic update in \eqref{eq:dec_stat_lin_2} has time-delayed parameter estimates and observations, i.e., delayed in the sense, in the ideal case the decision statistic update at a particular time instant, say $t$, would be using the parameter estimate at time $t$, but owing to the $k$ rounds of consensus, the algorithm uses parameter estimates which are delayed by $k$ time steps. Whenever, the $k$ rounds of consensus are done with, the algorithm incorporates its latest estimates and observations into decision statistics at respective agents. After the $k$ rounds of consensus, it is ensured that with inter-agent collaboration, the decision statistic at each agent attains more accuracy. Hence, there is an inherent trade-off between the performance (number of rounds of consensus) and the time delay. If the number of rounds of consensus is increased, the algorithm attains better detection performance asymptotically (the error probabilities have larger exponents), but at the same time the time lag in incorporating the latest sensed information into the decision statistic increases affecting possibly transient characteristics and vice-versa.
\end{Remark}

\noindent We make an assumption on $k$ which concerns with the number of rounds of consensus in the decision statistic update of $\mathcal{CIGLRT-L}$.
\begin{myassump}{B5}
	\label{bs:5}
	\emph{Recall $r$ as defined in \eqref{eq:dec_stat_lin_3}.
		The number of rounds $k$ of consensus between two updates of agent decision statistics satisfies
		\begin{align}
		\label{eq:k_bs4}
		k \geq 1+\left\lfloor\frac{-3\log N}{2\log r}\right\rfloor.
		\end{align}}
\end{myassump}

\noindent We make an assumption on $a$, which is in turn defined in \eqref{eq:ab_bs3}.
\begin{myassump}{B6}
	\label{bs:6}
\noindent Recall $a$ as defined in Assumption \ref{bs:3}.
We assume that $a$ satisfies
\begin{align}
\label{eq:bs5_1}
a \ge \frac{1}{2c_{1}}+2,
\end{align}
\noindent where $c_{1}$\footnote{We will later show that $c_{1}$ is strictly greater than zero.} is defined as
\begin{align}
\label{eq:bs5_2}
c_{1}=\min_{\left\|\mathbf{x}\right\|=1}\mathbf{x}^{\top}\left(\mathbf{L}\otimes\mathbf{I}_{M}+\mathbf{G}_{H}\mathbf{\Sigma}^{-1}\mathbf{G}_{H}^{\top}\right)\mathbf{x}=\lambda_{\mbox{\scriptsize{min}}}\left(\mathbf{L}\otimes\mathbf{I}_{M}+\mathbf{G}_{H}\mathbf{\Sigma}^{-1}\mathbf{G}_{H}^{\top}\right).
\end{align}
%whereas $t_{1}=\max\{t_{2},t_{3}\}$ and $t_{3}$ is given by,
%\begin{align}
%\label{eq:bs5_3}
%t_{3} = \left(\frac{N\lambda_{\mbox{\scriptsize{max}}}^{2}\left(\mathbf{G}_{H}\mathbf{\Sigma}^{-1}\mathbf{G}_{H}^{\top}\right)}{\beta_{0}\lambda_{\mbox{\scriptsize{min}}}\left(\mathbf{G}\right)\lambda_{2}\left(\mathbf{L}\right)}\right)^{\frac{1}{1-\delta_{1}}}-1,
%\end{align}
%$t_{2}$\footnote{Such a $t_{2}$ always exists as $\alpha_{t}, \beta_{t}\to 0$ with $t\to\infty$.} is such that, $\forall t\geq t_{2}$,
%\begin{align}
%\label{eq:bs5_4}
%\beta_{t}\lambda_{N}(\mathbf{L})+\alpha_{t}\lambda_{\mbox{\scriptsize{max}}}(\mathbf{G}_{H}\mathbf{\Sigma}^{-1}\mathbf{G}_{H}^{\top}) < 1,
%\end{align}
%where $\lambda_{\mbox{\scriptsize{max}}}\left(.\right)$ and $\lambda_{\mbox{\scriptsize{min}}}\left(.\right)$ represent the largest eigenvalue and the smallest eigenvalue respectively.
\end{myassump}

\section{Main Results}
\label{sec:main_res}

\noindent We formally state the main results in this section. We further divide this section into two subsections. The first subsection
caters to the consistency of the parameter estimate update and the analysis of the detection performance of algorithm $\mathcal{CIGLRT-NL}$, whereas the next subsection is concerned with the consistency of the parameter estimate update and the characterization of the large deviations exponent upper bounds for the algorithm $\mathcal{CIGLRT-L}$.
\subsection{Main Results : $\mathcal{CIGLRT-NL}$}
\noindent In this section, we provide the main results concerning the algorithm $\mathcal{CIGLRT-NL}$, while the proofs are provided in Section \ref{sec:proof_res_ciglrt}.
\label{subsec:main_res_ciglrt}
\begin{Theorem}
	\label{t1}
	\noindent Consider the $\mathcal{CIGLRT-NL}$ algorithm under Assumptions \ref{as:1}-\ref{as:5} with the additional assumption that $ac_{1}\geq 1$ with $c_{1}$ as defined in Assumption \ref{as:4}, and the sequence $\{\mathbf{\theta}(t)\}_{t\geq 0}$ generated according to \eqref{eq:est_upt2}. We then have
	\begin{align}
	\label{eq:t1_1}
	\mathbb{P}_{\mathbf{\theta}^{*}}\left(\lim_{t\to\infty}(t+1)^{\tau}\left\|\mathbf{\theta}_{n}(t)-\mathbf{\theta}^{*}\right\|=0, \forall 1\leq n \leq N\right)=1,
	\end{align}
	for all $\tau \in [0, 1/2)$.	
\end{Theorem}

\noindent To be specific, the estimate sequence $\{\mathbf{\theta}_{n}(t)\}_{t\geq 0}$ at agent $n$ is strongly consistent. Moreover, we also have that the convergence in Theorem \ref{t1} is order optimal, in the sense that results in estimation theory show that in general for the considered setup there is no centralized estimator $\{\hat{\mathbf{\theta}}(t)\}$ for which $(t+1)^{\tau}\left\|\hat{\mathbf{\theta}}(t)-\mathbf{\theta}^{*}\right\|\to 0$ a.s. as $t\to\infty$ for $\tau\geq1/2$. General nonlinear distributed parameter estimation procedures of the consensus + innovations form as in \eqref{eq:est_upt1} have been developed and investigated in~\cite{kar2014asymptotically}. The proof of Theorem \ref{t1} is inspired and follows similar arguments as in~\cite{kar2014asymptotically}, however, the specific state-dependent form of the innovation gains employed in \eqref{eq:est_upt1} requires a subtle modification of the arguments in~\cite{kar2014asymptotically}. The complete proof of Theorem \ref{t1} is provided in Section \ref{sec:proof_res_ciglrt}. In a sense, Theorem \ref{t1} extends the consensus + innovations framework~\cite{kar2014asymptotically} to the case of state-dependent innovation gains.

\noindent We, now state a result which characterizes the asymptotic normality of the decision statistic sequence $\{z_{n}(t)\}$ at every agent $n$.

\begin{Theorem}
	\label{t2}
	\noindent Consider the $\mathcal{CIGLRT-NL}$ algorithm under Assumptions \ref{as:1}-\ref{as:5}, and the sequence $\{\mathbf{z}(t)\}$ generated according to \eqref{eq:dec_upt3}. We then have under $\mathbb{P}_{\theta^{*}}$, for all $\left\|\mathbf{\theta}^{*}\right\| > 0$,
	\begin{align}
	\label{eq:t2_1}
	\sqrt{t+1}\left(z_{n}(t)-\frac{\mathbf{h}^{\top}\left(\mathbf{\theta}_{N}^{*}\right)\mathbf{\Sigma}^{-1}\mathbf{h}\left(\mathbf{\theta}_{N}^{*}\right)}{2N}\right)\overset{\mathcal{D}}{\Longrightarrow} \mathcal{N}\left(0, \frac{\mathbf{h}^{\top}\left(\mathbf{\theta}_{N}^{*}\right)\mathbf{\Sigma}^{-1}\mathbf{h}\left(\mathbf{\theta}_{N}^{*}\right)}{N^{2}}\right), \forall n
	\end{align}
	
	\noindent where $\mathbf{\theta}_{N}^{*}=\mathbf{1}_{N}\otimes\theta^{*}$, $\mathbf{h}\left(\mathbf{\theta}_{N}^{*}\right)=\left[\mathbf{h}_{1}^{\top}\left(\mathbf{\theta}^{*}\right)\cdots\mathbf{h}_{N}^{\top}\left(\mathbf{\theta}^{*}\right)\right]^{\top}$ and $\overset{\mathcal{D}}{\Longrightarrow}$ refers to convergence in distribution (weak convergence).
\end{Theorem}

\noindent The next result concerns with the characterization of thresholds which ensures the probability of miss and probability of false alarm as defined in \eqref{eq:dec_rule_2} go to zero asymptotically.
\begin{Theorem}
	\label{t3}
	Let the hypotheses of Theorem \ref{t2} hold. Consider the decision rule defined in \eqref{eq:dec_rule_1}.
	For all $\mathbf{\theta}^{*}$ which satisfy
	\begin{align}
	\label{eq:t3_2}
	\frac{\mathbf{h}^{\top}\left(\mathbf{\theta}_{N}^{*}\right)\mathbf{\Sigma}^{-1}\mathbf{h}\left(\mathbf{\theta}_{N}^{*}\right)}{2N} > \frac{\left(\frac{1}{N}+\sqrt{N}r\right)\sum_{n=1}^{N}M_{n}}{2},
	\end{align}
	
	we have the following choice of the thresholds
	\begin{align}
	\label{eq:t3_1}
	\frac{\left(\frac{1}{N}+\sqrt{N}r\right)\sum_{n=1}^{N}M_{n}}{2}<\eta<\frac{\mathbf{h}^{\top}\left(\mathbf{\theta}_{N}^{*}\right)\mathbf{\Sigma}^{-1}\mathbf{h}\left(\mathbf{\theta}_{N}^{*}\right)}{2N},
	\end{align}
	for which we have that $\mathbb{P}_{M,\theta^{*}}(t)\to 0$ and $\mathbb{P}_{FA}(t)\to 0$ as $t\to\infty$. Specifically, $\mathbb{P}_{FA}(t)$ decays to zero exponentially with the following large deviations exponent upper bound
	\begin{align}
	\label{eq:t3_3}
	\limsup_{t\to\infty}\frac{1}{t}\log\left(\mathbb{P}_{0}\left(z_{n}(t)>\eta\right)\right)\le -LE\left(\min\{\lambda^{*},1\}\right),
	\end{align}
	where $\theta_{N}^{*}=\mathbf{1}_{N}\otimes\theta^{*}$, $LE(.)$ and $\lambda^{*}$ are given by
	\begin{align}
	\label{eq:t3_4}
	& LE(\lambda)=\frac{\eta\lambda}{\frac{1}{N}+\sqrt{N}}+\left(\frac{\sum_{n=1}^{N}M_{n}}{2}\right)\log\left(1-\frac{\lambda\left(\frac{1}{N}+\sqrt{N}r\right)}{\frac{1}{N}+\sqrt{N}}\right),\nonumber\\
	& \lambda^{*}=\frac{\frac{1}{N}+\sqrt{N}}{\frac{1}{N}+\sqrt{N}r}-\frac{\left(\frac{1}{N}+\sqrt{N}\right)\sum_{n=1}^{N}M_{n}}{2\eta}.
	\end{align}
\end{Theorem}
\noindent We discuss how the above result can be used in practice to identify thresholds that lead to asymptotic decay of the probabilities of error (exponential decay for $\mathbb{P}_{FA}$). It is to be noted that as the observation parameters, i.e., $M_{n},  N$ and the connectivity of the communication graph, i.e., $r$ are known apriori, the threshold can be chosen to be $\frac{\left(\frac{1}{N}+\sqrt{N}r^{k-1}\right)\sum_{n=1}^{N}M_{n}}{2}+\epsilon$, where $\epsilon$ can be chosen to be arbitrarily small. This would guarantee exponential decay for the probability of false alarm. Further, from the feasible range of thresholds in \eqref{eq:t3_1}, a range on the $\theta^{*}$s' can be obtained in terms of $\left\|\mathbf{h}\left(\mathbf{1}_{N}\otimes\theta^{*}\right)\right\|$ such that under $\mathbb{H}_{1}$, as long as the true value $\theta^{*}$ of the parameter belongs to this range, the probability of miss is guaranteed to decay to zero asymptotically. It is important to note in this context that there exists some weak signals, i.e., signals with low $\left\|\mathbf{h}\left(\mathbf{1}_{N}\otimes\theta^{*}\right)\right\|$ (but non-zero), for which there may not exist a choice of thresholds to ensure asymptotically decaying probability of miss. The signals for which Theorem \ref{t3} is rendered to be inconclusive in the manner described above, can be categorized in terms of $\theta^{*}$. From, \eqref{eq:t3_2} we have that,
\begin{align}
\label{eq:t3_explain_0}
\lambda_{\mbox{\scriptsize{max}}}\left(\mathbf{\Sigma}^{-1}\right)\left\|\mathbf{h}\left(\mathbf{\theta}_{N}^{*}\right)\right\|^{2}\ge\mathbf{h}^{\top}\left(\mathbf{\theta}_{N}^{*}\right)\mathbf{\Sigma}^{-1}\mathbf{h}\left(\mathbf{\theta}_{N}^{*}\right) > \left(1+N\sqrt{N}r\right)\sum_{n=1}^{N}M_{n}.
\end{align}

\noindent Hence, a sufficient condition can be obtained in terms of $\theta^{*}$s' as follows

\begin{align}
\label{eq:t_3_explain}
\left\|\mathbf{h}\left(\mathbf{\theta}_{N}^{*}\right)\right\|^{2}<\frac{\left(1+N\sqrt{N}r\right)\sum_{n=1}^{N}M_{n}}{\lambda_{\mbox{\scriptsize{max}}}\left(\mathbf{\Sigma}^{-1}\right)},
\end{align}
where $\lambda_{\mbox{\scriptsize{max}}}(.)$ denotes the smallest eigenvalue, render Theorem \ref{t3} to be inconclusive.\\
\noindent Theorem \ref{t3} ensures that the $\mathcal{CIGLRT-NL}$ algorithm, in spite of being a sub-optimal algorithm with respect to the classical GLRT, has asymptotically decaying probability of errors in the large sample limit and also characterizes the feasible choice of thresholds for which such a decay is possible. The potential sub-optimality in the algorithm with respect to the classical GLRT is due to the possibly inaccurate initial estimates of the underlying parameter being incorporated into the decision statistic, while the classical GLRT uses the maximum likelihood estimate to generate its decision statistic. However, in order to incorporate the maximum likelihood estimate into the decision statistic, the classical GLRT is essentially an offline batch processing algorithm, while the $\mathcal{CIGLRT-NL}$ algorithm is an online algorithm. In the sequel, we analyze the algorithm $\mathcal{CIGLRT-L}$ which provides for the exponential decay of the probabilities of errors.
\subsection{Main Results : $\mathcal{CIGLRT-L}$}
\label{subsec:main_res_clirt}
\noindent In this section, we provide the main results concerning the consistency of the parameter estimate update and the characterization of large deviations exponent upper bounds of $\mathcal{CIGLRT-L}$, whereas the proofs are provided in Section \ref{sec:proof_res_cilrt}.
\noindent We first state the results concerning the consistency of the parameter estimation part of the $\mathcal{CIGLRT-L}$ algorithm, which essentially follows from \cite{KarMoura-LinEst-JSTSP-2011}.
\begin{Theorem}[\hspace{-0.5pt}\cite{KarMoura-LinEst-JSTSP-2011}]
	\label{th:b1}
	Let the Assumptions~\ref{bs:1}-\ref{bs:5} hold. Consider the parameter estimation part of the $\mathcal{CIGLRT-L}$ algorithm in \eqref{eq:est_upt_lin_1}. The algorithm generates a consistent estimate sequence $\{\theta_{n}(t)\}$ for each agent $n$, i.e.,
	\begin{align}
	\label{eq:th1_1}
	\mathbb{P}_{\theta^{*}}\left(\lim_{t\to\infty}\theta_{n}(t)=\theta^{*}\right)=1, \forall n.
	\end{align}
\end{Theorem}
\noindent The next theorem characterizes the large deviations exponent upper bound for the probability of miss and probability of false alarm related to the decision statistic sequence $\{z_{n}(t)\}$ generated at agent $n$, by the decision statistic update part of the $\mathcal{CIGLRT-L}$ algorithm. We define the following quantities which will play a crucial role in stating the next theorem:
let $c_{4}$ and $c_{4}^{*}$ be given by
\begin{align}
\label{eq:th_b2_4}
c_{4}= \frac{1}{\left\|\mathbf{G}_{H}\mathbf{\Sigma}^{-1}\mathbf{G}_{H}^{\top}\right\|\left(\sum_{v=0}^{t_{1}-1}\alpha_{v}^{2}\prod_{u=v+1}^{t_{1}-1}\left\|\mathbf{I}_{NM}-\beta_{u}\left(\mathbf{L}\otimes\mathbf{I}_{M}\right)-\alpha_{u}\mathbf{G}_{H}\mathbf{\Sigma}^{-1}\mathbf{G}_{H}^{\top}\right\|\frac{\left(t_{1}+1\right)^{2c_{1}\alpha_{0}}}{kt_{1}^{2c_{1}\alpha_{0}-1}}+\frac{\alpha_{0}^{2}}{kt_{1}}+\frac{\alpha_{0}^{2}}{2c_{1}\alpha_{0}-1}\right)},
\end{align}
and
\begin{align}
\label{eq:th_b2_5}
c_{4}^{*}=\frac{2c_{1}\alpha_{0}-1}{\alpha_{0}^{2}\left\|\mathbf{G}_{H}\mathbf{\Sigma}^{-1}\mathbf{G}_{H}^{\top}\right\|}-\frac{NM}{2\eta_{2}}
\end{align}
respectively, where $\eta_{2}$ is given by
\begin{align}
\label{eq:th_b2_6} \eta_{2}=\frac{-2N\eta+\left(\mathbf{\theta}^{*}\right)^{\top}\mathbf{G}\mathbf{\theta}^{*}\left(1-N\sqrt{N}r^{k-1}\right)}{4\left\|\mathbf{G}_{H}\mathbf{\Sigma}^{-1}\mathbf{G}_{H}^{\top}\right\|\left(1+N\sqrt{N}r^{k-1}\right)},
\end{align}
and $t_{1}$ defined as
\begin{align}
\label{eq:d131}
t_{1}=\max\{t_{2}, t_{3}\},
\end{align}
where $t_{3}$ is such that, $\forall t\geq t_{3}$,
\begin{align}
\label{eq:d21}
\lambda_{\mbox{\scriptsize{min}}}\left(\mathbf{L}\otimes\mathbf{I}_{M}+\mathbf{G}_{H}\mathbf{\Sigma}^{-1}\mathbf{G}_{H}^{\top}\right)\alpha_{t} < 1,
\end{align}
and $t_{2}$ is such that\footnote{It is to be noted that such $t_{2}$ and $t_{3}$ exist as $\alpha_{t}, \beta_{t} \to 0$ as $t\to\infty$.}, $\forall t\geq t_{2}$,
\begin{align}
\label{eq:ld31}
\beta_{t}\lambda_{N}\left(\mathbf{L}\right)+\alpha_{t}\lambda_{\mbox{\scriptsize{max}}}\left(\mathbf{G}_{H}\mathbf{\Sigma}^{-1}\mathbf{G}_{H}^{\top}\right)< 1.
\end{align}

\begin{Theorem}
	\label{th:b2}
	Let the Assumptions \ref{bs:1}-\ref{bs:6} hold. Consider, the decision statistic update of the $\mathcal{CIGLRT-L}$ algorithm in \eqref{eq:dec_stat_lin_1}.
	For all $\theta^{*}$, which satisfy the following condition
	\begin{align}
	\label{eq:th_b2_1}
	\frac{\left(\mathbf{\theta}^{*}\right)^{\top}\mathbf{G}\mathbf{\theta}^{*}\left(1-N\sqrt{N}r^{k-1}\right)}{2N} > \frac{M\alpha_{0}^{2}\left\|\mathbf{G}_{H}\mathbf{\Sigma}^{-1}\mathbf{G}_{H}^{\top}\right\|^{2}\left(1+N\sqrt{N}r^{k-1}\right)}{2c_{1}\alpha_{0}-1}+\frac{\left(\frac{1}{N}+\sqrt{N}r^{k-1}\right)\sum_{n=1}^{N}M_{n}}{2},
	\end{align}
	we have the following range of feasible thresholds,
	\begin{align}
	\label{eq:th_b2_0}
	\frac{\left(\frac{1}{N}+\sqrt{N}r^{k-1}\right)\sum_{n=1}^{N}M_{n}}{2} < \eta <\frac{\left(\mathbf{\theta}^{*}\right)^{\top}\mathbf{G}\mathbf{\theta}^{*}\left(1-N\sqrt{N}r^{k-1}\right)}{2N}-\frac{M\alpha_{0}^{2}\left\|\mathbf{G}_{H}\mathbf{\Sigma}^{-1}\mathbf{G}_{H}^{\top}\right\|^{2}\left(1+N\sqrt{N}r^{k-1}\right)}{2c_{1}\alpha_{0}-1},
	\end{align}
	
	for which we have the following large deviations upper bound characterization for the probability of false alarm $\mathbb{P}_{FA}$:
	\begin{align}
	\label{eq:th_b2}
	\limsup_{t\to\infty}\frac{1}{t}\log\left(\mathbb{P}_{0}\left(z_{n}(t) > \eta\right)\right)\le -\frac{\eta}{\frac{1}{N}+\sqrt{N}r^{k-1}}-\frac{\sum_{n=1}^{N}M_{n}}{2}\left(1+\log\frac{2\eta}{\left(\frac{1}{N}+\sqrt{N}r^{k-1}\right)\sum_{n=1}^{N}M_{n}}\right)=LD_{0}(\eta),
	\end{align}
	
	and the following large deviations upper bound characterization for the probability of miss $\mathbb{P}_{M}$:
	\begin{align}
	\label{eq:th_b2_2}
	&\limsup_{t\to\infty}\frac{1}{t}\log\left(\mathbb{P}_{1,\theta^{*}}\left(z_{n}(t) < \eta\right)\right)\nonumber\\
	&\le \max\left\{-\frac{\left(-\frac{\eta}{4}+\frac{\left(\mathbf{\theta}^{*}\right)^{\top}\mathbf{G}\mathbf{\theta}^{*}\left(\frac{1}{N}-\sqrt{N}r^{k-1}\right)}{8}\right)^{2}}{2\sum_{j=1}^{N}\left(\theta^{*}\right)^{\top}\mathbf{H}_{j}^{\top}\mathbf{\Sigma}_{j}^{-1}\mathbf{H}_{j}\theta^{*}\left(\frac{1}{N}+\sqrt{N}r^{k-1}\right)^{2}}, -LD\left(\min\left\{c_{4},c_{4}^{*}\right\}\right)\right\}=LD_{1}\left(\eta\right),
	\end{align}
	
	where,
	\begin{align}
	\label{eq:th_b2_3}
	LD(\lambda)=\lambda\eta_{2}+\frac{NM}{2}\log\left(1-\frac{\lambda\alpha_{0}^{2}\left\|\mathbf{G}_{H}\mathbf{\Sigma}^{-1}\mathbf{G}_{H}^{\top}\right\|}{2c_{1}\alpha_{0}-1}\right).
	\end{align}
	
\end{Theorem}

\noindent We discuss how the above result can be used in practice to identify thresholds that lead to exponential decay of the probabilities of error. It is to be noted that as the observation parameters, i.e., $M_{n},  N$ and the connectivity of the communication graph, i.e. $r$ are known apriori, the threshold can be chosen to be $\frac{\left(\frac{1}{N}+\sqrt{N}r^{k-1}\right)\sum_{n=1}^{N}M_{n}}{2}+\epsilon$, where $\epsilon$ can be chosen to be arbitrarily small. This would guarantee exponential decay for the probability of false alarm. Further, from the feasible range of thresholds in \eqref{eq:th_b2_0}, a range on the $\theta^{*}$s' can be obtained in terms of $\left\|\theta^{*}\right\|$ such that under $\mathbb{H}_{1}$, as long as the true value $\theta^{*}$ of the parameter belongs to this range, the probability of miss is guaranteed to decay to zero exponentially fast. It is important to note in this context that there exists some weak signals, i.e., signals with low $\left\|\theta^{*}\right\|$ (but non-zero), for which there may not exist a choice of thresholds for which exponential decay can be ensured for both the probability of miss and probability of false alarm. The signals for which Theorem \ref{th:b2} is rendered to be inconclusive in the manner described above, can be categorized in terms of $\theta^{*}$. Specifically, $\theta^{*}$s' which satisfy the following condition

\begin{align}
\label{eq:th_b2_explain}
	\left\|\mathbf{\theta}^{*}\right\|^{2}<\frac{\left(1+N\sqrt{N}r^{k-1}\right)\sum_{n=1}^{N}M_{n}}{\lambda_{\mbox{\scriptsize{min}}}(\mathbf{G})\left(1-N\sqrt{N}r^{k-1}\right)}+\frac{2MN\alpha_{0}^{2}\left\|\mathbf{G}_{H}\mathbf{\Sigma}^{-1}\mathbf{G}_{H}^{\top}\right\|^{2}\left(1+N\sqrt{N}r^{k-1}\right)}{\lambda_{\mbox{\scriptsize{min}}}(\mathbf{G})\left(2c_{1}\alpha_{0}-1\right)\left(1-N\sqrt{N}r^{k-1}\right)},
\end{align}
where $\lambda_{\mbox{\scriptsize{min}}}(.)$ denotes the minimum eigenvalue, render Theorem \ref{th:b2} to be inconclusive. For further clarification regarding the range of $\theta^{*}$'s for which Theorem \ref{th:b2} can ensure exponentially decaying probabilities of error, we point to Section \ref{subsec:illu_ex}.
 Furthermore, it can be seen that with better information exchange in the communication graph, i.e., with lower $r$, the exponents get better and hence there is a faster decay of probabilities of errors. It is also to be noted that the exponents get better with increasing $k$, which is due to more rounds of consensus, but at the cost of more inherent time delay in incorporating latest parameter estimates and observations into the decision statistic possibly affecting transient characteristics.
\section{Illustration of $\mathcal{CIGLRT-NL}$ and $\mathcal{CIGLRT-L}$}
\label{sec:illust_eg}
\subsection{Illustrative Example}
\label{subsec:illu_ex}
\noindent In this section, we explain the nuances of Theorem  \ref{th:b2} through an illustrative example. To give better intuition for the large deviations exponent upper bounds, we consider the following setup for the derivation of large deviations exponent upper bounds. We consider a scalar observation model, where the scaling for the parameter is $0$ for $N_{2}$ agents and is $h>0$ for $N_{1}$ agents, where $N_{1} > 0$ and $N_{2}=N-N_{1}$, i.e., $\mathbf{H}_{n}=0$ for $N_{2}$ agents and $\mathbf{H}_{n}=h$ for $N_{1}$ agents respectively according to the observation model given in \eqref{eq:sens_ob}. Technically speaking, $N_{1}$ agents observe scaled noisy versions of the parameter, while the other $N_{2}$ agents just observe noise. The noise power is $\sigma^{2}$ across all agents. Note the global observability condition Assumption~\ref{bs:1} reduces to $N_{1}$ being strictly positive in this context. We also note that, although the model is globally observable, the local models at the faulty agents are unobservable for the parameter. Finally, without loss of generality, assuming that agents $n=1,\cdots, N_{1}$ correspond to the set of $N_{1}$ agents that observe scaled noisy versions of the parameter, we have, $\mathbf{G}_{H}=diag\left[h\cdots h~0\cdots 0\right]$ and $\mathbf{\Sigma}=\sigma^{2}\mathbf{I}_{N}$.

\noindent We make an assumption on $a$ as defined in \eqref{bs:3} for the current model under consideration.
\begin{myassump}{B7}
	\label{bs:7}
	\emph{Recall $a$ as defined in Assumption \ref{bs:3}.
		The constant $a$ satisfies
		\begin{align}
		\label{eq:bs7_1}
		a \ge\frac{1}{2c_{1}}+2,
		\end{align}
		where $c_{1}$ is defined as
		\begin{align}
		\label{eq:bs7_2}
		c_{1}=\min_{\left\|\mathbf{x}\right\|=1}\mathbf{x}^{\top}\left(\mathbf{L}+\mathbf{G}_{H}\mathbf{\Sigma}^{-1}\mathbf{G}_{H}^{\top}\right)\mathbf{x}=\lambda_{\mbox{\scriptsize{min}}}\left(\mathbf{L}+\mathbf{G}_{H}\mathbf{\Sigma}^{-1}\mathbf{G}_{H}^{\top}\right).
		\end{align}
%		whereas $t_{1}=\max\{t_{2},t_{3}\}$ and $t_{3}$ is given by,
%		\begin{align}
%		\label{eq:bs7_3}
%		t_{3} =\left(\frac{Nh^{2}}{\beta_{0}N_{1}\sigma^{2}\lambda_{2}\left(\mathbf{L}\right)}\right)^{\frac{1}{1-\delta_{1}}}-1,
%		\end{align}
%		$t_{2}$ is such that, $\forall t\geq t_{2}$,
%		\begin{align}
%		\label{eq:bs7_4}
%		\beta_{t}\lambda_{N}(\mathbf{L})+\alpha_{t}\lambda_{\mbox{\scriptsize{max}}}(\mathbf{G}_{H}\mathbf{\Sigma}^{-1}\mathbf{G}_{H}^{\top}) < 1,
%		\end{align}
%		where $\lambda_{\mbox{\scriptsize{max}}}\left(.\right)$ and $\lambda_{\mbox{\scriptsize{min}}}\left(.\right)$ represent the largest eigenvalue and the smallest eigenvalue respectively.
}
\end{myassump}

\noindent In order to compare the large deviations exponent upper bounds of the proposed $\mathcal{CIGLRT-L}$ algorithm with the large deviations exponent upper bound of an optimal centralized detector, we consider a hypothetical fusion center which has access to all the observations, parameter estimates across all the agents at all times. The centralized parameter estimation scheme generates the sequence $\{\theta_{c}(t)\}$ at the fusion center as follows:
\begin{align}
\label{eq:cent_1}
\theta_{c}(t+1)=\theta_{c}(t)+\frac{\kappa_{t}}{N_{1}\sigma^{2}}\left(hy_{n}(t)-h^{2}\theta_{c}(t)\right),
\end{align}

\noindent where $\{\kappa_{t}\}$ is a weight sequence (to be specified shortly).

\noindent At the fusion center, the decision statistic sequence $\{z_{c}(t)\}$ at the fusion center as follows:
\begin{align}
\label{eq:cent_2}
z_{c}(t+1)=\frac{h\theta_{c}(t-1)}{N_{1}\sigma^{2}}\left(\sum_{j=1}^{N_{1}}s_{j}(t-1)-\frac{h\theta_{c}(t-1)}{2}\right),
\end{align}

\noindent where $s_{j}(t-1)$ is the time-average of all the observations made at agent $j$ until time $t-1$.

\noindent We state an assumption on the weight sequence for the centralized estimation scheme before proceeding to the main results.
\begin{myassump}{B8}
	\label{bs:8}
	\emph{The weight sequence $\{\kappa_{t}\}$ is of the form
		\begin{align}
		\label{eq:ab_bs7}
		\kappa_{t}=\frac{g}{(t+1)},
		\end{align}
		where $g > 0$ and $g$ satisfies
		\begin{align}
		\label{eq:ab_bs7_1}
		2h^{2}g>\sigma^{2}.
		\end{align}}
\end{myassump}

\noindent We formally state the result concerning the characterization of the large deviations exponent upper bounds of the probabilities of the errors pertaining to the distributed detector based on the scalar observation model in context with the proof relegated to Appendix \ref{a3}.\footnote{Note that the results obtained in Section \ref{subsec:main_res_ciglrt} for the $\mathcal{CIGLRT-L}$ algorithm for the general linear model apply to the current specific scalar case also. However, by exploiting the specifics of the scalar model, we derive tighter bounds in Appendix \ref{a3}.}

We define the following quantities which will be crucial in the statement of the next theorem : let $c_{4}$ and $c_{4}^{*}$ be constants given by,
	\begin{align}
	\label{eq:th_bs2_4}
	c_{4}= \frac{\sigma^{2}}{h^{2}\left(c_{3}\frac{\left(t_{1}+1\right)^{2c_{1}\alpha_{0}}}{kt_{1}^{2c_{1}\alpha_{0}-1}}+\frac{\alpha_{0}^{2}}{kt_{1}}+\frac{\alpha_{0}^{2}}{2c_{1}\alpha_{0}-1}\right)},
	\end{align}
	and
	\begin{align}
	\label{eq:th_bs2_5}
	c_{4}^{*}=\frac{\sigma^{2}(2c_{1}\alpha_{0}-1)}{\alpha_{0}^{2}h^{2}}-\frac{N}{2\eta_{2}}
	\end{align}
	respectively, where $\eta_{2}$ is given by
	\begin{align}
	\label{eq:th_bs3_6}
	\eta_{2}=\frac{-2N\sigma^{2}\eta+N_{1}h^{2}(\theta^{*})^{2}\left(1-N\sqrt{N}r^{k-1}\right)}{4h^{2}\left(1+N\sqrt{N}r^{k-1}\right)},
	\end{align}
	$c_{3}$ is defined as
	\begin{align}
	\label{eq:th_bs3_7}
	c_{3}=\sum_{v=0}^{t_{1}-1}\alpha_{v}^{2}\prod_{u=v+1}^{t_{1}-1}\left\|\mathbf{I}_{N}-\beta_{u}\mathbf{L}-\alpha_{u}\mathbf{G}_{H}\mathbf{\Sigma}^{-1}\mathbf{G}_{H}^{\top}\right\|,
	\end{align}
	and $t_{1}$ defined as
	\begin{align}
	\label{eq:d132}
	t_{1}=\max\{t_{2}, t_{3}\},
	\end{align}
	where $t_{3}$ is such that, $\forall t\geq t_{3}$,
	\begin{align}
	\label{eq:d22}
	\lambda_{\mbox{\scriptsize{min}}}\left(\mathbf{L}+\mathbf{G}_{H}\mathbf{\Sigma}^{-1}\mathbf{G}_{H}^{\top}\right)\alpha_{t} < 1,
	\end{align}
	and $t_{2}$ is such that\footnote{It is to be noted that $t_{2}$ and $t_{3}$ exist as $\alpha_{t}, \beta_{t} \to 0$.}, $\forall t\geq t_{2}$,
	\begin{align}
	\label{eq:ld32}
	\beta_{t}\lambda_{N}(\mathbf{L})+\alpha_{t}\frac{h^{2}}{\sigma^{2}} < 1.
	\end{align}

\begin{Theorem}
	\label{th:bs2}
	Let the Assumptions \ref{bs:1}-\ref{bs:5} and \ref{bs:7} hold. Consider, the decision statistic update of the $\mathcal{CIGLRT-L}$ algorithm in \eqref{eq:dec_stat_lin_1}.
	For all $\theta^{*}$ which satisfy the following condition
	\begin{align}
	\label{eq:th_bs2_00}
	|\theta^{*}|^{2}\geq \frac{1+N\sqrt{N}r^{k-1}}{1-N\sqrt{N}r^{k-1}}\left(\frac{N\sigma^{2}}{N_{1}h^{2}}+\frac{2N\alpha_{0}^{2}h^{2}}{N_{1}\sigma^{2}\left(2c_{1}\alpha_{0}-1\right)}\right),
	\end{align}
	we have the following range of feasible thresholds,
	\begin{align}
	\label{eq:th_bs2_0}
	&\frac{\left(\frac{1}{N}+\sqrt{N}r^{k-1}\right)N_{1}}{2} < \eta <\frac{N_{1}\left(h\theta^{*}\right)^{2}\left(1-N\sqrt{N}r^{k-1}\right)}{2N\sigma^{2}}\nonumber\\&-\frac{\alpha_{0}^{2}h^{4}\left(1+N\sqrt{N}r^{k-1}\right)}{\sigma^{4}\left(2c_{1}\alpha_{0}-1\right)},
	\end{align}
	
	for which we have the following large deviations upper bound characterization for the probability of false alarm $\mathbb{P}_{FA}$:
	\begin{align}
	\label{eq:th_bs2}
	&\limsup_{t\to\infty}\frac{1}{t}\log\left(\mathbb{P}_{0}\left(z_{n}(t) > \eta\right)\right)\le -\frac{\eta}{\frac{1}{N}+\sqrt{N}r^{k-1}}\nonumber\\&-\frac{N_{1}}{2}\left(1+\log\frac{2\eta}{\left(\frac{1}{N}+\sqrt{N}r^{k-1}\right)N_{1}}\right)\nonumber\\&=LD_{0}(\eta),
	\end{align}
	
	and the following large deviations upper bound characterization for the probability of miss $\mathbb{P}_{M}$:
	\begin{align}
	\label{eq:th_bs2_2}
	&\limsup_{t\to\infty}\frac{1}{t}\log\left(\mathbb{P}_{1,\theta^{*}}\left(z_{n}(t) < \eta\right)\right)\nonumber\\
	&\le \max\left\{-\left(\frac{\left(-\frac{\eta}{4}+\frac{N_{1}h^{2}(\mathbf{\theta}^{*})^{2}\left(\frac{1}{N}-\sqrt{N}r^{k-1}\right)}{8\sigma^{2}}\right)^{2}}{2N_{1}h^{2}(\theta^{*})^{2}\left(\frac{1}{N}+\sqrt{N}r^{k-1}\right)^{2}/\sigma^{2}}\right), -LD\left(\min\left\{c_{4},c_{4}^{*}\right\}\right)\right\}=LD_{1}\left(\eta\right),
	\end{align}
	
	where, $LD(\lambda)$ is given by
	\begin{align}
	\label{eq:th_bs2_3}
	LD(\lambda)=\lambda\eta_{2}+\frac{N}{2}\log\left(1-\frac{\lambda\alpha_{0}^{2}h^{2}}{\sigma^{2}(2c_{1}\alpha_{0}-1)}\right).
	\end{align}
\end{Theorem}

\noindent We now provide the result concerning the large deviations upper bounds of the probabilities of errors emanating from the centralized detection algorithm described in \eqref{eq:cent_1}-\eqref{eq:cent_2}. We skip the proof due to brevity. The proof follows in a very similar way to the proof of Theorem \ref{th:bs2}.
\begin{Theorem}
	\label{th:b3}
	Let Assumption \ref{bs:8} hold. Consider the centralized detection algorithm in \eqref{eq:cent_2}. For all $\theta^{*}$ which satisfy the following condition
	\begin{align}
	\label{eq:b3_1}
	\left|\theta^{*}\right|^{2}\ge \frac{2N_{1}h^{2}\alpha_{0}^{2}}{2h^{2}\kappa_{0}-\sigma^{2}}+\frac{N_{1}\sigma^{2}}{h^{2}},
	\end{align}
	we have the following range of feasible thresholds,
	\begin{align}
	\label{eq:b3_2}
	\frac{1}{2}<\eta<\frac{h^{2}(\theta^{*})^{2}}{2N_{1}\sigma^{2}}-\frac{2\kappa_{0}^{2}h^{4}}{N_{1}\sigma^{2}(h^{2}\kappa_{0}-\sigma^{2})},\end{align}
	for which we have the following large deviations upper bound characterization for the probability of false alarm
	\begin{align}
	\label{eq:b3_3}
	&\lim_{t\to\infty}\frac{1}{t}\log\left(\mathbb{P}_{0}\left(z_{c}(t)>\eta\right)\right)\le -N_{1}\eta-\frac{N_{1}}{2}\left(1+\log 2\eta\right)\nonumber\\&=LD_{0,c}(\eta),
	\end{align}
	and the following large deviations exponent upper bound characterization for the probability of miss
	\begin{align}
	\label{eq:b3_4}
	&\lim_{t\to\infty}\frac{1}{t}\log\left(\mathbb{P}_{1,\theta^{*}}\left(z_{c}(t)<\eta\right)\right)\nonumber\\
	&\le \max\left\{-LD_{c}(d_{1}^{*}), -\left(\frac{\sigma^{2}\left(-\frac{\eta}{4}+\frac{h^{2}(\theta^{*})^{2}}{8\sigma^{2}}\right)^{2}}{2h^{2}(\theta^{*})^{2}}\right)\right\}\nonumber\\&=LD_{1,c}\left(\eta\right),	
	\end{align}
	where
	\begin{align}
	\label{eq:b3_5}
	& LD_{c}(\lambda)=\lambda\eta_{c}+N_{1}\log\left(1-\frac{\lambda\kappa_{0}^{2}h^{2}}{2h^{2}\kappa_{0}-\sigma^{2}}\right),\nonumber\\ & d_{1}^{*}=\frac{2h^{2}N_{1}\kappa_{0}-N_{1}\sigma^{2}}{\kappa_{0}^{2}h^{2}}-\frac{N_{1}}{2\eta_{c}},\nonumber\\
	&\eta_{c}=\frac{N_{1}\sigma^{2}}{h^{2}}\left(-\frac{\eta}{2}+\frac{h^{2}(\theta^{*})^{2}}{4N_{1}\sigma^{2}}\right).
	\end{align}	
\end{Theorem}

\noindent The bounds derived for the range of parameter $\theta^{*}$ for which exponential decay of error probabilities can be ensured, for both the distributed $\mathcal{CIGLRT-L}$ detector and the centralized detector are conservative and hence might not be tight. With better network connectivity, the upper bounds of the large deviations exponent of the distributed detector approach the upper bounds of the large deviations exponents of that of the centralized detector. The range of $\theta^{*}$'s for which the distributed detector ensures exponential decay of error probabilities becomes bigger with better network connectivity\footnote{Intuitively, $r$ indicates how well a network is connected. For e.g. if a network is fully connected, i.e., has an all-to-all connected communication graph and hence $\mathbf{W}=\mathbf{J}$, $r=0$. In the absence of communication, $\mathbf{W}=\mathbf{I}$ and $r=1$. Hence, a lower value of $r$ indicates better connectivity of the graph.}, i.e., with smaller $r$. For our analysis of probability of errors and their respective decay rate characterizations, we considered a uniform~(i.e., agent and time independent) choice of thresholds. As far as the agent dependent thresholds are concerned, the thresholds for each agent could be functions of an individual agent's connectivity, thus allowing for more degrees of freedom in the design. Intuitively, an agent with \emph{better connectivity} would have a wider range of thresholds to choose from so as to do a more flexible trade-off between the probability of false alarm and the probability of miss. Though an individual agent dependent bounding would be ideal, but it makes the analysis seemingly intractable. For example, while bounding $\sum_{j=1}^{N}\phi_{n,j}(k-1)\frac{\left(\mathbf{\theta}^{*}\right)^{\top}\mathbf{H}_{j}^{\top}\mathbf{\Sigma}_{j}^{-1}\mathbf{H}_{j}\mathbf{\theta}^{*}}{2}$, where $\phi_{n,j}(k-1)=\left[\mathbf{W}^{k-1}\right]_{n,j}$, we use the same bound for each of $\phi_{n,j}(k-1)$ which is in turn given by $\phi_{n,j}(k-1)\ge \frac{1}{N}-\sqrt{N}r^{k-1}$. In such a setting, tracking the row corresponding to a particular agent in the weight matrix is seemingly intractable. Having said that, if individual agent dependent thresholds could have been used, the extent to which an agent can distinguish between different hypotheses would be different owing to different range of thresholds available to an agent to choose from.\\
The use of time dependent thresholds however does not seem to affect the zone of indifference. It is to be noted that the zone of the indifference is characterized in terms of the range of $\theta^{*}$'s under the alternate hypothesis for which exponential decay of both the probability of errors can be ensured. To be specific, in the analysis of the probability of the miss the crucial part of the analysis is the bound for the term $t\left\|\mathbf{P}_{t}\right\|$ which is given by
\begin{align*}
t\left\|\mathbf{P}_{t}\right\| \le c_{3}\frac{\left(t_{1}+1\right)^{2c_{1}\alpha_{0}}}{t^{2c_{1}\alpha_{0}-1}}+\frac{\alpha_{0}^{2}}{t}+\frac{\alpha_{0}^{2}}{2c_{1}\alpha_{0}-1},~~\forall t\geq t_{1}.
\end{align*}
If the thresholds are adjusted, so as to take into account the time-decaying terms $c_{3}\frac{\left(t_{1}+1\right)^{2c_{1}\alpha_{0}}}{t^{2c_{1}\alpha_{0}-1}}$ and  $\frac{\alpha_{0}^{2}}{t}$, both the terms decay to zero as $t\to\infty$ thereby ensuring that the large deviations exponent upper bound stays the same. However, it is to be noted that at any finite time, time dependent thresholds would give tighter probability of error bounds, thus improving transients of the approach.\\
The assumption regarding the inter-agent communication network which is instrumental in obtaining the results of this paper is the connectivity of the graph i.e., there exists a path between every two nodes. As long as the graph is connected, the results continue to hold. For example, for a network with $N$ agents, there needs to be at least $N-1$ links for the graph to be connected, which is very sparsely connected as out of the possible $N(N-1)/2$ possible links only $N-1$ links are present. Hence, the algorithms apply to very sparse (but connected) networks. However, it is important to note that the large deviations upper bounds might get worse with increasing sparsity of the inter-agent communication graph. The sparsity of the inter-agent communication graph is in turn reflected by the quantity\footnote{It is to be noted that as long as the graph is connected, $r < 1$.} $r=\left\|\mathbf{W}-\mathbf{J}\right\|$, where $\mathbf{W}$ and $\mathbf{J}$ represent the weight matrices associated with the inter-agent communication graph under consideration and a completely connected graph respectively. When $r=0$ it basically points to the case that $\mathbf{W}=\mathbf{J}$, i.e., the network under consideration is completely connected. The other extreme case $r=1$ is the case when $\mathbf{W}=\mathbf{I}$, i.e., when there are no links in the network. To sum it up, small values of $1-r$ reflect the sparseness of the network under consideration.\\
Furthermore, note that with increasing $k$, i.e., the time lag or equivalently the number of rounds of consensus between incorporating latest estimates (see \eqref{eq:dec_stat_lin_2}), the range of parameter $\theta^{*}$ for which exponential decay of error probabilities can be ensured increases, the large deviations upper bounds for the probabilities of miss and false alarm also increase. However, $k$ cannot be made arbitrarily large just based on improvement of the large deviations upper bounds, as large deviations analysis is essentially an asymptotic characterization and at the same time with increase in $k$ the inherent time delay in incorporating new estimates into the decision statistic also increases, and hence affecting the transient performance of the procedure. Recall from the decision statistic update in \eqref{eq:dec_stat_lin_2}, that the decision statistic update takes the value $z_{n}(kt-k)$ at all times $t\in\left[kt-k,kt-1\right]$. Thus, only at time instants which are of the form $kt$, the decision statistic has the minimum time-lag $k$ with respect to the latest information available in the multi-agent network which also makes the analysis more tractable. Moreover, from the perspective of a faulty agent, low $k$ would result in particularly bad detection performance as the dynamics of an accurate detection procedure at a faulty agent depends on the information it receives from its neighbors, which shows the necessity of inter-agent collaboration. In absence of a distributed mechanism characterized by a communication graph, a defective agent would fail to come up with a reasonable decision at all times, as the local sensed data at a defective agent is completely non-informative. Finally, no inference procedure is free of the \emph{curse of dimensionality}. It is to be noted that with increasing $M$, i.e., dimension of the underlying parameter $\theta^{*}$, the range of $\theta^{*}$ for which exponential decay of probabilities of errors can be ensured shrinks, the feasible range of thresholds also shrinks and finally the large deviations exponent upper bound for the probability of miss also decreases.

\subsection{Simulations: $\mathcal{CIGLRT-NL}$}
\label{subsec:sim_nl}
\begin{figure}[!h]
	\centering
	\captionsetup{justification=centering}
	\includegraphics[width=90mm]{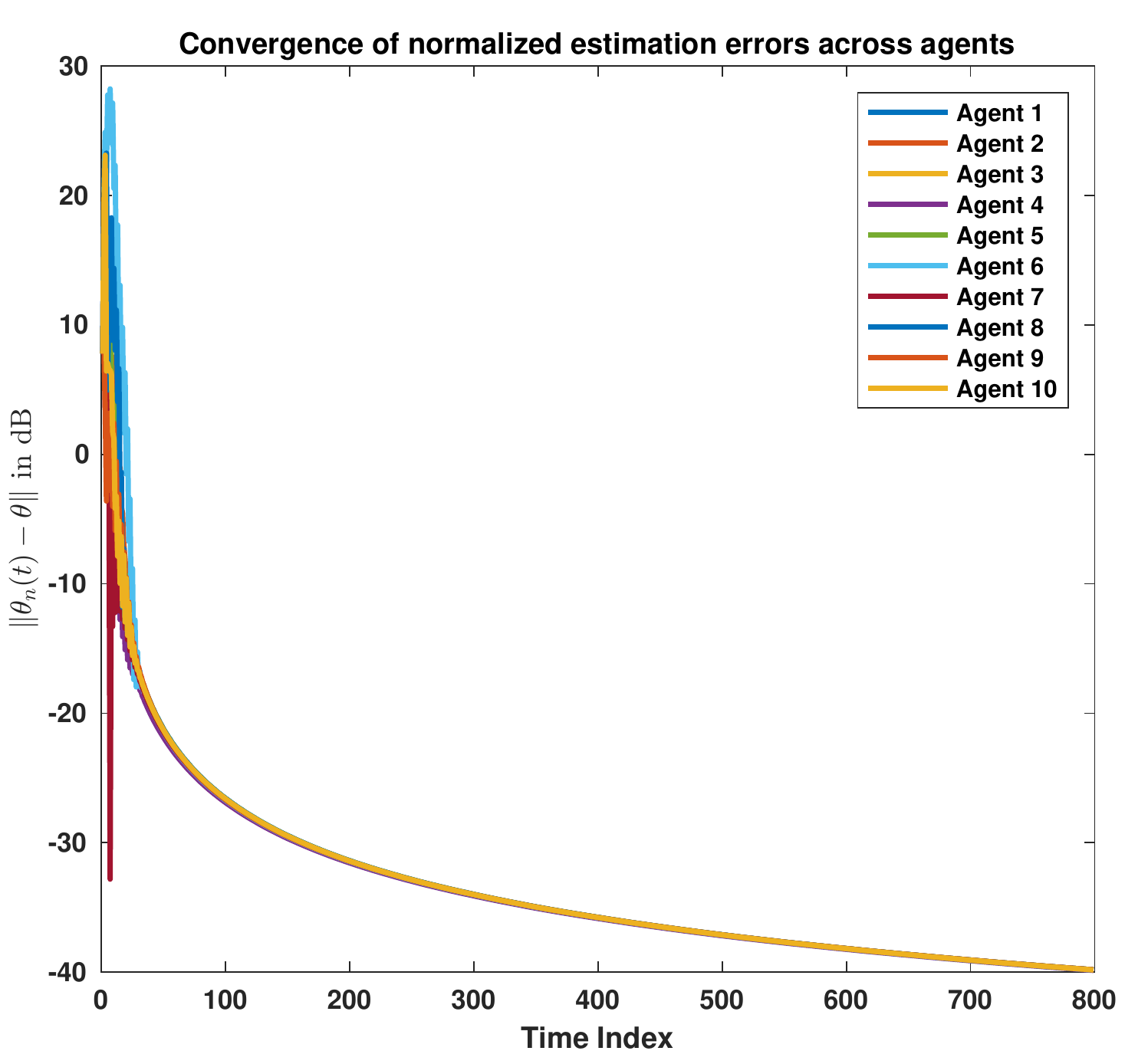}
	\caption{Convergence of estimation error at each agent}\label{fig:nl_1}
\end{figure}

\begin{figure}[!h]
	\centering
	\captionsetup{justification=centering}
	\includegraphics[width=90mm]{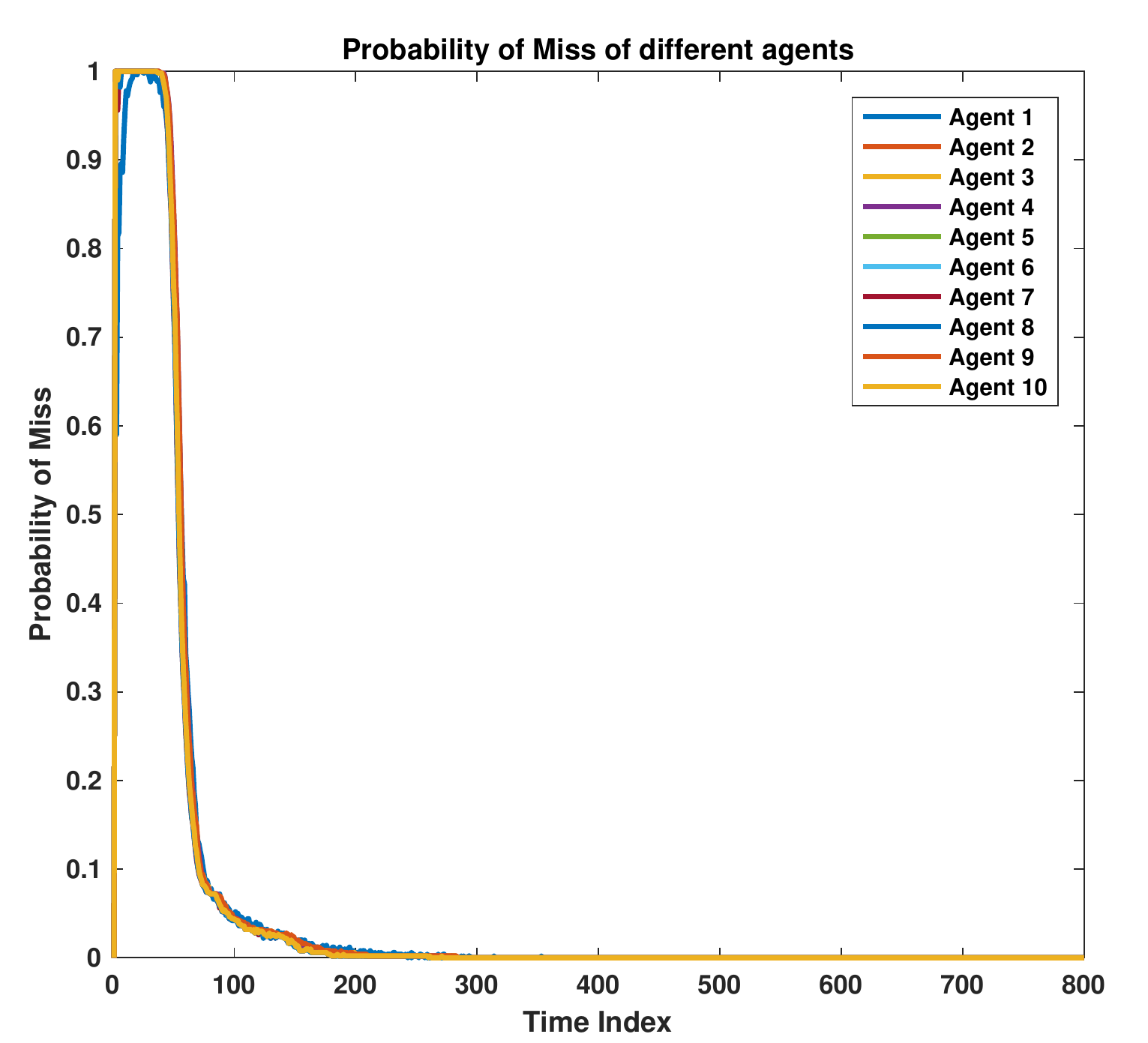}
	\caption{Probability of miss of the agents}\label{fig:nl_2}
\end{figure}

\noindent  We generate a random geometric network of $10$ agents. The location of agents are generated by sampling the $x$ coordinates and the $y$ coordinates from a uniform distribution on the interval $[0,1]$. We link two vertices by an edge if the distance between them is less than or equal to $g=0.4$. We go on iterating this procedure until we get a connected graph. The network connectivity expressed in terms of $r=\left\|\mathbf{W}-\mathbf{J}\right\|$ is given by $r=0.3904$. We consider the underlying parameter to be a $5$-dimensional parameter, i.e., $M=5$ and $\theta^{*}=[\pi/6~-\pi/4~\pi/4~-\pi/5~\pi/6]$. For the nonlinear observation model, we consider trigonometric sensing functions which are given by, $\mathbf{f}_{1}(\btheta)=5\sin(\theta_{1}+\theta_{2}), \mathbf{f}_{2}(\btheta)=5\sin(\theta_{3}+\theta_{2}), \mathbf{f}_{3}(\btheta)=5\sin(\theta_{3}+\theta_{4}), \mathbf{f}_{4}(\btheta)=5\sin(\theta_{4}+\theta_{5}), \mathbf{f}_{5}(\btheta)=5\sin(\theta_{1}+\theta_{5}), \mathbf{f}_{6}(\btheta)=5\sin(\theta_{1}+\theta_{3}), \mathbf{f}_{7}(\btheta)=5\sin(\theta_{4}+\theta_{2}), \mathbf{f}_{8}(\btheta)=5\sin(\theta_{3}+\theta_{5}), \mathbf{f}_{9}(\btheta)=5\sin(\theta_{1}+\theta_{4})$ and  $\mathbf{f}_{10}(\btheta)=5\sin(\theta_{1}+\theta_{5})$, where the underlying parameter is $5$ dimensional, $\btheta=\left[\theta_{1},~\theta_{2},~\theta_{3},~\theta_{4},~\theta_{5}\right]$. However, we restrict the values of $\btheta$ to the set $\left[-\frac{\pi}{4}, \frac{\pi}{4}\right]^{5}\in\mathbb{R}^{5}$. Note that, in spite of restricting the parameter to a set, the alternate hypothesis is still parameterized by vector parameters from a continuous set. The local sensing models are unobservable, but collectively they are globally observable since, $\sin(\cdot)$ is one-to-one on the set $\left[-\frac{\pi}{4}, \frac{\pi}{4}\right]^{5}$ and the set of linear combinations of the $\btheta$ components corresponding to the arguments of the $\sin(\cdot)$'s constitute a full-rank system for $\btheta$. The agents make noisy scalar observations where the observation noise process is Gaussian and the noise covariance is given by $\mathbf{R}=2\mathbf{I}_{10}$. It is readily seen that the sensing model with the restriction that the parameter can take values from the set $\left[-\frac{\pi}{4}, \frac{\pi}{4}\right]^{5}$ satisfy Assumptions \ref{as:1}-\ref{as:4}.
We carry out $2000$ Monte-Carlo simulations for analyzing the convergence of the parameter estimate sequences. The estimates are initialized to be $0$, i.e., $\btheta_{n}(0)=\mathbf{0}$ for $n=1,\cdots, 10$. The normalized error for the $n$-th agent at time $t$ is given by the quantity $\left\|\btheta_{n}(t)-\btheta^{*}\right\|$. Figure \ref{fig:nl_1} shows the estimation error at every agent against the time index $t$.
For the analysis of the probability of miss, we run the algorithm for $2000$ sample paths so as to empirically estimate the probability of miss. Figure \ref{fig:nl_2} verifies the assertion in Theorem \ref{t3}. The threshold is chosen to be equal to $7$. It is to be noted that, from Figure \ref{fig:nl_2} the probability of miss starts decaying even before the parameter estimates get reasonably close to the true underlying parameter, which further indicates the \emph{online} nature of the proposed algorithm $\mathcal{CIGLRT-NL}$. The decay of the probability of the miss can be attributed to the fact that, in order to reach the correct decision, the decision statistics of the agents need to cross the threshold which is achieved even before the agents' parameter estimates reach close to the true underlying parameter.

\subsection{Simulations: $\mathcal{CIGLRT-L}$}
\label{subsec:sim_l}
\begin{figure}
	\centering
	\captionsetup{justification=centering}
	\includegraphics[width=90mm]{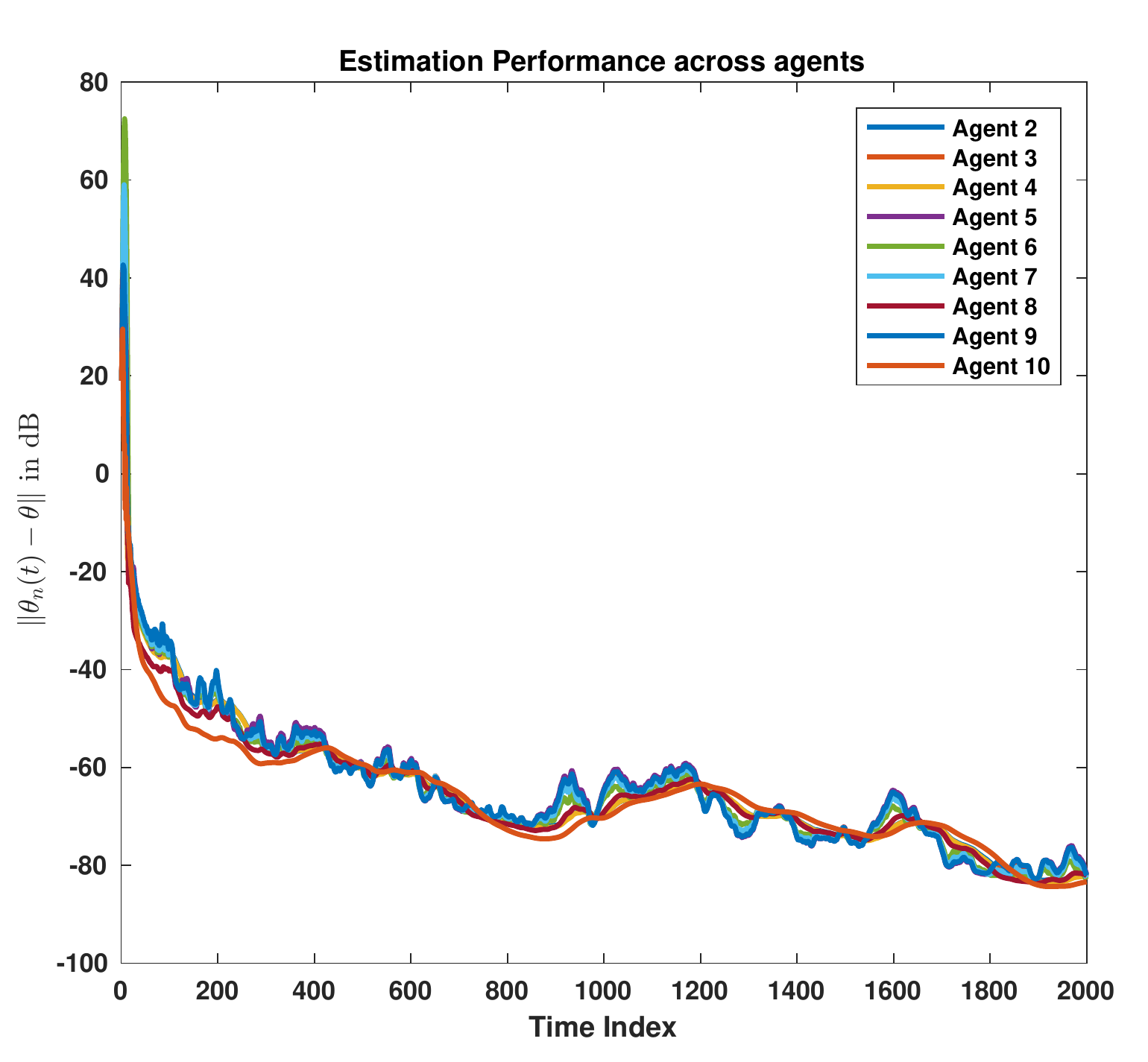}
	\caption{Convergence of estimation error at each agent}\label{fig:1}
\end{figure}

\noindent We generate a planar ring network of $10$ agents, where every agent has exactly two neighbors. We consider the underlying parameter to be a $5$-dimensional parameter, i.e., $M=5$ and $\theta^{*}=\theta^{*}=[1~0.9~1.2~1.1~1.5]$. The observation matrices for the agents are of the dimension $5\times 1$, i.e., $M_{n}=1$, $\forall n$. Specifically the $\mathbf{H}_{n}$'s are given by $H_{1}=[1~1~0~0~0], H_{2}=[0~1~1~0~0], H_{3}=[0~0~1~1~0], H_{4}=[0~0~0~1~1], H_{5}=[1~0~0~0~1], H_{6}=[1~0~1~0~0], H_{7}=[0~1~0~1~0], H_{8}=[0~0~1~0~1], H_{9}=[1~0~0~1~0], H_{10}=[0~1~0~0~1]$.  The noise covariance matrix $\mathbf{\Sigma}$ is taken to be $3\mathbf{I}_{10}$. We emphasize that the above design ensures global observability (in the sense of Assumption \ref{bs:1}), as the matrix $G$ is invertible, but at the same time the parameter of interest is locally unobservable at all agents. The network is poorly connected which in turn is reflected by the quantity $r=\left\|\mathbf{W}-\mathbf{J}\right\|$ and is given by $0.8404$. In particular, for the parameter estimation algorithm, $a=9.1$ and $\delta_{2}=0.4$, where $a,\delta_{2}$ are as defined in Assumption \ref{bs:3}. The time-lag $k$ is taken to be $k=20$.  Figures \ref{fig:1} shows the convergence of the parameter estimates of the agents to the underlying parameter in different dimensions which in turn demonstrates the consistency of the parameter estimation scheme.
\begin{figure}
	\centering
	\captionsetup{justification=centering}
	\includegraphics[width=90mm]{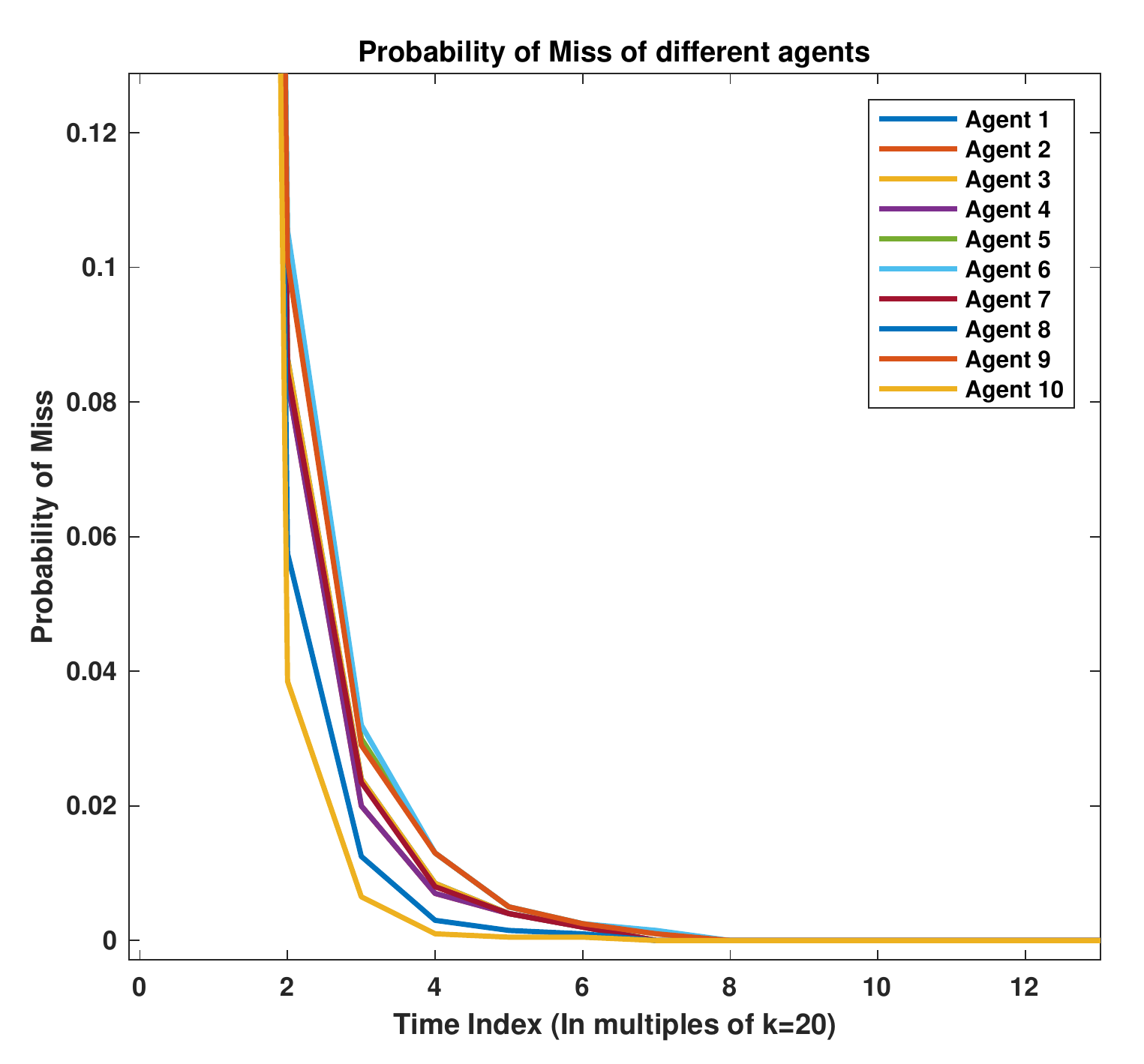}
	\caption{Probability of Miss at each agent}\label{fig:6}
\end{figure}
\begin{figure}
	\centering
	\captionsetup{justification=centering}
	\includegraphics[width=90mm]{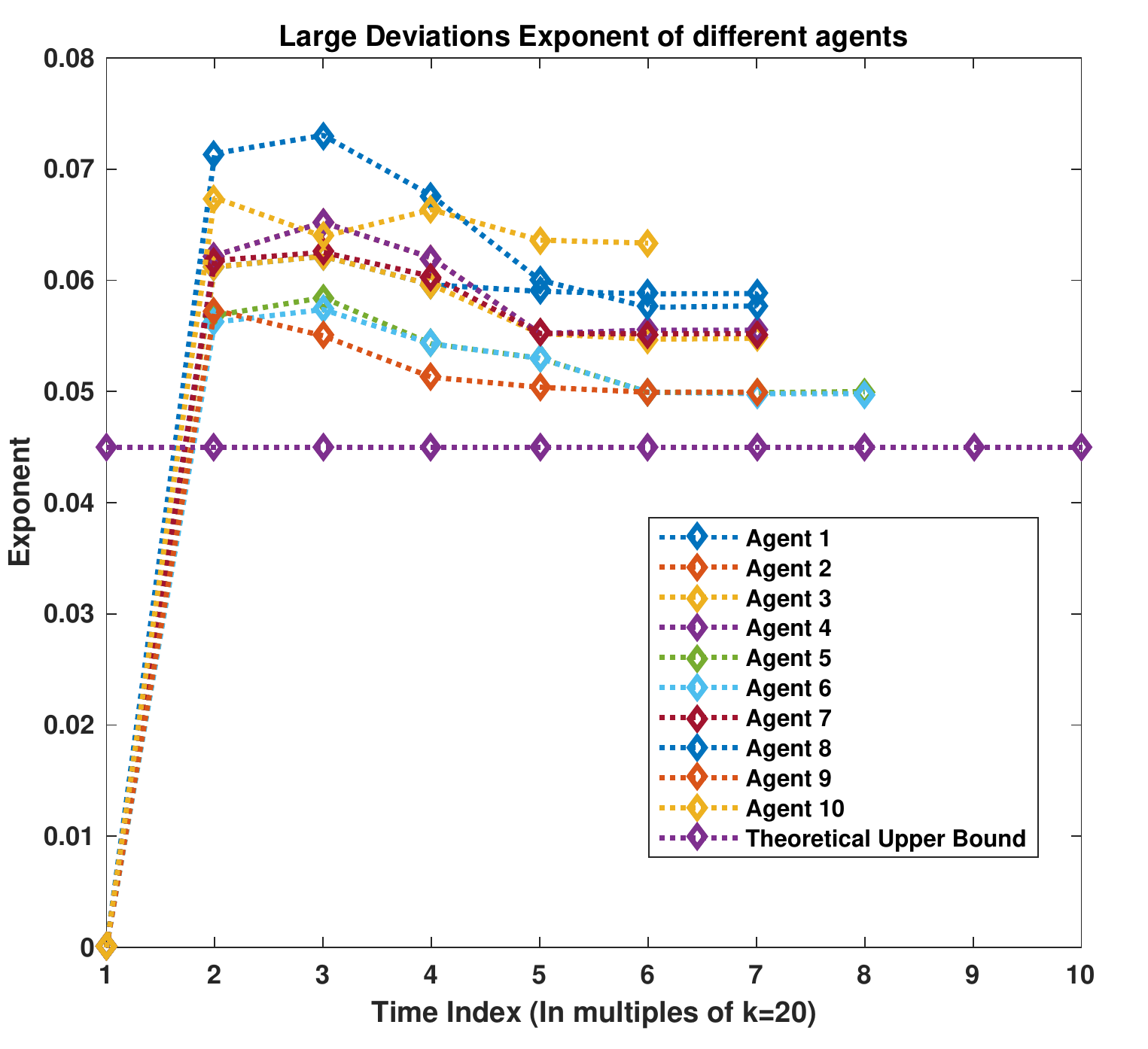}
	\caption{Large Deviations Exponent Upper bounds}\label{fig:7}
\end{figure}

\noindent For the analysis of the probability of miss, we run the algorithm for $2000$ sample paths. The threshold is chosen as $\eta=\frac{\left(\frac{1}{N}+\sqrt{N}r^{k-1}\right)\sum_{n=1}^{N}M_{n}}{2}+0.01=0.8280$. The evolution of the test statistic can be closely seen in Figure \ref{fig:6} with the probability of miss decaying exponentially and thus verifying the assertion in Theorem \ref{th:b2}. It is to be noted that, from Figure \ref{fig:6} the probability of miss starts decaying even before the parameter estimates get reasonably close to the true underlying parameter, which further indicates the \emph{online} nature of the proposed algorithm $\mathcal{CIGLRT-L}$. The large deviations exponent across different agents are plotted in Figure \ref{fig:7}. The theoretical upper bound in \eqref{eq:th_b2_2} is found to be $0.045$. The simulation results verify that the empirically estimated large deviations exponent upper bound for different agents is upper bounded by the theoretically derived upper bound\footnote{It is an upper bound if the quantity of interest is $\frac{1}{t}\log\left(\mathbb{P}(.)\right)$. It is a lower bound if the quantity of interest is the positive exponent, i.e., $-\frac{1}{t}\log\left(\mathbb{P}(.)\right)$.}

\section{Proof of Main Results : $\mathcal{CIGLRT-NL}$}
\label{sec:proof_res_ciglrt}

\subsection{Proof of Theorem \ref{t1}}
\label{subsec:th_t1}
\begin{IEEEproof}
\noindent The proof of Theorem \ref{t1} is accomplished in steps, the key ingredients being Lemma \ref{l2} and Lemma \ref{l32} which concern the boundedness of the processes $\{\mathbf{\theta}_{n}(t)\}$, $n=1,\cdots, N$ and subsequently the consistency of the agent estimate sequences respectively. To this end, we follow the basic idea developed in~\cite{kar2014asymptotically}, but with subtle modifications to take into account the state-dependent nature of the innovation gains. We state Lemma \ref{l2} and Lemma \ref{l32} here, with the proofs relegated to Appendix \ref{a1}.
\begin{Lemma}
	Let the hypothesis of Theorem \ref{t1} hold. Then, for each $n$ and $\forall \mathbf{\theta}^{*}$ the process $\{\mathbf{\theta}_{n}(t)\}$
	\label{l2}
	\begin{align}
	\label{eq:7.1}
	\mathbb{P}_{\theta^{*}}\left(\sup_{t\ge 0} \left\|\mathbf{\theta}_{n}(t)\right\| < \infty\right) =1.
	\end{align}
\end{Lemma}
\begin{Lemma}
	\label{l32}
	Let the hypotheses of Theorem \ref{t1} hold. Then, for each $n$ and $\forall \mathbf{\theta}^{*}$, we have,
	\begin{align}
	\label{eq:l32_1}
	\mathbb{P}_{\theta^{*}}\left(\lim_{t\to\infty}\mathbf{\theta}_{n}(t)=\mathbf{\theta}^{*}\right)=1.
	\end{align}
\end{Lemma}

\noindent In the sequel, we analyze the rate of convergence of the parameter estimate sequence to the true parameter.

\noindent We will use the following approximation result (Lemma \ref{l33}) and the generalized convergence criterion (Lemma \ref{l34}) for the proof of Theorem \ref{t1}.
\begin{Lemma}[Lemma 4.3 in \cite{Fabian-1}]
	\label{l33}
	Let $\{b_{t}\}$ be a scalar sequence satisfying
	\begin{align}
	\label{eq:l33_1}
	b_{t+1}\le \left(1-\frac{c}{t+1}\right)b_{t}+d_{t}(t+1)^{-\tau},
	\end{align}
	where $c > \tau, \tau > 0$, and the sequence $\{d_{t}\}$ is summable. Then, we have,
	\begin{align}
	\label{eq:l33_2}
	\limsup_{t\to\infty}~(t+1)^{\tau}b_{t}<\infty.
	\end{align}
\end{Lemma}

\begin{Lemma}[Lemma 10 in \cite{dubins1965sharper}]
	\label{l34}
	Let $\{J(t)\}$ be an $\mathbb{R}$-valued $\{\mathcal{F}_{t+1}\}$-adapted process such that $\mathbb{E}\left[J(t)|\mathcal{F}_{t}\right]=0$ a.s. for each $t\geq 1$. Then the sum $\sum_{t\geq 0}J(t)$ exists and is finite a.s. on the set where $\sum_{t\geq 0}\mathbb{E}\left[J^{2}(t)|\mathcal{F}_{t}\right]$ is finite.
	
\end{Lemma}
\noindent We now return to the proof of Theorem 4.1.

\noindent {\bf Proof of Theorem \ref{t1}}. We follow closely the corresponding development in Lemma 5.9 of~\cite{kar2013distributed}. Define $\bar{\tau}\in[0,1/2)$ such that,
\begin{align}
\label{eq:t1_pr_1}
\mathbb{P}_{\theta^{*}}\left(\lim_{t\to\infty}(t+1)^{\bar{\tau}}\left\|\mathbf{x}(t)\right\|=0\right)=1,
\end{align}
\noindent where $\mathbf{x}(t)=\mathbf{\theta}(t)-\mathbf{1}_{N}\otimes\theta^{*}$. Note that such a $\bar{\tau}$ exists by Lemma \ref{l32} (in particular, by taking $\bar{\tau}=0$). We now analyze and finally show that there exists a $\tau$ such that $\bar{\tau} < \tau <1/2$ for which the claim holds. Now, choose a $\hat{\tau} \in (\bar{\tau}, 1/2)$ and let $\mu=(\hat{\tau}+\bar{\tau})/2$. By standard algebraic manipulations, it can be readily seen that the recursion for $\{\mathbf{x}(t)\}$ satisfies
\begin{align}
\label{eq:t1_pr_2}
&\left\|\mathbf{x}(t+1)\right\|^{2}=\left\|\mathbf{x}(t)\right\|^{2}-2\beta_{t}\mathbf{x}^{\top}(t)\left(\mathbf{L} \otimes \mathbf{I}_{M}\right)\mathbf{x}(t) -2\alpha_{t}\mathbf{x}^{\top}(t)\mathbf{G}\left(\mathbf{\theta}(t)\right)\mathbf{\Sigma}^{-1}\left(\mathbf{h}\left(\mathbf{\theta}(t)\right)-\mathbf{h}\left(\mathbf{\theta}^{*}\right)\right)\nonumber\\
&+\beta_{t}^{2}\mathbf{x}^{\top}(t)\left(\mathbf{L} \otimes \mathbf{I}_{M}\right)^{2}\mathbf{x}(t)+2\alpha_{t}\beta_{t}\mathbf{x}^{\top}(t)\left(\mathbf{L} \otimes \mathbf{I}_{M}\right)\mathbf{G}\left(\mathbf{\theta}(t)\right)\mathbf{\Sigma}^{-1}\left(\mathbf{h}\left(\mathbf{\theta}(t)\right)-\mathbf{h}\left(\mathbf{\theta}^{*}\right)\right)\nonumber\\
&+\alpha_{t}^{2}\left(\mathbf{y}(t)-\mathbf{h}\left(\theta^{*}\right)\right)^{\top}\mathbf{\Sigma}^{-1}\mathbf{G}^{\top}\left(\mathbf{\theta}(t)\right)\mathbf{G}\left(\mathbf{\theta}(t)\right)\mathbf{\Sigma}^{-1}\left(\mathbf{y}(t)-\mathbf{h}\left(\theta^{*}\right)\right)\nonumber\\&+\alpha_{t}^{2}\left(\mathbf{h}\left(\mathbf{\theta}(t)\right)-\mathbf{h}\left(\mathbf{\theta}^{*}\right)\right)^{\top}\mathbf{\Sigma}^{-1}\mathbf{G}^{\top}\left(\mathbf{\theta}(t)\right)\mathbf{G}\left(\mathbf{\theta}(t)\right)\mathbf{\Sigma}^{-1}\left(\mathbf{h}\left(\mathbf{\theta}(t)\right)-\mathbf{h}\left(\mathbf{\theta}^{*}\right)\right)
\nonumber\\ &+2\alpha_{t}\mathbf{x}^{\top}(t)\mathbf{G}\left(\mathbf{\theta}(t)\right)\mathbf{\Sigma}^{-1}\left(\mathbf{y}(t)-\mathbf{h}(\mathbf{\theta}^{*})\right).
\end{align}
\noindent Let $\mathbf{J}(t)=\mathbf{G}\left(\mathbf{\theta}(t)\right)\mathbf{\Sigma}^{-1}\left(\mathbf{y}(t)-\mathbf{h}\left(\theta^{*}\right)\right)$.
\noindent From Assumption \ref{as:3}, we have that $\left\|\nabla \mathbf{h}_{n}\left(\mathbf{\theta}_{n}(t)\right)\right\|$ is uniformly bounded from above by $k_{n}$ for all $n$. Hence, we have that $\left\|\mathbf{G}\left(\mathbf{\theta}(t)\right)\right\|\le \max_{n=1,\cdots,N}k_{n}$. Now, we consider the term $\alpha_{t}^{2}\left\|\mathbf{J}(t)\right\|^{2}$. Since, the noise process under consideration is a temporally independent Gaussian sequence and $2\mu < 1$, we have,
\begin{align}
\label{eq:t1_pr_3}
\sum_{t\geq 0}(t+1)^{2\mu}\alpha_{t}^{2}\left\|\mathbf{J}(t)\right\|^{2} < \infty~\mbox{a.s.}
\end{align}

\noindent Let $\mathbf{W}(t)=\alpha_{t}\mathbf{x}^{\top}(t)\mathbf{G}\left(\mathbf{\theta}(t)\right)\mathbf{\Sigma}^{-1}\left(\mathbf{y}(t)-\mathbf{h}(\mathbf{\theta}^{*})\right)$. It follows that $\mathbb{E}_{\theta^{*}}\left[\mathbf{W}(t)|\mathcal{F}_{t}\right]=0$.

We also have that $\mathbb{E}_{\theta^{*}}\left[\mathbf{W}^{2}(t)|\mathcal{F}_{t}\right]\le\alpha_{t}^{2}\left\|\mathbf{x}(t)\right\|^{2}\left\|\mathbf{J}(t)\right\|^{2}$. Noting, that the noise under consideration is temporally independent with finite second moment, we have,
\begin{align}
\label{eq:t1_pr_4}
\mathbb{E}_{\theta^{*}}\left[\mathbf{W}^{2}(t)|\mathcal{F}_{t}\right]=o\left((t+1)^{-2-2\bar{\tau}}\right)
\end{align} and hence,
\begin{align}
\label{eq:t1_pr_5}
\mathbb{E}_{\theta^{*}}\left[(t+1)^{4\mu}\mathbf{W}^{2}(t)|\mathcal{F}_{t}\right]=o\left((t+1)^{-2+2\hat{\tau}}\right).
\end{align}

\noindent Hence, by Lemma \ref{l34}, we conclude that $\sum_{t\geq 0}(t+1)^{2\mu}\mathbf{W}(t)$ exists and is finite, as $2\hat{\tau} < 1$ and hence the left hand side (L.H.S) in \eqref{eq:t1_pr_5} is summable. Using all the inequalities derived in \eqref{eq:l2_pr_9}-\eqref{eq:l2_pr_14}, we have,
\begin{align}
\label{eq:t1_pr_6}
\left\|\mathbf{x}(t+1)\right\|^{2}\le\left(1-c_{1}\alpha_{t}+c_{5}\left(\alpha_{t}\beta_{t}+\alpha_{t}^{2}\right)\right)\left\|\mathbf{x}(t)\right\|^{2}-c_{6}(\beta_{t}-\beta^{2}_{t})\left\|\mathbf{x}_{C\perp}(t)\right\|^{2}+\alpha_{t}^{2}\left\|\mathbf{J}(t)\right\|^{2}+2\mathbf{W}(t).
\end{align}

\noindent Finally, noting that $c_{1}\alpha_{t}$ dominates $c_{5}\left(\alpha_{t}\beta_{t}+\alpha_{t}^{2}\right)$ and $\beta_{t}$ dominates $\beta_{t}^{2}$, we obtain
\begin{align}
\label{eq:t1_pr_7}
\left\|\mathbf{x}(t+1)\right\|^{2}\le\left(1-c_{1}\alpha_{t}\right)\left\|\mathbf{x}(t)\right\|^{2}+\alpha_{t}^{2}\left\|\mathbf{J}(t)\right\|^{2}+2\mathbf{W}(t).
\end{align}

\noindent Now, using the analysis in \eqref{eq:t1_pr_3}-\eqref{eq:t1_pr_5}, we have, from \eqref{eq:t1_pr_7}
\begin{align}
\label{eq:t1_pr_8}
\left\|\mathbf{x}(t+1)\right\|^{2}\le\left(1-c_{1}\alpha_{t}\right)\left\|\mathbf{x}(t)\right\|^{2}+d_{t}(t+1)^{-2\mu},
\end{align}
\noindent where
\begin{align}
\label{eq:t1_pr_9}
d_{t}(t+1)^{-2\mu}=\alpha_{t}^{2}\left\|\mathbf{J}(t)\right\|^{2}+2\mathbf{W}(t).
\end{align}
\noindent Finally, noting that $c_{1}\alpha_{t}(t+1)\geq 1>2\mu$, an immediate application of Lemma \ref{l33} gives
\begin{align}
\label{eq:t1_pr_10}
\limsup_{t\to\infty}(t+1)^{2\mu}\left\|\mathbf{x}(t)\right\|^{2} < \infty~ \mbox{a.s.}
\end{align}

\noindent So, we have that, there exists a $\tau$ with $\bar{\tau}<\tau<\mu$ for which $(t+1)^{\tau}\left\|\mathbf{x}(t)\right\|\to 0$ as $t\to\infty$. Thus for every $\bar{\tau}$ for which \eqref{eq:t1_1} holds, there exists $\tau\in(\bar{\tau},1/2)$ for which the result in \eqref{eq:t1_1} continues to hold. We thus conclude that the result holds for all $\tau \in [0,1/2)$.

\end{IEEEproof}

\subsection{Proof of Theorem \ref{t2}}
\label{subsec:th_t2}

\noindent \begin{IEEEproof}
\noindent The proof of Theorem \ref{t2} needs the following Lemma from \cite{Fabian-2} (stated in a form suitable to our needs) concerning the asymptotic normality of non-Markov stochastic recursions and an intermediate result which concerns with the asymptotic normality of the averaged decision statistic.
\begin{Lemma}[Theorem 2.2 in \cite{Fabian-2}]
\label{main_res_l0}
\noindent Let $\{\mathbf{z}_{t}\}$ be an $\mathbb{R}^{k}$-valued $\{\mathcal{F}_{t}\}$-adapted process that satisfies
\begin{align}
\label{eq:l0_1}
\mathbf{z}_{t+1}=\left(\mathbf{I}_{k}-\frac{1}{t+1}\mathbf{\Gamma}_{t}\right)\mathbf{z}_{t}+(t+1)^{-1}\mathbf{\Phi}_{t}\mathbf{V}_{t}+(t+1)^{-3/2}\mathbf{T}_{t},
\end{align}
\noindent where the stochastic processes $\{\mathbf{V}_{t}\}, \{\mathbf{T}_{t}\} \in \mathbb{R}^{k}$ while $\{\mathbf{\Gamma}_{t}\}, \{\mathbf{\Phi}_{t}\} \in \mathbb{R}^{k\times k}$. Moreover, for each $t$, $\mathbf{V}_{t-1}$ and $\mathbf{T}_{t}$ are $\mathcal{F}_{t}$ -adapted, whereas the processes $\{\mathbf{\Gamma}_{t}\}$, $\{\mathbf{\Phi}_{t}\}$ are $\{\mathcal{F}_{t}\}$ adapted.

\noindent Also, assume that
\begin{align}
\label{eq:l0_2}
\mathbf{\Gamma}_{t}\to\mathbf{I}_{k}, ~\mathbf{\Phi}_{t}\to\mathbf{\Phi}~ \mbox{and} ~\mathbf{T}_{t}\to 0 ~\mbox{a.s. as}~ t\to\infty.
\end{align}
\noindent Furthermore, let the sequence $\{\mathbf{V}_{t}\}$ satisfy $\mathbb{E}\left[\mathbf{V}_{t}|\mathcal{F}_{t}\right]=0$ for each $t$ and suppose there exists a positive constant $C$ and a matrix $\Sigma$ such that $C > \left\|\mathbb{E}\left[\mathbf{V}_{t}\mathbf{V}_{t}^{\top}|\mathcal{F}_{t}\right]-\Sigma\right\|\to 0$ a.s. as $t\to\infty$ and with $\sigma_{t,r}^{2}=\int_{\left\|\mathbf{V}_{t}\right\|^{2} \ge r(t+1)}\left\|\mathbf{V}_{t}\right\|^{2}d\mathbb{P}$, let $\lim_{t\to\infty}\frac{1}{t+1}\sum_{s=0}^{t}\sigma_{s,r}^{2}=0$ for every $r > 0$.

\noindent Then, we have,
\begin{align}
\label{eq:l0_3}
(t+1)^{1/2}\mathbf{z}_{t}\overset{\mathcal{D}}{\Longrightarrow}\mathcal{N}\left(\mathbf{0}, \mathbf{\Phi}\Sigma\mathbf{\Phi}^{\top}\right).
\end{align}
\end{Lemma}
We state the lemma concerning the asymptotic normality of the averaged decision statistic here, while the proof is relegated to Appendix \ref{a1}.
\begin{Lemma}
\label{main_res_l1}
\noindent Let the hypotheses of Theorem \ref{t2} hold. Consider the averaged decision statistic sequence, $\{z_{\mbox{\scriptsize{avg}}}(t)\}$, defined as $z_{\mbox{\scriptsize{avg}}}(t)=\frac{1}{N}\sum_{n=1}^{N}z_{n}(t)$. Then, we have, under $\mathbb{P}_{\mathbf{\theta}^{*}}$ for all $\left\|\mathbf{\theta}^{*}\right\| > 0$,
\begin{align}
\label{eq:main_res_11_1}
\sqrt{t+1}\left(z_{\mbox{\scriptsize{avg}}}(t)-\frac{\mathbf{h}^{\top}\left(\mathbf{\theta}_{N}^{*}\right)\mathbf{\Sigma}^{-1}\mathbf{h}\left(\mathbf{\theta}_{N}^{*}\right)}{2N}\right)\overset{\mathcal{D}}{\Longrightarrow} \mathcal{N}\left(0, \frac{\mathbf{h}^{\top}\left(\mathbf{\theta}_{N}^{*}\right)\mathbf{\Sigma}^{-1}\mathbf{h}\left(\mathbf{\theta}_{N}^{*}\right)}{N^{2}}\right), \forall n.
\end{align}
\end{Lemma}\noindent We now use a lemma which establishes that the sequences $\{z_{\mbox{\scriptsize{avg}}}(t)\}$ and $\{z_{n}(t)\}$ are indistinguishable in the $\sqrt{t}$ time scale. We state the lemma here, while the proof is relegated to Appendix \ref{a1}.
\begin{Lemma}
	\label{int_res_2}
	\noindent Given the averaged decision statistic sequence, $\{z_{\mbox{\scriptsize{avg}}}(t)\}$, for each $\delta_{0}\in [0, 1)$ we have
	\begin{align}
	\label{eq:int_res_2_2}
	\mathbb{P}_{\theta^{*}}(\lim_{t\to\infty}(t+1)^{\delta_{0}}(\mathbf{z}(t)-\mathbf{1}_{N}\otimes z_{\mbox{\scriptsize{avg}}}(t)=\mathbf{0})=1.
	\end{align}
\end{Lemma}

We now return to the proof of Theorem~\ref{t2}.

\noindent {\bf Proof of Theorem~\ref{t2}}. Note that as $\delta_{0}$ in Lemma \ref{int_res_2} can be chosen to be greater than $\frac{1}{2}$, we have for all $n$,
\begin{align}
\label{eq:main_t2_pr1}
&\mathbb{P}_{\theta^{*}}\left(\lim_{t\to\infty}\left\|\sqrt{t+1}\left(z_{n}(t)-\frac{\mathbf{h}^{\top}\left(\mathbf{\theta}_{N}^{*}\right)\mathbf{\Sigma}^{-1}\mathbf{h}\left(\mathbf{\theta}_{N}^{*}\right)}{2N}\right)-\sqrt{t+1}\left(z_{\mbox{\scriptsize{avg}}}(t)-\frac{\mathbf{h}^{\top}\left(\mathbf{\theta}_{N}^{*}\right)\mathbf{\Sigma}^{-1}\mathbf{h}\left(\mathbf{\theta}_{N}^{*}\right)}{2N}\right)\right\|=0\right)\nonumber\\
&=\mathbb{P}_{\theta^{*}}\left(\lim_{t\to\infty}\left\|\sqrt{t+1}\left(z_{n}(t)-z_{\mbox{\scriptsize{avg}}}(t)\right)\right\|=0\right)\nonumber\\
&=\mathbb{P}_{\theta^{*}}\left(\lim_{t\to\infty}\left\|(t+1)^{0.5-\delta_{0}}(t+1)^{\delta_{0}}\left(z_{n}(t)-z_{\mbox{\scriptsize{avg}}}(t)\right)\right\|=0\right)=1,
\end{align}
\noindent where the last step follows from Lemma \ref{int_res_2} and the fact that $\delta_{0}>1/2$. Thus, the difference of the sequences $\left\{\sqrt{t+1}\left(z_{n}(t)-\frac{\mathbf{h}^{\top}\left(\mathbf{\theta}_{N}^{*}\right)\mathbf{\Sigma}^{-1}\mathbf{h}\left(\mathbf{\theta}_{N}^{*}\right)}{2N}\right)\right\}$ and $\left\{\sqrt{t+1}\left(z_{\mbox{\scriptsize{avg}}}(t)-\frac{\mathbf{h}^{\top}\left(\mathbf{\theta}_{N}^{*}\right)\mathbf{\Sigma}^{-1}\mathbf{h}\left(\mathbf{\theta}_{N}^{*}\right)}{2N}\right)\right\}$ converges a.s. to zero and hence we have,
\begin{align}
\label{eq:main_t2_pr2}
\sqrt{t+1}\left(z_{n}(t)-\frac{\mathbf{h}^{\top}\left(\mathbf{\theta}_{N}^{*}\right)\mathbf{\Sigma}^{-1}\mathbf{h}\left(\mathbf{\theta}_{N}^{*}\right)}{2N}\right)\overset{\mathcal{D}}{\Longrightarrow} \mathcal{N}\left(0,\frac{\mathbf{h}^{\top}\left(\mathbf{\theta}_{N}^{*}\right)\mathbf{\Sigma}^{-1}\mathbf{h}\left(\mathbf{\theta}_{N}^{*}\right)}{N^{2}}\right).
\end{align}
\end{IEEEproof}

\subsection{Proof of Theorem \ref{t3}}
\label{subsec:th_t3}

\begin{IEEEproof}
\noindent From \eqref{eq:dec_rule_2}, we have,
\begin{align}
\label{eq:t3_pr_1}
&\mathbb{P}_{M,\theta^{*}}(t)=\mathbb{P}_{1,\theta^{*}}\left(z_{n}(t) < \eta\right)\nonumber\\
&=\mathbb{P}_{1,\theta^{*}}\left(z_{n}(t)-\frac{\mathbf{h}^{\top}\left(\mathbf{\theta}_{N}^{*}\right)\mathbf{\Sigma}^{-1}\mathbf{h}\left(\mathbf{\theta}_{N}^{*}\right)}{2N}<\eta-\frac{\mathbf{h}^{\top}\left(\mathbf{\theta}_{N}^{*}\right)\mathbf{\Sigma}^{-1}\mathbf{h}\left(\mathbf{\theta}_{N}^{*}\right)}{2N}\right)\nonumber\\
&=\mathbb{P}_{1,\theta^{*}}\left(\sqrt{t+1}\left(z_{n}(t)-\frac{\mathbf{h}^{\top}\left(\mathbf{\theta}_{N}^{*}\right)\mathbf{\Sigma}^{-1}\mathbf{h}\left(\mathbf{\theta}_{N}^{*}\right)}{2N}\right)<\sqrt{t+1}\left(\eta-\frac{\mathbf{h}^{\top}\left(\mathbf{\theta}_{N}^{*}\right)\mathbf{\Sigma}^{-1}\mathbf{h}\left(\mathbf{\theta}_{N}^{*}\right)}{2N}\right)\right).
\end{align}

\noindent Now, invoking Theorem \ref{t2}, where we have established the asymptotic normality for the decision statistic sequence $\{z_{n}(t)\}$, we have,
\begin{align}
\label{eq:t3_pr_2}
&\lim_{t\to\infty}\mathbb{P}_{1,\theta^{*}}\left(\sqrt{t+1}\left(z_{n}(t)-\frac{\mathbf{h}^{\top}\left(\mathbf{\theta}_{N}^{*}\right)\mathbf{\Sigma}^{-1}\mathbf{h}\left(\mathbf{\theta}_{N}^{*}\right)}{2N}\right)<\sqrt{t+1}\left(\eta-\frac{\mathbf{h}^{\top}\left(\mathbf{\theta}_{N}^{*}\right)\mathbf{\Sigma}^{-1}\mathbf{h}\left(\mathbf{\theta}_{N}^{*}\right)}{2N}\right)\right)\nonumber\\
&=\mathbb{P}_{1,\theta^{*}}(z < -\infty)=0,
\end{align}
\noindent where $z$ is a normal random variable with $z\sim\mathcal{N}\left(0,\frac{\mathbf{h}^{\top}\left(\mathbf{\theta}_{N}^{*}\right)\mathbf{\Sigma}^{-1}\mathbf{h}\left(\mathbf{\theta}_{N}^{*}\right)}{N^{2}}\right)$. In the derivation of \eqref{eq:t3_pr_2} we have used the Portmanteau characterization for weak convergence and the fact that
\begin{align}
\label{eq:t3_pr_21}
\eta < \frac{\mathbf{h}^{\top}\left(\mathbf{\theta}_{N}^{*}\right)\mathbf{\Sigma}^{-1}\mathbf{h}\left(\mathbf{\theta}_{N}^{*}\right)}{2N}.
\end{align}
\noindent Hence, we have, from \eqref{eq:t3_pr_1} and \eqref{eq:t3_pr_2}
\begin{align}
\label{eq:t3_pr_3}
\lim_{t\to\infty}\mathbb{P}_{M,\theta^{*}}(t)=0
\end{align}
as long as \eqref{eq:t3_pr_21} holds.

\noindent For the null hypothesis $\mathcal{H}_{0}$, from \eqref{eq:dec_rule_2} and with $0<\lambda <1$, we have,
\begin{align}
\label{eq:t3_pr_4}
&\mathbb{P}_{FA}(t)=\mathbb{P}_{0}\left(z_{n}(t) > \eta\right)\nonumber\\
&=\mathbb{P}_{0}\left(\frac{1}{t}\sum_{s=0}^{t-1}\mathbf{e}_{n}^{\top}\mathbf{W}^{t-1-s}\mathbf{h}^{\top}\left(\mathbf{\theta}(s)\right)\mathbf{\Sigma}^{-1}\left(\mathbf{y}(s)-\frac{\mathbf{h}\left(\mathbf{\theta}(s)\right)}{2}\right) > \eta\right)\nonumber\\
&=\mathbb{P}_{0}\left(\frac{1}{t}\sum_{s=0}^{t-1}\sum_{j=1}^{N}\phi_{n,j}(s,t-1)\left(\mathbf{h}_{j}^{\top}\left(\mathbf{\theta}_{j}(s)\right)\mathbf{\Sigma}_{j}^{-1}\mathbf{\gamma}_{j}(s)-\frac{\mathbf{h}_{j}^{\top}\left(\mathbf{\theta}_{j}(s)\right)\mathbf{\Sigma}_{j}^{-1}\mathbf{h}_{j}\left(\mathbf{\theta}_{j}(s)\right)}{2}\right)>\eta\right)\nonumber\\
&=\mathbb{P}_{0}\left(\frac{1}{t}\sum_{s=0}^{t-1}\sum_{j=1}^{N}\phi_{n,j}(s,t-1)\left(\frac{\mathbf{\gamma}_{j}^{\top}(s)\mathbf{\Sigma}_{j}^{-1}\mathbf{\gamma}_{j}(s)}{2}-\frac{\left(\mathbf{\gamma}_{j}(s)-\mathbf{h}_{j}\left(\mathbf{\theta}_{j}(s)\right)\right)^{\top}\mathbf{\Sigma}_{j}^{-1}\left(\mathbf{\gamma}_{j}(s)-\mathbf{h}_{j}\left(\mathbf{\theta}_{j}(s)\right)\right)}{2}\right)>\eta\right)\nonumber\\
&\le \mathbb{P}_{0}\left(\frac{1}{t}\sum_{s=0}^{t-1}\sum_{j=1}^{N}\phi_{n,j}(s,t-1)\frac{\mathbf{\gamma}_{j}^{\top}(s)\mathbf{\Sigma}_{j}^{-1}\mathbf{\gamma}_{j}(s)}{2}>\eta\right)\nonumber\\
&\overset{(a)}{\le} \mathbb{P}_{0}\left(\frac{1}{t}\sum_{s=0}^{t-1}\sum_{j=1}^{N}\left(\frac{1}{N}+\sqrt{N}r^{t-1-s}\right)\frac{\mathbf{\gamma}_{j}^{\top}(s)\mathbf{\Sigma}_{j}^{-1}\mathbf{\gamma}_{j}(s)}{2}>\eta\right)\nonumber\\
&\le\exp\left(-\frac{t\eta\lambda}{\frac{1}{N}+\sqrt{N}}\right)\prod_{j=1}^{N}\prod_{s=0}^{t-1}\mathbb{E}_{0}\left[\exp\left(\lambda\frac{\frac{1}{N}+\sqrt{N}r^{t-1-s}}{\frac{2}{N}+2\sqrt{N}}\mathbf{\gamma}_{j}(s)^{\top}\mathbf{\Sigma}_{j}^{-1}\mathbf{\gamma}_{j}(s)\right)\right]\nonumber\\
&\overset{(b)}{=}\exp\left(-\frac{t\eta\lambda}{\frac{1}{N}+\sqrt{N}}\right)\exp\left(-\sum_{s=0}^{t-1}\left(\frac{\sum_{n=1}^{N}M_{n}}{2}\right)\log\left(1-\frac{\lambda\left(\frac{1}{N}+\sqrt{N}r^{t-1-s}\right)}{\frac{1}{N}+\sqrt{N}}\right)\right)\nonumber\\
&\le \exp\left(-\frac{t\eta\lambda}{\frac{1}{N}+\sqrt{N}}\right)\exp\left(-\left(\frac{\sum_{n=1}^{N}M_{n}}{2}\right)\log\left(1-\lambda\right)\right)\exp\left(-\left(t-1\right)\left(\frac{\sum_{n=1}^{N}M_{n}}{2}\right)\log\left(1-\frac{\lambda\left(\frac{1}{N}+\sqrt{N}r\right)}{\frac{1}{N}+\sqrt{N}}\right)\right),
\end{align}

\noindent where $\phi_{n,j}\left(s,t-1\right)$ denotes the $(n,j)$-th element of $\mathbf{W}^{t-1-s}$, $(a)$ follows due to $\|\phi_{n,j}\left(s,t-1\right)-\frac{1}{N}\|\le \sqrt{N}r^{t-1-s}$ and $(b)$ follows due to the fact that the random variable $\mathbf{\gamma}_{j}(s)^{\top}\mathbf{\Sigma}_{j}^{-1}\mathbf{\gamma}_{j}(s)$ is a chi-squared random variable with $M_{j}$ degrees of freedom and the associated moment generating function exists since $\lambda < 1$.

\noindent Now, taking limits on both sides of the equation \eqref{eq:t3_pr_4}, we have,
\begin{align}
\label{eq:t3_pr_41}
&\frac{1}{t}\log\left(\mathbb{P}_{0}\left(z_{n}(t) > \eta \right)\right)\nonumber\\
&\le -\frac{\eta\lambda}{\frac{1}{N}+\sqrt{N}}-\left(\frac{\sum_{n=1}^{N}M_{n}}{2t}\right)\log\left(1-\lambda\right)-\frac{t-1}{t}\left(\frac{\sum_{n=1}^{N}M_{n}}{2}\right)\log\left(1-\frac{\lambda\left(\frac{1}{N}+\sqrt{N}r\right)}{\frac{1}{N}+\sqrt{N}}\right)\nonumber\\
&\Rightarrow\limsup_{t\to\infty}\frac{1}{t}\log\left(\mathbb{P}_{0}\left(z_{n}(t) > \eta \right)\right)\nonumber\\
&\le -\frac{\eta\lambda}{\frac{1}{N}+\sqrt{N}}-\left(\frac{\sum_{n=1}^{N}M_{n}}{2}\right)\log\left(1-\frac{\lambda\left(\frac{1}{N}+\sqrt{N}r\right)}{\frac{1}{N}+\sqrt{N}}\right)=-LE(\lambda).
\end{align}
\noindent First we note that, as \eqref{eq:t3_pr_41} holds for all $\lambda \in (0,1)$, we have that
\begin{align}
\label{eq:t3_pr_411}
\limsup_{t\to\infty}\frac{1}{t}\log\left(\mathbb{P}_{0}\left(z_{n}(t) > \eta \right)\right)\le-LE(1-\epsilon),
\end{align}
where $\epsilon\in (0,1)$. Moreover, as $LE(\lambda)$ is a continuous function of $\lambda$ in the interval $\lambda \in (0,1]$, we can force $\epsilon$ to zero and thereby conclude that
\begin{align}
\label{eq:t3_pr_412}
\limsup_{t\to\infty}\frac{1}{t}\log\left(\mathbb{P}_{0}\left(z_{n}(t) > \eta \right)\right)\le-LE(1).
\end{align}
\noindent Now consider $\lambda^{*}$ which is given by
\begin{align}
\label{eq:t3_pr_6}
\lambda^{*}=\frac{\frac{1}{N}+\sqrt{N}}{\frac{1}{N}+\sqrt{N}r}-\frac{\left(\frac{1}{N}+\sqrt{N}\right)\sum_{n=1}^{N}M_{n}}{2\eta}.
\end{align}
\noindent It is to be noted that $\lambda^{*}$ is positive when
\begin{align}
\label{eq:t3_pr_7}
\eta > \frac{\left(\frac{1}{N}+\sqrt{N}r\right)\sum_{n=1}^{N}M_{n}}{2}.
\end{align}
\noindent Furthermore, $LE(\lambda)$ is maximixed at $\lambda=\lambda^{*}$ when $\lambda^{*}\in(0,1)$. Hence, in the case when $\lambda^{*}\in(0,1)$, we have
\begin{align}
\label{eq:t3_pr_413}
\limsup_{t\to\infty}\frac{1}{t}\log\left(\mathbb{P}_{0}\left(z_{n}(t) > \eta \right)\right)\le-LE(\lambda^{*}).
\end{align}

\noindent It is to be noted that $LE(\lambda)$ is an increasing function of $\lambda$ in the interval $(0,\lambda^{*})$ and hence in the case when $\lambda^{*} > 1$, we have that $LE(\lambda)$ is non-negative and increasing in the interval $(0,1)$ and we have the exponent as $LE(1)$ from \eqref{eq:t3_pr_412}.
Finally, combining \eqref{eq:t3_pr_412} and \eqref{eq:t3_pr_413}, we have,
\begin{align}
\label{eq:t3_pr_8}
\limsup_{t\to\infty}\frac{1}{t}\log\left(\mathbb{P}_{0}\left(z_{n}(t)>\eta\right)\right)\le -LE\left(\min\{\lambda^{*},1\}\right).
\end{align}
%Hence, for $LE(\lambda)$, we obtain,
%\begin{align}
%\label{eq:t3_pr_5}
%\limsup_{t\to\infty}\frac{1}{t}\log\left(\mathbb{P}_{0}\left(z_{n}(t)>\eta\right)\right)\le -\frac{\eta}{\frac{1}{N}+\sqrt{N}r}-\frac{\sum_{n=1}^{N}M_{n}}{2}\left(\frac{1}{N}+\sqrt{N}+\log\frac{2\eta}{\left(\frac{1}{N}+\sqrt{N}r\right)\sum_{n=1}^{N}M

%\noindent Combining the threshold choices obtained in \eqref{eq:t3_pr_21} and \eqref{eq:t3_pr_7}, we have,
%\begin{align}
%\label{eq:t3_pr_9}
%\frac{\left(\frac{1}{N}+\sqrt{N}r\right)\sum_{n=1}^{N}M_{n}}{2}<\eta<\frac{\mathbf{h}^{\top}\left(\mathbf{\theta}_{N}^{*}\right)\mathbf{\Sigma}^{-1}\mathbf{h}\left(\mathbf{\theta}_{N}^{*}\right)}{2N}.
%\end{align}

\noindent Finally, the above arguments and the threshold choices obtained in \eqref{eq:t3_pr_21} and \eqref{eq:t3_pr_7} establish that as long as the true $\theta^{*}$ satisfies the following condition
\begin{align}
\label{eq:t3_pr_10}
\frac{\mathbf{h}^{\top}\left(\mathbf{\theta}_{N}^{*}\right)\mathbf{\Sigma}^{-1}\mathbf{h}\left(\mathbf{\theta}_{N}^{*}\right)}{2N} > \frac{\left(\frac{1}{N}+\sqrt{N}r\right)\sum_{n=1}^{N}M_{n}}{2},
\end{align}
any $\eta$ satisfying
\begin{align}
\label{eq:t3_pr_9}
\frac{\left(\frac{1}{N}+\sqrt{N}r\right)\sum_{n=1}^{N}M_{n}}{2}<\eta<\frac{\mathbf{h}^{\top}\left(\mathbf{\theta}_{N}^{*}\right)\mathbf{\Sigma}^{-1}\mathbf{h}\left(\mathbf{\theta}_{N}^{*}\right)}{2N}.
\end{align}
would guarantee $\mathbb{P}_{M,\theta^{*}}(t), \mathbb{P}_{FA}(t) \to 0$ as $t\to\infty$. Hence, the assertion is proved.
\end{IEEEproof}	

\section{Proof of Main Results : $\mathcal{CIGLRT-L}$}
	\label{sec:proof_res_cilrt}

\subsection{Proof of Theorem \ref{th:b1}}
\label{subsec:th_b1}
\noindent\begin{IEEEproof}
	\noindent The proof follows from the proof of Theorem $12$ in \cite{KarMoura-LinEst-JSTSP-2011}. To be specific, with $\gamma_{0}=0$, $\mathbf{L}(i)=\mathbf{L}, \forall i$ ($i$ denotes time in \cite{KarMoura-LinEst-JSTSP-2011})~and $\mathbf{K}=\mathbf{I}$ in \cite{KarMoura-LinEst-JSTSP-2011}, the algorithm $\mathcal{GLU}$ in \cite{KarMoura-LinEst-JSTSP-2011} exactly reduces to the parameter estimate update of the $\mathcal{CIGLRT-L}$ algorithm as defined in \eqref{eq:est_upt_lin3}.
\end{IEEEproof}

\subsection{Proof of Theorem \ref{th:b2}}
\label{subsec:th_b2}
\noindent\begin{IEEEproof}
	The following result which characterizes $\left\|\mathbf{I}_{NM}-\beta_{t}\left(\mathbf{L}\otimes\mathbf{I}_{M}\right)-\alpha_{t}\mathbf{G}_{H}\mathbf{\Sigma}^{-1}\mathbf{G}_{H}^{\top}\right\|$, will be crucial for the subsequent analysis.
	\noindent We state the result here, while the proof is relegated to Appendix \ref{a2}.
	 \begin{Lemma}
	 	\label{le:det1}
	 	Let the Assumptions~\ref{bs:1}-\ref{bs:3} hold. Consider the parameter estimate update of the $\mathcal{CIGLRT-L}$ algorithm in \eqref{eq:est_upt_lin3}. Then, we have,
	 	\begin{align}
	 	\label{eq:d1}
	 	\left\|\mathbf{I}_{NM}-\beta_{t}\left(\mathbf{L}\otimes\mathbf{I}_{M}\right)-\alpha_{t}\mathbf{G}_{H}\mathbf{\Sigma}^{-1}\mathbf{G}_{H}^{\top}\right\|\le 1-c_{1}\alpha_{t},~~\forall t \ge t_{1},
	 	\end{align}
	 	where
	 	\begin{align}
	 	\label{eq:d12}
	 	c_{1}=\min_{\left\|x\right\|=1}x^{\top}\left(\mathbf{L}\otimes\mathbf{I}_{M}+\mathbf{G}_{H}\mathbf{\Sigma}^{-1}\mathbf{G}_{H}^{\top}\right)x=\lambda_{\mbox{\scriptsize{min}}}\left(\mathbf{L}\otimes\mathbf{I}_{M}+\mathbf{G}_{H}\mathbf{\Sigma}^{-1}\mathbf{G}_{H}^{\top}\right),
	 	\end{align}
	 	\begin{align}
	 	\label{eq:d13}
	 	t_{1}=\max\{t_{2}, t_{3}\},
	 	\end{align}
	 	and $t_{2}, t_{3}$ are positive constants~(integers) chosen such that $\forall t\geq t_{2}$,
	 	\begin{align}
	 	\label{eq:ld3}
	 	\beta_{t}\lambda_{N}\left(\mathbf{L}\right)+\alpha_{t}\lambda_{\mbox{\scriptsize{max}}}\left(\mathbf{G}_{H}\mathbf{\Sigma}^{-1}\mathbf{G}_{H}^{\top}\right)\le 1,
	 	\end{align}
	 	 and $\forall t\geq t_{3}$,
	 	\begin{align}
	 	\label{eq:d2}
		\alpha_{t}\lambda_{\mbox{\scriptsize{min}}}\left(\mathbf{L}\otimes\mathbf{I}_{M}+\mathbf{G}_{H}\mathbf{\Sigma}^{-1}\mathbf{G}_{H}^{\top}\right) < 1
	 	\end{align}
	respectively.
	 \end{Lemma}
	
	\noindent Under the null hypothesis, we have, for all $\lambda \in (0,1)$,
	\begin{align}
	\label{eq:f1}
	z_{n}(kt)=\mathbf{e}_{n}^{\top}\mathbf{W}^{k-1}\mathbf{G}_{\theta}(k(t-1))\mathbf{\Sigma}^{-1}\left(\mathbf{s}(k(t-1))-\frac{\mathbf{G}_{H}^{\top}\mathbf{\theta}(k(t-1))}{2}\right).
	\end{align}
	\noindent From \eqref{eq:f1}, we have,
	\begin{align}
	\label{eq:f2}
	&\mathbb{P}_{0}\left(z_{n}(kt) > \eta\right)\le e^{-\frac{(k(t-1)+1)\eta\lambda}{\frac{1}{N}+\sqrt{N}r^{k-1}}}\mathbb{E}_{0}\left[e^{\frac{(k(t-1)+1)}{\frac{1}{N}+\sqrt{N}r^{k-1}}\lambda z_{n}(kt)}\right]\nonumber\\
	&\overset{(a)}{=} e^{-\frac{(k(t-1)+1)\eta\lambda}{\frac{1}{N}+\sqrt{N}r^{k-1}}}\mathbb{E}_{0}\left[\exp\left(\frac{\lambda}{\frac{1}{N}+\sqrt{N}r^{k-1}}\sum_{i=0}^{k(t-1)}\sum_{j=1}^{N}\phi_{n,j}(k-1)\left(\frac{\mathbf{\gamma}_{j}^{\top}(i)\mathbf{\Sigma}_{j}^{-1}\mathbf{\gamma}_{j}(i)}{2}\right.\right.\right.\nonumber\\&\left.\left.\left.-\frac{\left(\gamma_{j}(i)-\mathbf{H}_{j}\mathbf{\theta}_{j}(t-1)\right)^{\top}\mathbf{\Sigma}_{j}^{-1}\left(\gamma_{j}(i)-\mathbf{H}_{j}\mathbf{\theta}_{j}(t-1)\right)}{2}\right)\right)\right]\nonumber\\
	&\overset{(b)}{\le} e^{-\frac{(k(t-1)+1)\eta\lambda}{\frac{1}{N}+\sqrt{N}r^{k-1}}}\mathbb{E}_{0}\left[\exp\left(\frac{\lambda}{\frac{1}{N}+\sqrt{N}r^{k-1}}\sum_{j=1}^{N}\phi_{n,j}(k-1)\sum_{i=0}^{k(t-1)}\frac{\mathbf{\gamma}_{j}^{\top}(i)\mathbf{\Sigma}_{j}^{-1}\mathbf{\gamma}_{j}(i)}{2}\right)\right]\nonumber\\
	&\overset{(c)}{\le} e^{-\frac{(k(t-1)+1)\eta\lambda}{\frac{1}{N}+\sqrt{N}r^{k-1}}}\mathbb{E}_{0}\left[\exp\left(\lambda\sum_{j=1}^{N}\sum_{i=0}^{k(t-1)}\frac{\mathbf{\gamma}_{j}^{\top}(i)\mathbf{\Sigma}_{j}^{-1}\mathbf{\gamma}_{j}(i)}{2}\right)\right]\nonumber\\
	&\overset{(d)}{=} e^{-\frac{(k(t-1)+1)\eta\lambda}{\frac{1}{N}+\sqrt{N}r^{k-1}}}\prod_{j=1}^{N}\prod_{i=0}^{k(t-1)}\mathbb{E}_{0}\left[\exp\left(\frac{\lambda\mathbf{\gamma}_{j}^{\top}(i)\mathbf{\Sigma}_{j}^{-1}\mathbf{\gamma}_{j}(i)}{2}\right)\right]\nonumber\\
	&\overset{(e)}{=}\exp\left(-\frac{\lambda\eta(k(t-1)+1)}{\frac{1}{N}+\sqrt{N}r^{k-1}}-\frac{(k(t-1)+1)\sum_{n=1}^{N}M_{n}}{2}\log(1-\lambda)\right),
	\end{align}
	
	\noindent where $\phi_{n,j}(k-1)$ denotes the $(n,j)$-th entry of $\mathbf{W}^{k-1}$ and $r$ denotes $\|\mathbf{W}-\mathbf{J}\|$. It is to be noted that $(a)$ follows due to the fact that under the null hypothesis the observations made at the agents are of the form $\mathbf{y}_{n}(t)=\mathbf{\gamma}_{n}(t)$, $(b)$ follows due to the fact that the inverse covariances are positive definite and hence the quadratic forms are positive, $(c)$ follows due to $|\phi_{n,j}(k-1)-\frac{1}{N}|\le \sqrt{N}r^{k-1}$, $(d)$ follows due to the independence of the noise processes over time and space and $(e)$ follows due to the fact that for each $i, j$ the random variable $\gamma_{j}(i)^{\top}\Sigma_{j}^{-1}\gamma_{j}(i)$ corresponds to a standard chi-squared random variable with $M_{j}$ degrees of freedom and the associated moment generating functions\footnote{The moment generating function $\mathbb{E}\left[\exp(\rho z)\right]$ of a chi-squared random variable $z$ with $M_{n}$ degrees of freedom exists and is given by $(1-2\rho)^{-\frac{M_{n}}{2}}$ for all $\rho < 1/2$.} exists since $\lambda < 1$.
	
	\noindent Taking limits on both sides, we have,
	\begin{align}
	\label{eq:f3}
	\limsup_{t\to\infty}\frac{1}{kt}\log\left(\mathbb{P}_{0}\left(z_{n}(kt) > \eta\right)\right)\le -\frac{\lambda\eta}{\frac{1}{N}+\sqrt{N}r^{k-1}}-\frac{\sum_{n=1}^{N}M_{n}}{2}\log(1-\lambda),
	\end{align}
	\noindent which holds for all $\lambda$ with $0 < \lambda < 1$. Now, supposing that
		\begin{align}
		\label{eq:f5}
		\eta > \frac{\left(\frac{1}{N}+\sqrt{N}r^{k-1}\right)\sum_{n=1}^{N}M_{n}}{2},
		\end{align}
		it can be shown that the right-hand side (RHS) of \eqref{eq:f3} is minimized at $\lambda^{*}=1-\frac{\left(\frac{1}{N}+\sqrt{N}r^{k-1}\right)\sum_{n=1}^{N}M_{n}}{2\eta}$. It is to be noted that with the condition in \eqref{eq:f5} in force, $\lambda^{*}\in(0,1)$. Hence, by substituting $\lambda=\lambda^{*}$ in \eqref{eq:f3} we have,
	\begin{align}
	\label{eq:f4}
	\limsup_{t\to\infty}\frac{1}{kt}\log\left(\mathbb{P}_{0}\left(z_{n}(kt) > \eta\right)\right)\le -\frac{\eta}{\frac{1}{N}+\sqrt{N}r^{k-1}}-\frac{\sum_{n=1}^{N}M_{n}}{2}\left(1+\log\frac{2\eta}{\left(\frac{1}{N}+\sqrt{N}r^{k-1}\right)\sum_{n=1}^{N}M_{n}}\right).
	\end{align}

	\noindent We specifically focused on the sub-sequence $\{z_{n}(kt)\}$ for the derivation of large deviations\footnote{By large deviations exponent, we mean the exponent associated with our large deviations upper bound.} exponent in this proof. It can be readily seen that other time-shifted sub-sequences (with constant time-shifts upto $k$ units) also inherit a similar large deviations upper bound as by construction, (see \eqref{eq:dec_stat_lin_12} for example), the decision statistic $z_{n}(kt)$ stays constant on the time interval $\left[kt, kt+k-1\right]$. Hence, the large deviations upper bound can be extended as a large deviations upper bound for the sequence $\{z_{n}(t)\}$.\\
	
	 \noindent For notational simplicity we denote $\mathbf{1}_{N}\otimes\mathbf{\theta}^{*}$ as $\mathbf{\theta}_{N}^{*}$. Before analyzing the probability of miss $\mathbb{P}_{1,\mathbf{\theta}^{*}}\left(z_{n}(kt) < \eta\right)$ and its error exponent, we first analyze the term $\left\|\mathbf{G}_{H}^{\top}\left(\mathbf{\theta}(t)-\tn\right)\right\|^{2}$. We have,
	 \begin{align}
	 \label{eq:e2}
	 \left\|\mathbf{G}_{H}^{\top}\left(\mathbf{\theta}(t)-\tn\right)\right\|\le\left\|\mathbf{G}_{H}\right\|\left\|\mathbf{\theta}(t)-\tn\right\|.
	 \end{align}
	 \noindent From \eqref{eq:est_upt_lin3}, we have that,
	 \begin{align}
	 \label{eq:e3}
	 \mathbf{\theta}(t+1)-\tn=\underbrace{\left(\mathbf{I}_{NM}-\beta_{t}\left(\mathbf{L}\otimes\mathbf{I}_{M}\right)-\alpha_{t}\mathbf{G}_{H}\mathbf{\Sigma}^{-1}\mathbf{G}_{H}^{\top}\right)}_{\text{\bf{A}(t)}}\left(\mathbf{\theta}(t)-\tn\right)+\alpha_{t}\mathbf{G}_{H}\mathbf{\Sigma}^{-1}\mathbf{\gamma}(t).
	 \end{align}
	 \noindent Let
	 \begin{align}
	 \label{eq:e31}
	 \gg(t)=\mathbf{G}_{H}\mathbf{\Sigma}^{-1}\mathbf{\gamma}(t).
	 \end{align}
	 \noindent Then, we have,
	 \begin{align}
	 \label{eq:e4}
	 &\left\|\mathbf{\theta}(t)-\tn\right\|^{2}=\left(\mathbf{\theta}(t)-\tn\right)^{\top}\left(\mathbf{\theta}(t)-\tn\right)\nonumber\\
	 &=\sum_{i=0}^{t-1}\sum_{j=0}^{t-1}\alpha_{i}\alpha_{j}\gg(i)^{\top}\alpha_{i}\alpha_{j}\prod_{u=0}^{t-2-i}\mathbf{A}(t-1-u)\prod_{v=j+1}^{t-1}\mathbf{A}(v)\gg(j)\nonumber\\
	 &=\mathbf{\gamma}_{G,t}^{\top}\mathbf{P}_{t}\mathbf{\gamma}_{G,t}=\operatorname{tr}\left(\mathbf{P}_{t}\mathbf{\gamma}_{G,t}\mathbf{\gamma}_{G,t}^{\top}\right),
	 \end{align}
	 \noindent where
	 \begin{align}
	 \label{eq:e41}
	 \mathbf{\gamma}_{G,t}=[\gg^{\top}(0)~\gg^{\top}(1)~\cdots~\gg^{\top}(t-1)]^{\top}
	 \end{align}
	 \noindent and $\mathbf{P}_{t}$ is a block matrix of dimension $NMt\times NMt$, whose $(i,j)$-th block $i,j=0,\cdots,t-1$ is given as follows:
	 \begin{align}
	 \label{eq:e5}
	 \left[\mathbf{P}_{t}\right]_{ij}=\alpha_{i}\alpha_{j}\prod_{u=0}^{t-2-i}\mathbf{A}(t-1-u)\prod_{v=j+1}^{t-1}\mathbf{A}(v).
	 \end{align}
	 \noindent First, note that the $\mathbf{A}(i)$'s commute and are symmetric and hence the individual blocks $\left[\mathbf{P}_{t}\right]_{ij}$-s and $\mathbf{P}_{t}$ is symmetric. We also note that, $\mathbf{P}_{t}$ is positive semi-definite, as using an expansion similar to \eqref{eq:e4} it can be shown that any quadratic form of $\mathbf{P}_{t}$ is non-negative.
	
	 \noindent Before, characterizing the large deviation exponents, we state the following lemma, the proof of which is provided in Appendix \ref{a2}.
	 \begin{Lemma}
	 	\label{wish}
	 	Let Assumptions~\ref{bs:1}-\ref{bs:4} and \ref{bs:6} hold. Given, the block matrix $\mathbf{P}_{t}$ as defined in \eqref{eq:e5}, we have the following upper bound,
	 	\begin{align}
	 	\label{eq:e7}
	 	t\left\|\mathbf{P}_{t}\right\| \le c_{3}\frac{\left(t_{1}+1\right)^{2c_{1}\alpha_{0}}}{t^{2c_{1}\alpha_{0}-1}}+\frac{\alpha_{0}^{2}}{t}+\frac{\alpha_{0}^{2}}{2c_{1}\alpha_{0}-1},~~\forall t\geq t_{1},
	 	\end{align}
	 	where $t_{1}$ is as defined in \eqref{eq:d13}-\eqref{eq:d2} and $c_{3}=\sum_{v=0}^{t_{1}-1}\alpha_{v}^{2}\prod_{u=v+1}^{t_{1}-1}\left\|\mathbf{A}(u)\right\|$.
	 \end{Lemma}
	
	\noindent For $\mathcal{H}_{1}$, we have,
	\begin{align}
	\label{eq:g1}
	& z_{n}(kt)=\frac{1}{(k(t-1)+1)}\sum_{j=1}^{N}\phi_{n,j}(k-1)\sum_{i=0}^{k(t-1)}\mathbf{\theta}_{j}^{\top}(k(t-1))\mathbf{H}_{j}^{\top}\mathbf{\Sigma}_{j}^{-1}\gamma_{j}(i)\nonumber\\&-\frac{\left(\mathbf{H}_{j}\left(\mathbf{\theta}_{j}(k(t-1))-\mathbf{\theta}^{*}\right)\right)^{\top}\mathbf{\Sigma}_{j}^{-1}\left(\mathbf{H}_{j}\left(\mathbf{\theta}_{j}(k(t-1))-\mathbf{\theta}^{*}\right)\right)}{2}+\frac{\left(\mathbf{\theta}^{*}\right)^{\top}\mathbf{H}_{j}^{\top}\mathbf{\Sigma}_{j}^{-1}\mathbf{H}_{j}\mathbf{\theta}^{*}}{2}.
	\end{align}
	
	\noindent For notational simplicity, we denote,
	\begin{align}
	\label{eq:g21} \eta_{2}=\frac{-2N\eta+\left(\mathbf{\theta}^{*}\right)^{\top}\mathbf{G}\mathbf{\theta}^{*}\left(1-N\sqrt{N}r^{k-1}\right)}{4\left\|\mathbf{G}_{H}\mathbf{\Sigma}^{-1}\mathbf{G}_{H}^{\top}\right\|\left(1+N\sqrt{N}r^{k-1}\right)}.
	\end{align}
	
	\noindent Moreover, supposing that
	\begin{align}
	\label{eq:g22}
	\eta < \frac{\left(\mathbf{\theta}^{*}\right)^{\top}\mathbf{G}\mathbf{\theta}^{*}\left(1-N\sqrt{N}r^{k-1}\right)}{2N},
	\end{align}
	
	we have $\eta_{2} > 0$, and the probability of miss can be characterized as follows:
	\begin{align}
	\label{eq:g2}
	&\mathbb{P}_{1,\theta^{*}}\left(z_{n}(kt) < \eta\right)\nonumber\\
	&=\mathbb{P}_{1,\theta^{*}}\left(\frac{1}{(k(t-1)+1)}\sum_{j=1}^{N}\phi_{n,j}(k-1)\sum_{i=0}^{k(t-1)}\mathbf{\theta}_{j}^{\top}(k(t-1))\mathbf{H}_{j}^{\top}\mathbf{\Sigma}_{j}^{-1}\gamma_{j}(i)\right.\nonumber\\&\left.-\frac{\left(\mathbf{H}_{j}\left(\mathbf{\theta}_{j}(k(t-1))-\mathbf{\theta}^{*}\right)\right)^{\top}\mathbf{\Sigma}_{j}^{-1}\left(\mathbf{H}_{j}\left(\mathbf{\theta}_{j}(k(t-1))-\mathbf{\theta}^{*}\right)\right)}{2}+\frac{\left(\mathbf{\theta}^{*}\right)^{\top}\mathbf{H}_{j}^{\top}\mathbf{\Sigma}_{j}^{-1}\mathbf{H}_{j}\mathbf{\theta}^{*}}{2}< \eta\right)\nonumber\\
	&\overset{(a)}{\le} \mathbb{P}_{1,\theta^{*}}\left(\frac{1}{(k(t-1)+1)}\sum_{j=1}^{N}\phi_{n,j}(k-1)\sum_{i=0}^{k(t-1)}\mathbf{\theta}_{j}^{\top}(k(t-1))\mathbf{H}_{j}^{\top}\mathbf{\Sigma}_{j}^{-1}\gamma_{j}(i)\right.\nonumber\\&\left.-\frac{\left(\mathbf{H}_{j}\left(\mathbf{\theta}_{j}(k(t-1))-\mathbf{\theta}^{*}\right)\right)^{\top}\mathbf{\Sigma}_{j}^{-1}\left(\mathbf{H}_{j}\left(\mathbf{\theta}_{j}(k(t-1))-\mathbf{\theta}^{*}\right)\right)}{2} < \eta-\frac{\left(\mathbf{\theta}^{*}\right)^{\top}\mathbf{G}\mathbf{\theta}^{*}\left(\frac{1}{N}-\sqrt{N}r^{k-1}\right)}{2}\right)\nonumber\\
	&\overset{(b)}{\le}\mathbb{P}_{1,\theta^{*}}\left(\sum_{j=1}^{N}\phi_{n,j}(k-1)\frac{\left(\mathbf{H}_{j}\left(\mathbf{\theta}_{j}(k(t-1))-\mathbf{\theta}^{*}\right)\right)^{\top}\mathbf{\Sigma}_{j}^{-1}\left(\mathbf{H}_{j}\left(\mathbf{\theta}_{j}(k(t-1))-\mathbf{\theta}^{*}\right)\right)}{2}\right.\nonumber\\&\left.> -\frac{\eta}{2}+\frac{\left(\mathbf{\theta}^{*}\right)^{\top}\mathbf{G}\mathbf{\theta}^{*}\left(\frac{1}{N}-\sqrt{N}r^{k-1}\right)}{4}\right)\nonumber\\&+\mathbb{P}_{1,\theta^{*}}\left(\frac{1}{(k(t-1)+1)}\sum_{j=1}^{N}\phi_{n,j}(k-1)\sum_{i=0}^{k(t-1)}\mathbf{\theta}_{j}^{\top}(k(t-1))\mathbf{H}_{j}^{\top}\mathbf{\Sigma}_{j}^{-1}\gamma_{j}(i)< \frac{\eta}{2}-\frac{\left(\mathbf{\theta}^{*}\right)^{\top}\mathbf{G}\mathbf{\theta}^{*}\left(\frac{1}{N}-\sqrt{N}r^{k-1}\right)}{4}\right)\nonumber\\
	&\overset{(c)}{\le}\underbrace{\mathbb{P}_{1,\theta^{*}}\left(\left\|\mathbf{G}_{H}\mathbf{\Sigma}^{-1}\mathbf{G}_{H}^{\top}\right\|\frac{\left\|\mathbf{\theta}(k(t-1))-\tn\right\|^{2}}{2}> \frac{-2N\eta+\left(\mathbf{\theta}^{*}\right)^{\top}\mathbf{G}\mathbf{\theta}^{*}\left(1-N\sqrt{N}r^{k-1}\right)}{4\left(1+N\sqrt{N}r^{k-1}\right)}\right)}_\text{(t1)}\nonumber\\&+\underbrace{\mathbb{P}_{1,\theta^{*}}\left(\frac{1}{(k(t-1)+1)}\sum_{j=1}^{N}\phi_{n,j}(k-1)\sum_{i=0}^{k(t-1)}\left(\mathbf{\theta}_{j}(k(t-1))-\mathbf{\theta}^{*}\right)^{\top}\mathbf{H}_{j}^{\top}\mathbf{\Sigma}_{j}^{-1}\gamma_{j}(i)< \frac{\eta}{4}-\frac{\left(\mathbf{\theta}^{*}\right)^{\top}\mathbf{G}\mathbf{\theta}^{*}\left(\frac{1}{N}-\sqrt{N}r^{k-1}\right)}{8}\right)}_\text{(t2)}\nonumber\\&+\underbrace{\mathbb{P}_{1,\theta^{*}}\left(\frac{1}{(k(t-1)+1)}\sum_{j=1}^{N}\phi_{n,j}(k-1)\sum_{i=0}^{k(t-1)}\left(\mathbf{\theta}^{*}\right)^{\top}\mathbf{H}_{j}^{\top}\mathbf{\Sigma}_{j}^{-1}\gamma_{j}(i)< \frac{\eta}{4}-\frac{\left(\mathbf{\theta}^{*}\right)^{\top}\mathbf{G}\mathbf{\theta}^{*}\left(\frac{1}{N}-\sqrt{N}r^{k-1}\right)}{8}\right)}_\text{(t3)},
	\end{align}
	
	\noindent where $(a)$ follows from $|\phi_{n,j}(k-1)-\frac{1}{N}|\le \sqrt{N}r^{k-1}$, $(b)$ follows from the union bound and $(c)$ follows from the union bound and the inequality $|\phi_{n,j}(k-1)-\frac{1}{N}|\le \sqrt{N}r^{k-1}$. Note that, Assumption \ref{bs:5} ensures that $\frac{1}{N}-\sqrt{N}r^{k-1}$ is positive.
	
	\noindent First, we analyze the term (t1) in \eqref{eq:g2}. We first note that, if $\lambda$ is chosen to be $ \lambda \le c_{4}$, where
	\begin{align}
	\label{eq:g31}
	c_{4}= \frac{1}{\left\|\mathbf{G}_{H}\mathbf{\Sigma}^{-1}\mathbf{G}_{H}^{\top}\right\|\left(c_{3}\frac{\left(t_{1}+1\right)^{2c_{1}\alpha_{0}}}{kt_{1}^{2c_{1}\alpha_{0}-1}}+\frac{\alpha_{0}^{2}}{kt_{1}}+\frac{\alpha_{0}^{2}}{2c_{1}\alpha_{0}-1}\right)},
	\end{align}
	we have that $kt\lambda\left\|\mathbf{P}_{kt}\left(\mathbf{I}_{kt}\otimes\mathbf{G}_{H}\mathbf{\Sigma}^{-1}\mathbf{G}_{H}^{\top}\right)\right\|<1$.
	Hence, we finally have that $\forall t \ge t_{1}$, with $t_{1}$ as defined in \eqref{eq:d13}
	\begin{align}
		\label{eq:g4}
		\det\left(\mathbf{I}_{NMkt}-kt\lambda\mathbf{P}_{kt}\left(\mathbf{I}_{kt}\otimes\mathbf{G}_{H}\mathbf{\Sigma}^{-1}\mathbf{G}_{H}^{\top}\right)\right) \ge (1-kt\lambda\left\|\mathbf{P}_{kt}\right\|\left\|\mathbf{G}_{H}\mathbf{\Sigma}^{-1}\mathbf{G}_{H}^{\top}\right\|)^{NMkt},
	\end{align}
	which ensures the existence of the moment generating function of the Wishart distribution under consideration (to be specified shortly).
	\noindent We have,
	\begin{align}
	\label{eq:g3}
	&\mathbb{P}_{1,\theta^{*}}\left(\frac{\left\|\mathbf{\theta}(k(t-1))-\tn\right\|^{2}}{2}> \eta_{2}\right)\nonumber\\
	&\le e^{-\lambda\eta_{2}kt}\mathbb{E}_{1,\mathbf{\theta}^{*}}\left[\exp\left(kt\lambda\left\|\mathbf{\theta}(k(t-1))-\tn\right\|^{2}\right)\right]\nonumber\\
	&\overset{(a)}{=}e^{-\lambda\eta_{2}kt}\mathbb{E}_{1, \mathbf{\theta}^{*}}\left[\exp\left(kt\lambda \operatorname{tr}\left(\frac{\mathbf{P}_{kt}}{2}\mathbf{\gamma}_{G,kt}\mathbf{\gamma}_{G,kt}^{\top}\right)\right)\right]\nonumber\\
	&\overset{(b)}{=}e^{-\lambda\eta_{2}kt}\times \left(\det\left(\mathbf{I}_{NMkt}-kt\lambda\mathbf{P}_{kt}\left(\mathbf{I}_{kt}\otimes\mathbf{G}_{H}\mathbf{\Sigma}^{-1}\mathbf{G}_{H}^{\top}\right)\right)\right)^{-1/2},
	\end{align}
	\noindent where in $(a)$, we use the definition of $\mathbf{P}_{kt}$ and $\mathbf{\gamma}_{G,kt}$ as defined in \eqref{eq:e5} and \eqref{eq:e41} respectively and in $(b)$ we use the moment generating function of the Wishart distribution (see, for example, \cite{anderson1946non})~as $\mathbf{\gamma}_{G,kt}\mathbf{\gamma}_{G,kt}^{\top}$ follows a Wishart distribution.
	\noindent Moreover, from \eqref{eq:e13}, we have that,
	\begin{align}
	\label{eq:g5}
	\limsup_{t\to\infty}kt\left\|\mathbf{P}_{kt}\right\|\le \frac{\alpha_{0}^{2}}{2c_{1}\alpha_{0}-1}.
	\end{align}	
	\noindent Now, on using \eqref{eq:g5} and \eqref{eq:g4} in \eqref{eq:g3}, we have,
	\begin{align}
	\label{eq:g6}
	&\mathbb{P}_{1,\theta^{*}}\left(\left\|\mathbf{\theta}(k(t-1))-\tn\right\|^{2}> \eta_{2}\right)\nonumber\\
	&\le e^{-\lambda\eta_{2}kt}\times \left(\det\left(\mathbf{I}_{NMkt}-kt\lambda\mathbf{P}_{kt}\left(\mathbf{I}_{kt}\otimes\mathbf{G}_{H}\mathbf{\Sigma}^{-1}\mathbf{G}_{H}^{\top}\right)\right)\right)^{-1/2}\nonumber\\
	&\le e^{-\lambda\eta_{2}kt}\times \left(1-kt\lambda\left\|\mathbf{P}_{kt}\right\|\left\|\mathbf{G}_{H}\mathbf{\Sigma}^{-1}\mathbf{G}_{H}^{\top}\right\|\right)^{-NMkt/2}\nonumber\\
	&\Rightarrow \frac{1}{kt}\log\left(\mathbb{P}_{1,\theta^{*}}\left(\left\|\mathbf{\theta}(k(t-1))-\tn\right\|^{2}> \eta_{2}\right)\right)\nonumber\\
	&\le -\lambda\eta_{2}-\frac{NM}{2}\log\left(1-kt\lambda\left\|\mathbf{P}_{kt}\right\|\left\|\mathbf{G}_{H}\mathbf{\Sigma}^{-1}\mathbf{G}_{H}^{\top}\right\|\right)\nonumber\\
	&\Rightarrow \limsup_{t\to\infty}\frac{1}{kt}\log\left(\mathbb{P}_{1,\theta^{*}}\left(\left\|\mathbf{\theta}(k(t-1))-\tn\right\|^{2}> \eta_{2}\right)\right)\nonumber\\
	&\le -\lambda\eta_{2}-\frac{NM}{2}\log\left(1-\frac{\lambda\alpha_{0}^{2}\left\|\mathbf{G}_{H}\mathbf{\Sigma}^{-1}\mathbf{G}_{H}^{\top}\right\|}{2c_{1}\alpha_{0}-1}\right).
	\end{align}	
	\noindent Let $LD(\lambda)=\lambda\eta_{2}+\frac{NM}{2}\log\left(1-\frac{\lambda\alpha_{0}^{2}\left\|\mathbf{G}_{H}\mathbf{\Sigma}^{-1}\mathbf{G}_{H}^{\top}\right\|}{2c_{1}\alpha_{0}-1}\right)$.	
	\noindent We first note that $LD(0)=0$. In order to ensure that the term $(t1)$ decays exponentially, the function $LD(.)$ needs to be increasing in an interval of the form, $\left[0,c_{5}\right]$, where $0 < c_{4} \le c_{5}$, with $c_{4}$ as defined in \eqref{eq:g31} which is formalized as follows:
	\begin{align}
	\label{eq:g61}
	\lambda < \frac{2c_{1}\alpha_{0}-1}{\alpha_{0}^{2}\left\|\mathbf{G}_{H}\mathbf{\Sigma}^{-1}\mathbf{G}_{H}^{\top}\right\|}-\frac{NM}{2\eta_{2}}=c_{4}^{*},
	\end{align}
	\noindent with $\eta_{2}$ as defined in \eqref{eq:g21}.	
	\noindent In order to have a positive large deviations upper bound, the RHS of \eqref{eq:g61} needs to be positive and hence, we require,
	\begin{align}
	\label{eq:g62}
	&\frac{2c_{1}\alpha_{0}-1}{\alpha_{0}^{2}\left\|\mathbf{G}_{H}\mathbf{\Sigma}^{-1}\mathbf{G}_{H}^{\top}\right\|}-\frac{NM}{2\eta_{2}} > 0\nonumber\\
	&\Rightarrow \eta <\frac{\left(\mathbf{\theta}^{*}\right)^{\top}\mathbf{G}\mathbf{\theta}^{*}\left(1-N\sqrt{N}r^{k-1}\right)}{2N}-\frac{M\alpha_{0}^{2}\left\|\mathbf{G}_{H}\mathbf{\Sigma}^{-1}\mathbf{G}_{H}^{\top}\right\|^{2}\left(1+N\sqrt{N}r^{k-1}\right)}{2c_{1}\alpha_{0}-1}.
	\end{align} 	
	\noindent We note that the condition derived in \eqref{eq:g62} is tighter than \eqref{eq:g22}. Now, combining the threshold condition derived above in \eqref{eq:g62} and the one derived in \eqref{eq:f5}, we have the following condition on the parameter $\mathbf{\theta}^{*}$
	\begin{align}
	\label{eq:g63}
	\frac{\left(\mathbf{\theta}^{*}\right)^{\top}\mathbf{G}\mathbf{\theta}^{*}\left(1-N\sqrt{N}r^{k-1}\right)}{2N} > \frac{M\alpha_{0}^{2}\left\|\mathbf{G}_{H}\mathbf{\Sigma}^{-1}\mathbf{G}_{H}^{\top}\right\|^{2}\left(1+N\sqrt{N}r^{k-1}\right)}{2c_{1}\alpha_{0}-1}+\frac{\left(\frac{1}{N}+\sqrt{N}r^{k-1}\right)\sum_{n=1}^{N}M_{n}}{2}
	\end{align}
	\noindent  which ensures the exponential decay of the term $(t1)$.
	\noindent Now, when we analyze (t2) and (t3) in \eqref{eq:g2}, we note that (t2) involves an additional time-decaying term, i.e., $\theta_{j}(k(t-1))
	-\theta^{*}$ which contributes to the large deviations upper bound as well. Hence, the exponent which will dominate among (t2) and (t3), would be the exponent of their sum.
	\noindent Using the condition derived in \eqref{eq:g22} and the union bound on (t3), we have,
	\begin{align}
	\label{eq:g7}
	&\mathbb{P}_{1,\theta^{*}}\left(\frac{1}{(k(t-1)+1)}\sum_{j=1}^{N}\phi_{n,j}(k-1)\sum_{i=0}^{k(t-1)}\left(\mathbf{\theta}^{*}\right)^{\top}\mathbf{H}_{j}^{\top}\mathbf{\Sigma}_{j}^{-1}\gamma_{j}(i)< \frac{\eta}{4}-\frac{\left(\mathbf{\theta}^{*}\right)^{\top}\mathbf{G}\mathbf{\theta}^{*}\left(\frac{1}{N}-\sqrt{N}r^{k-1}\right)}{8}\right)\nonumber\\
%	&\le \sum_{j=1}^{N}\mathbb{P}_{1,\theta^{*}}\left(\frac{1}{(k(t-1)+1)}\phi_{n,j}(k-1)\sum_{i=0}^{k(t-1)}\left(-\mathbf{\theta}^{*}\right)^{\top}\mathbf{H}_{j}^{\top}\mathbf{\Sigma}_{j}^{-1}\gamma_{j}(i) > -\frac{\eta}{4N}+\frac{\left(\mathbf{\theta}^{*}\right)^{\top}\mathbf{G}\mathbf{\theta}^{*}\left(\frac{1}{N}-\sqrt{N}r^{k-1}\right)}{8N}\right)\nonumber\\
%	&\le\sum_{j=1}^{N}\mathbb{Q}\left(\frac{-\frac{\eta\sqrt{k(t-1)+1}}{4N}+\frac{\sqrt{k(t-1)+1}\left(\mathbf{\theta}^{*}\right)^{\top}\mathbf{G}\mathbf{\theta}^{*}\left(\frac{1}{N}-\sqrt{N}r^{k-1}\right)}{8N}}{\phi_{n,j}(k-1)\sqrt{\left(\theta^{*}\right)^{\top}\mathbf{H}_{j}^{\top}\mathbf{\Sigma}_{j}^{-1}\mathbf{H}_{j}\theta^{*}}}\right)\nonumber\\
	&\le\mathbb{Q}\left(\frac{-\frac{\eta\sqrt{k(t-1)+1}}{4}+\frac{\sqrt{k(t-1)+1}\left(\mathbf{\theta}^{*}\right)^{\top}\mathbf{G}\mathbf{\theta}^{*}\left(\frac{1}{N}-\sqrt{N}r^{k-1}\right)}{8}}{\sqrt{\sum_{j=1}^{N}\left(\theta^{*}\right)^{\top}\mathbf{H}_{j}^{\top}\mathbf{\Sigma}_{j}^{-1}\mathbf{H}_{j}\theta^{*}}\left(\frac{1}{N}+\sqrt{N}r^{k-1}\right)}\right)\nonumber\\
	&\Rightarrow \limsup_{t\to\infty}\frac{1}{kt}\log\left(\mathbb{P}_{1,\theta^{*}}\left(\frac{1}{(k(t-1)+1)}\sum_{j=1}^{N}\phi_{n,j}(k-1)\sum_{i=0}^{k(t-1)}\left(\mathbf{\theta}^{*}\right)^{\top}\mathbf{H}_{j}^{\top}\mathbf{\Sigma}_{j}^{-1}\gamma_{j}(i)< \frac{\eta}{4}-\frac{\left(\mathbf{\theta}^{*}\right)^{\top}\mathbf{G}\mathbf{\theta}^{*}\left(\frac{1}{N}-\sqrt{N}r^{k-1}\right)}{8}\right)\right)\nonumber\\
	&\le -\left(\frac{\left(-\frac{\eta}{4}+\frac{\left(\mathbf{\theta}^{*}\right)^{\top}\mathbf{G}\mathbf{\theta}^{*}\left(\frac{1}{N}-\sqrt{N}r^{k-1}\right)}{8}\right)^{2}}{2\sum_{j=1}^{N}\left(\theta^{*}\right)^{\top}\mathbf{H}_{j}^{\top}\mathbf{\Sigma}_{j}^{-1}\mathbf{H}_{j}\theta^{*}\left(\frac{1}{N}+\sqrt{N}r^{k-1}\right)^{2}}\right).
	\end{align}
	
	\noindent Combining \eqref{eq:g7} and \eqref{eq:g6}, we have,
	\begin{align}
	\label{eq:g8}
	&\limsup_{t\to\infty}\frac{1}{kt}\log\left(\mathbb{P}_{1,\theta^{*}}\left(z_{n}(kt) < \eta\right)\right)\nonumber\\
	&\le \max\left\{-\frac{\left(-\frac{\eta}{4}+\frac{\left(\mathbf{\theta}^{*}\right)^{\top}\mathbf{G}\mathbf{\theta}^{*}\left(\frac{1}{N}-\sqrt{N}r^{k-1}\right)}{8}\right)^{2}}{2\sum_{j=1}^{N}\left(\theta^{*}\right)^{\top}\mathbf{H}_{j}^{\top}\mathbf{\Sigma}_{j}^{-1}\mathbf{H}_{j}\theta^{*}\left(\frac{1}{N}+\sqrt{N}r^{k-1}\right)^{2}}, -LD\left(\min\left\{c_{4},c_{4}^{*}\right\}\right)\right\}=LD_{1}\left(\eta\right),
	\end{align}
	
	We specifically focused on the sub-sequence $\{z_{n}(kt)\}$ for the derivation of large deviations\footnote{By large deviations exponent, we mean the exponent associated with out large deviations upper bound.} exponent in this proof. It can be readily seen that other time-shifted sub-sequences (with constant time-shifts upto $k$ units) also inherit a similar large deviations upper bound as by construction, (see \eqref{eq:dec_stat_lin_12} for example), the decision statistic $z_{n}(kt)$ stays constant on the time interval $\left[kt, kt+k-1\right]$. Hence, the large deviations upper bound can be extended as a large deviations upper bound for the sequence $\{z_{n}(t)\}$.
\end{IEEEproof}

\section{Conclusion}
\label{sec:conc}
\noindent In this paper, we have considered the problem of a recursive composite hypothesis testing in a network of sparsely interconnected agents where the objective is to test a simple null hypothesis against a composite alternative concerning the state of the field, modeled as a vector of (continuous) unknown parameters determining the parametric family of probability measures induced on the agents' observation spaces under the hypotheses. We have proposed two $\emph{consensus}+\emph{innovations}$ type algorithms, $\mathcal{CIGLRT-NL}$ and $\mathcal{CIGLRT-L}$, in which every agent updates its parameter estimate and decision statistic by simultaneous processing of neighborhood information and local newly sensed information and in which the inter-agent collaboration is restricted to a possibly sparse but connected communication graph. For linear observation models, we have established the consistency of the parameter estimate sequences and characterized the large deviations exponent upper bounds of the probabilities of errors pertaining to the detection scheme for the algorithm $\mathcal{CIGLRT-L}$. We have established consistency of the parameter estimate sequences and the existence of appropriate algorithm parameters which ensure asymptotically decaying probabilities of errors in the large sample limit for the algorithm $\mathcal{CIGLRT-NL}$, under a general non-linear sensing model satisfying a global observability condition. Moreover, for both the algorithms proposed in this work, the parameter estimation scheme and the decision statistic update schemes run in a parallel fashion and thus making the algorithms, recursive \emph{online} algorithms. The tools developed in this paper are of independent interest and might be applicable or extended to other recursive online distributed inference algorithms. A natural direction for future research consists of considering models with non-Gaussian noise. We also intend to develop extensions of the $\mathcal{CIGLRT-NL}$ in which the parameter domain is restricted to constrained domains such as convex subsets of the Euclidean space or manifolds.

\appendices
\section{Proofs of Lemmas in Section \ref{sec:proof_res_ciglrt}}
\label{a1}
\begin{IEEEproof}[Proof of Lemma \ref{l2}]
	\noindent The proof follows similarly as the proof of Lemma IV.1 in \cite{kar2014asymptotically} with appropriate modifications to take into account the state-dependent nature of the innovation gains. Define the process $\{\mathbf{x}(t)\}$ as $\mathbf{x}(t)=\mathbf{\theta}(t)-\mathbf{1}_{N} \otimes \mathbf{\theta}^{*}$ where $\theta^{*}$ denotes the true but unknown parameter.
	The process $\left\{\mathbf{x}(t)\right\}$ satisfies the following recursion:
	\begin{align}
	\label{eq:l2_pr_1}
	\mathbf{x}(t+1)=\mathbf{x}(t)-\beta_{t}(\mathbf{L} \otimes \mathbf{I}_{M})x(t)+\alpha_{t}\mathbf{G}\left(\mathbf{\theta}(t)\right)\mathbf{\Sigma}^{-1}\left(\mathbf{y}(t)-\mathbf{h}(\mathbf{\theta}(t))\right),
	\end{align}\\
	which implies that,
	\begin{align}
	\mathbf{x}(t+1)=\mathbf{x}(t)-\beta_{t}(\mathbf{L} \otimes \mathbf{I}_{M})x(t)+\alpha_{t}\mathbf{G}\left(\mathbf{\theta}(t)\right)\mathbf{\Sigma}^{-1}\left(\mathbf{y}(t)-\mathbf{h}\left(\theta_{N}^{*}\right)\right)-\alpha_{t}\mathbf{G}\left(\mathbf{\theta}(t)\right)\mathbf{\Sigma}^{-1}\left(\mathbf{h}\left(\mathbf{\theta}(t)\right)-\mathbf{h}\left(\theta_{N}^{*}\right)\right).
	\end{align}
	\noindent It follows from basic properties of the Laplacian $\mathbf{L}$, that
	\begin{align}
	\label{eq:l2_pr_2}
	\left(\mathbf{L} \otimes \mathbf{I}_{M}\right)\left(\mathbf{1}_{N} \otimes \mathbf{\theta}^{*}\right)
	=\left(\mathbf{L}\mathbf{1}_{N}\right)\otimes\left(\mathbf{I}_{M}\mathbf{\theta}^{*}\right)
	=\mathbf{0}.
	\end{align}
	
	\noindent Taking norms of both sides of \eqref{eq:l2_pr_1}, we have,
	\begin{align}
	\label{eq:l2_pr_3}
	&\left\|\mathbf{x}(t+1)\right\|^{2}=\left\|\mathbf{x}(t)\right\|^{2}-2\beta_{t}\mathbf{x}^{\top}(t)\left(\mathbf{L} \otimes \mathbf{I}_{M}\right)\mathbf{x}(t) -2\alpha_{t}\mathbf{x}^{\top}(t)\mathbf{G}\left(\mathbf{\theta}(t)\right)\mathbf{\Sigma}^{-1}\left(\mathbf{h}\left(\mathbf{\theta}(t)\right)-\mathbf{h}\left(\mathbf{\theta}_{N}^{*}\right)\right)\nonumber\\
	&+\beta_{t}^{2}\mathbf{x}^{\top}(t)\left(\mathbf{L} \otimes \mathbf{I}_{M}\right)^{2}\mathbf{x}(t)+2\alpha_{t}\beta_{t}\mathbf{x}^{\top}(t)\left(\mathbf{L} \otimes \mathbf{I}_{M}\right)\mathbf{G}\left(\mathbf{\theta}(t)\right)\mathbf{\Sigma}^{-1}\left(\mathbf{h}\left(\mathbf{\theta}(t)\right)-\mathbf{h}\left(\mathbf{\theta}_{N}^{*}\right)\right)\nonumber\\
	&-2\alpha_{t}\beta_{t}\mathbf{x}^{\top}(t)\left(\mathbf{L} \otimes \mathbf{I}_{M}\right)\mathbf{G}\left(\mathbf{\theta}(t)\right)\mathbf{\Sigma}^{-1}\left(\mathbf{y}(t)-\mathbf{h}\left(\mathbf{\theta}_{N}^{*}\right)\right)\nonumber\\
	&+\alpha_{t}^{2}\left(\mathbf{y}(t)-\mathbf{h}\left(\theta_{N}^{*}\right)\right)^{\top}\mathbf{\Sigma}^{-1}\mathbf{G}^{\top}\left(\mathbf{\theta}(t)\right)\mathbf{G}\left(\mathbf{\theta}(t)\right)\mathbf{\Sigma}^{-1}\left(\mathbf{y}(t)-\mathbf{h}\left(\theta_{N}^{*}\right)\right)\nonumber\\&+\alpha_{t}^{2}\left(\mathbf{h}\left(\mathbf{\theta}(t)\right)-\mathbf{h}\left(\mathbf{\theta}_{N}^{*}\right)\right)^{\top}\mathbf{\Sigma}^{-1}\mathbf{G}^{\top}\left(\mathbf{\theta}(t)\right)\mathbf{G}\left(\mathbf{\theta}(t)\right)\mathbf{\Sigma}^{-1}\left(\mathbf{h}\left(\mathbf{\theta}(t)\right)-\mathbf{h}\left(\mathbf{\theta}_{N}^{*}\right)\right)
	+2\alpha_{t}\mathbf{x}^{\top}(t)\mathbf{G}\left(\mathbf{\theta}(t)\right)\mathbf{\Sigma}^{-1}\left(\mathbf{y}(t)-\mathbf{h}(\mathbf{\theta}_{N}^{*})\right)\nonumber\\
	&-2\alpha_{t}^{2}\left(\mathbf{y}(t)-\mathbf{h}\left(\mathbf{\theta}_{N}^{*}\right)\right)^{\top}\mathbf{\Sigma}^{-1}\mathbf{G}^{\top}\left(\mathbf{\theta}(t)\right)\mathbf{G}\left(\mathbf{\theta}(t)\right)\mathbf{\Sigma}^{-1}\left(\mathbf{h}\left(\mathbf{\theta}(t)\right)-\mathbf{h}\left(\mathbf{\theta}_{N}^{*}\right)\right).
	\end{align}
	
	\noindent Consider the orthogonal decomposition,
	\begin{align}
	\label{eq:l2_pr_4}
	\mathbf{x}=\mathbf{x}_{c}+\mathbf{x}_{c \perp},
	\end{align}
	
	\noindent where $\mathbf{x}_{c}$ denotes the projection of $\mathbf{x}$ to the consensus subspace $\mathcal{C}$ with
	\begin{align}
	\label{eq:l2_pr_5}
	\mathcal{C}=\{\mathbf{x} \in \mathbb{R}^{MN} ~|~\mathbf{x}=1_{N}\otimes a, \textit{for~~some~~}a \in \mathbb{R}^{M} \}.
	\end{align}
	
	\noindent From, \eqref{eq:1}, we have that,
	\begin{align}
	\label{eq:l2_pr_6}
	\mathbb{E}_{\mathbf{\theta}^{*}}\left[\mathbf{y}(t)-\mathbf{h}\left(\mathbf{\theta}_{N}^{*}\right)\right]=\mathbf{0}.
	\end{align}
	
	\noindent Consider the process
	\begin{align}
	\label{eq:l2_pr_7}
	V_{2}(t)=\left\|\mathbf{x}(t)\right\|^{2}.
	\end{align}

	\noindent Using conditional independence properties we have,
	\begin{align}
	\label{eq:l2_pr_8}
	&\mathbb{E}_{\theta^{*}}[V_{2}(t+1)|\mathcal{F}_{t}]=V_{2}(t) + \beta_{t}^{2}\mathbf{x}^{\top}(t)\left(\mathbf{L} \otimes \mathbf{I}_{M}\right)^{2}\mathbf{x}(t)\nonumber\\
	&+\alpha_{t}^{2}\mathbb{E}_{\theta^{*}}\left[\left(\mathbf{y}(t)-\mathbf{h}\left(\mathbf{\theta}_{N}^{*}\right)\right)^{\top}\mathbf{\Sigma}^{-1}\mathbf{G}^{\top}\left(\mathbf{\theta}(t)\right)\mathbf{G}\left(\mathbf{\theta}(t)\right)\mathbf{\Sigma}^{-1}\left(\mathbf{y}(t)-\mathbf{h}\left(\mathbf{\theta}_{N}^{*}\right)\right)\right]-2\beta_{t}\mathbf{x}^{\top}(t)\left(\mathbf{L}\otimes \mathbf{I}_{M}\right)\mathbf{x}(t)\nonumber\\
	&-2\alpha_{t}\mathbf{x}^{\top}(t)\mathbf{G}\left(\mathbf{\theta}(t)\right)\mathbf{\Sigma}^{-1}\left(\mathbf{h}\left(\mathbf{\theta}(t)\right)-\mathbf{h}\left(\mathbf{\theta}_{N}^{*}\right)\right)+2\alpha_{t}\beta_{t}\mathbf{x}^{\top}(t)\left(\mathbf{L} \otimes \mathbf{I}_{M}\right)\mathbf{G}\left(\mathbf{\theta}(t)\right)\mathbf{\Sigma}^{-1}\left(\mathbf{h}\left(\mathbf{\theta}(t)\right)-\mathbf{h}\left(\mathbf{\theta}_{N}^{*}\right)\right) \nonumber\\
	&+\alpha_{t}^{2}\left\|\left(\mathbf{h}\left(\mathbf{\theta}(t)\right)-\mathbf{h}\left(\mathbf{\theta}_{N}^{*}\right)\right)^{\top}\mathbf{G}^{\top}\left(\mathbf{\theta}(t)\right)\mathbf{\Sigma}^{-1}\right\|^{2}.
	\end{align}

	\noindent We use the following inequalities $\forall t \ge t_{1}$,
	\begin{align}
	\label{eq:l2_pr_9}
	&\mathbf{x}^{\top}(t)\left(\mathbf{L} \otimes \mathbf{I}_{M}\right)^{2}\mathbf{x}(t) \overset{(q1)}{\le}\lambda_{N}^{2}(\mathbf{L})||\mathbf{x}_{C\perp}(t)||^{2};\nonumber\\
	&\mathbf{x}^{\top}(t)\mathbf{G}\left(\mathbf{\theta}(t)\right)\mathbf{\Sigma}^{-1}\left(\mathbf{h}\left(\mathbf{\theta}(t)\right)-\mathbf{h}\left(\mathbf{\theta}_{N}^{*}\right)\right)\ge c_{1}||\mathbf{x}(t)||^{2} \overset{(q2)}{\ge}0;\nonumber\\
	&\mathbf{x}^{\top}(t)\left(\mathbf{L} \otimes \mathbf{I}_{M}\right)\mathbf{x}(t) \overset{(q3)}{\ge} \lambda_{2}(\mathbf{L})\left\|\mathbf{x}_{C\perp}(t)\right\|^{2};\nonumber\\
	&\mathbf{x}^{\top}(t)\left(\mathbf{L} \otimes \mathbf{I}_{M}\right)\mathbf{G}\left(\mathbf{\theta}(t)\right)\mathbf{\Sigma}^{-1}\left(\mathbf{h}\left(\mathbf{\theta}(t)\right)-\mathbf{h}\left(\mathbf{\theta}_{N}^{*}\right)\right) \overset{(q4)}{\le} c_{2}\left\|\mathbf{x}(t)\right\|^{2},
	\end{align}
	\noindent for $c_{1}$ as defined in Assumption \ref{as:4}, and a positive constant $c_{2}$, where $(q2)$ follows from Assumption \ref{as:4} and $(q4)$ follows from Assumption \ref{as:3} by which we have that $\left\|\nabla \mathbf{h}_{n}\left(\mathbf{\theta}_{n}(t)\right)\right\|$ is uniformly bounded from above by $k_{n}$ for all $n$, and hence, we have that $\left\|\mathbf{G}\left(\mathbf{\theta}(t)\right)\right\|\le \max_{n=1,\cdots,N}k_{n}$. We also have
	
	\begin{align}
	\label{eq:l2_pr_13}
	\mathbb{E}_{\theta^{*}}\left[\left(\mathbf{y}(t)-\mathbf{h}\left(\mathbf{\theta}_{N}^{*}\right)\right)^{\top}\mathbf{\Sigma}^{-1}\mathbf{G}^{\top}(\mathbf{\theta}(t))\mathbf{G}\left(\mathbf{\theta}(t)\right)\mathbf{\Sigma}^{-1}\left(\mathbf{y}(t)-\mathbf{h}\left(\theta_{N}^{*}\right)\right)\right] \le c_{4},
	\end{align}
	\noindent for some constant $c_{4} > 0$. In \eqref{eq:l2_pr_13}, we use the fact that the noise process under consideration is Gaussian and hence has finite moments. We also use the fact that $\left\|\mathbf{G}\left(\mathbf{\theta}(t)\right)\right\|\le \max_{n=1,\cdots,N}k_{n}$, which in turn follows from Assumption \ref{as:3}.

	\noindent We further have that,
	\begin{align}
	\label{eq:l2_pr_14}
	\left(\mathbf{h}\left(\mathbf{\theta}(t)\right)-\mathbf{h}\left(\mathbf{\theta}_{N}^{*}\right)\right)^{\top}\mathbf{\Sigma}^{-1}\mathbf{G}^{\top}\left(\mathbf{\theta}(t)\right)\mathbf{G}\left(\mathbf{\theta}(t)\right)\mathbf{\Sigma}^{-1}\left(\mathbf{h}\left(\mathbf{\theta}(t)\right)-\mathbf{h}\left(\mathbf{\theta}_{N}^{*}\right)\right)\le c_{3}\left\|\mathbf{x}(t)\right\|^{2},
	\end{align}
	\noindent where $c_{3} > 0$ is a constant.
	\noindent It is to be noted that \eqref{eq:l2_pr_14} follows from the Lipschitz continuity in Assumption \ref{as:3} and the fact that $\left\|\mathbf{G}\left(\mathbf{\theta}(t)\right)\right\|\le \max_{n=1,\cdots,N}k_{n}$.

	\noindent Using \eqref{eq:l2_pr_8}-\eqref{eq:l2_pr_14}, we have,
	\begin{align}
	\label{eq:l2_pr_15}
	\mathbb{E}_{\theta^{*}}[V_{2}(t+1)|\mathcal{F}_{t}]\le \left(1+c_{5}\left(\alpha_{t}\beta_{t}+\alpha_{t}^{2}\right)\right)V_{2}(t)-c_{6}(\beta_{t}-\beta^{2}_{t})||\mathbf{x}_{C\perp}(t)||^{2}+c_{4}\alpha_{t}^{2},
	\end{align}
	\noindent for some positive constants $c_{5}$ and $c_{6}$.
	\noindent As $\beta_{t}^{2}$ goes to zero faster than $\beta_{t}$, $\exists t_{2}$ such that $\forall t\ge t_{2}$, $\beta_{t} \ge \beta^{2}_{t}$.
	\noindent Hence $\exists t_{2}$ and $\exists \tau_{1}, \tau_{2} > 1$  such that for all $t \ge t_{2}$
	\begin{align}
	\label{eq:l2_pr_16}
	c_{5}\left(\alpha_{t}\beta_{t}+\alpha_{t}^{2}\right) \le \frac{c_{7}}{(t+1)^{\tau_{1}}}=\gamma_{t},~
	c_{4}\alpha_{t}^{2}\le\frac{c_{8}}{(t+1)^{\tau_{2}}}=\hat{\gamma}_{t}
	\end{align}
	\noindent where $c_{7}, c_{8} > 0$ are constants.
	
	\noindent By the above construction we obtain, $\forall t\geq t_{2}$,
	\begin{align}
	\label{eq:l2_pr_17}
	\mathbb{E}_{\theta^{*}}[V_{2}(t+1) | \mathcal{F}_{t}] \le (1+\gamma_{t})V_{2}(t)+\hat{\gamma}_{t},
	\end{align}
	
	\noindent where the positive weight sequences $\{\gamma_{t}\}$ and $\{\hat{\gamma}_{t}\}$ are summable, i.e.,
	\begin{align}
	\label{eq:l2_pr_18}
	\sum_{t\ge0}\gamma_{t} < \infty, 	~\sum_{t\ge0}\hat{\gamma}_{t} < \infty.
	\end{align}
	\noindent By \eqref{eq:l2_pr_18}, the product $\prod_{s=t}^{\infty}(1+\gamma_{s})$ exists for all $t$. Now let $\{W(t)\}$ be such that
	\begin{align}
	\label{eq:l2_pr_19}
	W(t)=\left(\prod_{s=t}^{\infty}(1+\gamma_{s})\right)V_{2}(t)+\sum_{s=t}^{\infty}\hat{\gamma}_{s},~~~~\forall t\geq t_{2}.
	\end{align}
	
	\noindent By \eqref{eq:l2_pr_19}, it can be shown that $\{W(t)\}$ satisfies,
	\begin{align}
	\label{eq:l2_pr_20}
	\mathbb{E}_{\theta^{*}}[W(t+1) | \mathcal{F}_{t}] \le W(t).
	\end{align}
	
	\noindent Hence,  $\{W(t)\}$ is a non-negative super martingale and converges a.s. to a bounded random variable $W^{*}$ as $t\to\infty$. It then follows from \eqref{eq:l2_pr_19} that $V_{2}(t)\to W^{*}$ as $t\to\infty$. Thus, we conclude that the sequences $\{\mathbf{\theta}_{n}(t)\}$ are bounded for all $n$.
	
\end{IEEEproof}
\begin{IEEEproof}[Proof of Lemma \ref{l32}]
	\noindent The proof follows exactly the development in theorem IV.1 of~\cite{kar2014asymptotically}. Let $\mathbf{x}(t)$ denote the residual $\mathbf{\theta}(t)-\mathbf{1}_{N}\otimes\mathbf{\theta}^{*}$.
	
	\noindent For $\epsilon\in (0,1)$, define the set $\Gamma_{\epsilon}$
	\begin{align}
	\label{eq:l32_pr_71}
	\Gamma_{\epsilon}=\left\{\mathbf{\theta}\in\mathbb{R}^{NM} : \epsilon\leq \left\|\mathbf{\theta}-\mathbf{1}_{N}\otimes\mathbf{\theta}^{*}\right\|\le\frac{1}{\epsilon}\right\}.
	\end{align}
	
	\noindent Let $\rho_{\epsilon}$ denote the $\{\mathcal{F}_{t}\}$ stopping time
	\begin{align}
	\label{eq:l32_pr_72}
	\rho_{\epsilon}=\inf\{t\geq 0: \mathbf{\theta}(t)\notin \Gamma_{\epsilon}\},
	\end{align}
	
	\noindent where $\Gamma_{\epsilon}$ is defined in \eqref{eq:l32_pr_71}. Let $\{V^{\epsilon}(t)\}$ denote the stopped process
	\begin{align}
	\label{eq:l32_pr_73}
	V^{\epsilon}(t)=V_{2}(\max\{t,\rho_{\epsilon}\}), \forall t,
	\end{align}
	
	\noindent with $V_{2}(t)$ as defined in \eqref{eq:l2_pr_7}.
	
	\noindent Then, we have,
	\begin{align}
	\label{eq:l32_pr_74}
	V^{\epsilon}(t+1)=V_{2}(t+1)\mathbb{I}\left(\rho_{\epsilon}>t\right)+V_{2}(\rho_{\epsilon})\mathbb{I}\left(\rho_{\epsilon}\leq t\right),
	\end{align}
	\noindent where $\mathbb{I}(\cdot)$ denotes the indicator function. Due to the fact that $\mathbb{I}\left(\rho_{\epsilon}>t\right)$ and $V_{2}(\rho_{\epsilon})\mathbb{I}\left(\rho_{\epsilon}\leq t\right)$ are adapted to $\mathcal{F}_{t}$ for all $t$, we have,
	\begin{align}
	\label{eq:l32_pr_75}
	\mathbb{E}_{\theta^{*}}\left[V^{\epsilon}(t+1)|\mathcal{F}_{t}\right]=\mathbb{E}_{\theta^{*}}\left[V_{2}(t+1)\right]\mathbb{I}\left(\rho_{\epsilon}>t\right)+V_{2}(\rho_{\epsilon})\mathbb{I}\left(\rho_{\epsilon}\leq t\right),
	\end{align}
	for all $t$.
	
	\noindent First, noting the inequality derived in \eqref{eq:l2_pr_9} in $(q2)$ and rewriting it as,
	\begin{align}
	\label{eq:l32_pr_76}
	-\mathbf{x}(t)^{T}\mathbf{G}\left(\mathbf{\theta}(t)\right)\mathbf{\Sigma}^{-1}\left(\mathbf{h}\left(\mathbf{\theta}(t)\right)-\mathbf{h}\left(\mathbf{\theta}_{N}^{*}\right)\right)\leq -c_{1}||\mathbf{x}(t)||^{2},
	\end{align}
	
	\noindent we have with a slight rearrangement of terms from the expansion in \eqref{eq:l2_pr_8},
	\begin{align}
	\label{eq:l32_pr_77}
	&\mathbb{E}_{\theta^{*}}[V_{2}(t+1)|\mathcal{F}_{t}]=V_{2}(t) + \beta_{t}^{2}\mathbf{x}^{T}(t)\left(\mathbf{L} \otimes \mathbf{I}_{M}\right)^{2}\mathbf{x}(t)\nonumber\\
	&+\alpha_{t}^{2}\mathbb{E}_{\theta^{*}}\left[\left(\mathbf{y}(t)-\mathbf{h}\left(\mathbf{\theta}_{N}^{*}\right)\right)^{\top}\mathbf{\Sigma}^{-1}\mathbf{G}^{\top}\left(\mathbf{\theta}(t)\right)\mathbf{G}\left(\mathbf{\theta}(t)\right)\mathbf{\Sigma}^{-1}\left(\mathbf{y}(t)-\mathbf{h}\left(\mathbf{\theta}_{N}^{*}\right)\right)\right]-2\beta_{t}\mathbf{x}^{\top}(t)\left(\mathbf{L}\otimes \mathbf{I}_{M}\right)\mathbf{x}(t)\nonumber\\
	&-2\alpha_{t}\mathbf{x}^{\top}(t)\mathbf{G}\left(\mathbf{\theta}(t)\right)\mathbf{\Sigma}^{-1}\left(\mathbf{h}\left(\mathbf{\theta}(t)\right)-\mathbf{h}\left(\mathbf{\theta}_{N}^{*}\right)\right)+2\alpha_{t}\beta_{t}\mathbf{x}^{\top}(t)\left(\mathbf{L} \otimes \mathbf{I}_{M}\right)\mathbf{G}\left(\mathbf{\theta}(t)\right)\mathbf{\Sigma}^{-1}\left(\mathbf{h}\left(\mathbf{\theta}(t)\right)-\mathbf{h}\left(\mathbf{\theta}_{N}^{*}\right)\right) \nonumber\\
	&+\alpha_{t}^{2}\left\|\left(\mathbf{h}\left(\mathbf{\theta}(t)\right)-\mathbf{h}^{\top}\left(\mathbf{\theta}_{N}^{*}\right)\right)\mathbf{G}^{\top}\left(\mathbf{\theta}(t)\right)\mathbf{\Sigma}^{-1}\right\|^{2}.
	\end{align}
	
	\noindent Now, using \eqref{eq:l32_pr_76} in \eqref{eq:l32_pr_77} and the inequalities derived in \eqref{eq:l2_pr_9}-\eqref{eq:l2_pr_14}, we have,
	\begin{align}
	\label{eq:l32_pr_78}
	\mathbb{E}_{\theta^{*}}[V_{2}(t+1)|\mathcal{F}_{t}]\le \left(1-c_{1}\alpha_{t}+c_{5}\left(\alpha_{t}\beta_{t}+\alpha_{t}^{2}\right)\right)V_{2}(t)-c_{6}(\beta_{t}-\beta^{2}_{t})||\mathbf{x}_{C\perp}(t)||^{2}+c_{4}\alpha_{t}^{2},
	\end{align}
	where $c_{5}, c_{6}, c_{4}$ are appropriately chosen constants.
	
	\noindent Now, by choosing a large enough $t_{\epsilon}$, such that for all $t \geq t_{\epsilon}$, we can assert that,
	\begin{align}
	\label{eq:l32_pr_79}
	&\beta_{t}-\beta_{t}^{2}\geq 0,\nonumber\\
	& c_{1}\alpha_{t}-c_{5}\left(\alpha_{t}\beta_{t}+\alpha_{t}^{2}\right)\geq c_{7}\alpha_{t}.
	\end{align}
	\noindent Thus, we have for $t\geq t_{\epsilon}$,
	\begin{align}
	\label{eq:l32_pr_80}
	\mathbb{E}_{\theta^{*}}[V_{2}(t+1)|\mathcal{F}_{t}]\le \left(1-c_{1}\alpha_{t}\right)V_{2}(t)+c_{4}\alpha_{t}^{2}.
	\end{align}
	
	\noindent Furthermore, by the definition of $\Gamma_{\epsilon}$, we have,
	\begin{align}
	\label{eq:l32_pr_81}
	\left\|\mathbf{x}(t)\right\|^{2}\geq \epsilon^{2}~\mbox{on}~\left\{\mathbf{x}(t)\in\Gamma_{\epsilon}\right\},
	\end{align}
	\noindent and hence by the definition of $V_{2}(t)$, we have that there exists a constant $c_{7}\left(\epsilon\right) > 0$ such that
	\begin{align}
	\label{eq:l32_pr_82}
	V_{2}(t) \geq c_{7}(\epsilon)~\mbox{on}~\left\{\mathbf{x}(t)\in\Gamma_{\epsilon}\right\}.
	\end{align}
	
	\noindent Using the above relation in \eqref{eq:l32_pr_80}, we then have for all $t\geq t_{\epsilon}$,
	\begin{align}
	\label{eq:l32_pr_83}
	\mathbb{E}_{\theta^{*}}[V_{2}(t+1)|\mathcal{F}_{t}]\mathbb{I}\left(\rho_{\epsilon} > t\right)\le \left[V_{2}(t)-c_{8}(\epsilon)\alpha_{t}+c_{4}\alpha_{t}^{2}\right]\mathbb{I}\left(\rho_{\epsilon} > t\right),
	\end{align}
	\noindent where $c_{8}(\epsilon) > 0$ is an appropriately chosen constant. Finally, the observation that $\alpha_{t} > \alpha_{t}^{2}$ establishes that
	\begin{align}
	\label{eq:l32_pr_84}
	\mathbb{E}_{\theta^{*}}[V_{2}(t+1)|\mathcal{F}_{t}]\mathbb{I}\left(\rho_{\epsilon} > t\right)\le \left[V_{2}(t)-c_{9}(\epsilon)\alpha_{t}\right]\mathbb{I}\left(\rho_{\epsilon} > t\right),
	\end{align}
	\noindent where $c_{9}(\epsilon) > 0$ is an appropriately chosen constant
	\noindent Finally, from \eqref{eq:l32_pr_75}, we have that
	\begin{align}
	\label{eq:l32_pr_85}
	&\mathbb{E}_{\theta^{*}}[V^{\epsilon}(t+1)|\mathcal{F}_{t}]\le V_{2}(t)\mathbb{I}\left(\rho_{\epsilon} > t\right)+V_{2})(\rho_{\epsilon})\mathbb{I}\left(\rho_{\epsilon} \leq t\right)-c_{9}(\epsilon)\alpha_{t}\mathbb{I}\left(\rho_{\epsilon} > t\right)\nonumber\\
	&=V^{\epsilon}(t)-c_{9}(\epsilon)\alpha_{t}\mathbb{I}\left(\rho_{\epsilon} > t\right).
	\end{align}
	
	\noindent It is to be noted that $\{V^{\epsilon}(t)\}_{t\geq t_{\epsilon}}$ satisfies $\mathbb{E}_{\theta^{*}}[V^{\epsilon}(t+1)|\mathcal{F}_{t}]\leq V^{\epsilon}(t)$ for all $t\geq t_{\epsilon}$, which being a non-negative supermartingale, there exists an a.s. finite $V^{\epsilon}$ such that $V^{\epsilon}(t+1)\to V^{\epsilon}$ a.s. as $t\to\infty$.
	To this end, define the process $\{V_{1}^{\epsilon}(t)\}$ given by
	\begin{align}
	\label{eq:l32_pr_86}
	V_{1}^{\epsilon}(t)=V^{\epsilon}(t)+c_{9}(\epsilon)\sum_{s=0}^{t-1}\alpha_{s}\mathbb{I}(\rho_{\epsilon} > s),
	\end{align}
	and by \eqref{eq:l32_pr_85} we have that
	\begin{align}
	\label{eq:l32_pr_87}
	\mathbb{E}_{\theta^{*}}[V_{1}^{\epsilon}(t+1)|\mathcal{F}_{t}]\le V^{\epsilon}(t)-c_{9}(\epsilon)\alpha_{t}\mathbb{I}\left(\rho_{\epsilon} > t\right)+c_{9}(\epsilon)\sum_{s=0}^{t-1}\alpha_{s}\mathbb{I}(\rho_{\epsilon} > s)=V_{1}^{\epsilon}(t),
	\end{align}
	\noindent for all $t\geq t_{\epsilon}$.
	\noindent Hence, we have that $\{V_{1}^{\epsilon}(t)\}_{t\geq t_{\epsilon}}$ is a non-negative supermartingale and there exists a finite random variable $V_{1}^{\epsilon}$ such that $V_{1}^{\epsilon}(t)\to V_{1}^{\epsilon}$ a.s. as $t\to\infty$. From the definition in \eqref{eq:l32_pr_86}, we have that the following limit exists:
	\begin{align}
	\label{eq:l32_pr_88}
	\lim_{t\to\infty}c_{9}(\epsilon)\sum_{s=0}^{t-1}\alpha_{s}\mathbb{I}(\rho_{\epsilon} > s)=V_{1}^{\epsilon}-V^{\epsilon} < \infty~\mbox{a.s.}
	\end{align}
	\noindent We also have that as $t \to\infty$, $\sum_{s=0}^{t-1}\alpha_{s}\to\infty$, the limit condition in \eqref{eq:l32_pr_88} is satisfied only if $\rho_{\epsilon} < \infty$ a.s.
	
	\noindent Let's define the sequence $\{\mathbf{x}(\rho_{1/p})\}$, by choosing $\epsilon=1/p$, for each positive integer $p > 1$. By definition, we have,
	\begin{align}
	\label{eq:l32_pr_89}
	\left\|\mathbf{x}(\rho_{1/p})\right\|\in [1, 1/p)\cup (p, \infty)~\mbox{a.s.}
	\end{align}
	
	\noindent We also have from Lemma \ref{l2} that
	\begin{align}
	\label{eq:l32_pr_90}
	\mathbb{P}_{\theta^{*}}\left(\left\|\mathbf{x}(\rho_{1/p})\right\| > p ~\mbox{i.o.}\right)=0,
	\end{align}
	\noindent where i.o. denotes infinitely often as $p\to\infty$. Hence, by \eqref{eq:l32_pr_89} we have that there exists a finite integer valued random variable $p^{*}$ such that $\left\|\mathbf{x}(\rho_{1/p})\right\| < 1/p^{*}$, $\forall p\geq p^{*}$, which in turn implies that $\left\|\mathbf{x}(\rho_{1/p})\right\| \to 0$ as $p\to\infty$. Finally, we have that
	\begin{align}
	\label{eq:l32_pr_91}
	\mathbb{P}_{\theta^{*}}\left(\liminf_{t\to\infty}\left\|\mathbf{x}(\rho_{1/p})\right\| =0\right)=1.
	\end{align}
	\noindent With the above development in place we have from \eqref{eq:l2_pr_7} that $\liminf_{t\to\infty}V_{2}(t)=0$ a.s.
	Noting that the limit of $\{V_{2}(t)\}$ exists, we have that $V_{2}(t)\to 0$ as $t\to\infty$ a.s. and again from \eqref{eq:l2_pr_7}, we have that $\mathbf{x}(t) \to 0$ as $t\to\infty$ a.s.
\end{IEEEproof}
\noindent\begin{IEEEproof}[Proof of Lemma \ref{main_res_l1}]
	\noindent Define, the process $\{\hat{z}_{\mbox{\scriptsize{avg}}}(t)\}$ as follows :
	\begin{align}
	\label{eq:main_l1_1}
	\hat{z}_{\mbox{\scriptsize{avg}}}(t)=z_{\mbox{\scriptsize{avg}}}(t)-\frac{\mathbf{h}^{\top}\left(\mathbf{\theta}_{N}^{*}\right)\mathbf{\Sigma}^{-1}\mathbf{h}\left(\mathbf{\theta}_{N}^{*}\right)}{2N}.
	\end{align}
	\noindent The recursion for $\{\hat{z}_{\mbox{\scriptsize{avg}}}(t)\}$ can then be represented as
	\begin{align}
	\label{eq:main_l1_2}
	&\hat{z}_{\mbox{\scriptsize{avg}}}(t+1)=\left(1-\frac{1}{t+1}\right)\hat{z}_{\mbox{\scriptsize{avg}}}(t)+\frac{1}{N\left(t+1\right)}\sum_{n=1}^{N}\mathbf{h}_{n}^{\top}\left(\mathbf{\theta}_{n}(t)\right)\mathbf{\Sigma}_{n}^{-1}\left(\mathbf{y}_{n}(t)-\mathbf{h}_{n}\left(\mathbf{\theta}^{*}\right)\right)\nonumber\\&-\frac{1}{2N(t+1)}\sum_{n=1}^{N}\left(\mathbf{h}_{n}\left(\mathbf{\theta}_{n}(t)\right)-\mathbf{h}_{n}\left(\mathbf{\theta}^{*}\right)\right)^{\top}\mathbf{\Sigma}_{n}^{-1}\left(\mathbf{h}_{n}\left(\mathbf{\theta}_{n}(t)\right)-\mathbf{h}_{n}\left(\mathbf{\theta}^{*}\right)\right)\nonumber\\
	&=\left(1-\frac{1}{t+1}\right)\hat{z}_{\mbox{\scriptsize{avg}}}(t)+\frac{1}{N\left(t+1\right)}\mathbf{h}^{\top}\left(\mathbf{\theta}(t)\right)\mathbf{\Sigma}^{-1}\left(\mathbf{y}(t)-\mathbf{h}\left(\mathbf{\theta}_{N}^{*}\right)\right)\nonumber\\&-\frac{1}{2N(t+1)}\left(\mathbf{h}\left(\mathbf{\theta}(t)\right)-\mathbf{h}\left(\mathbf{\theta}_{N}^{*}\right)\right)^{\top}\mathbf{\Sigma}^{-1}\left(\mathbf{h}\left(\mathbf{\theta}(t)\right)-\mathbf{h}\left(\mathbf{\theta}_{N}^{*}\right)\right).
	\end{align}
	\noindent In order to apply Lemma \ref{main_res_l0} to the process $\{\hat{z}_{\mbox{\scriptsize{avg}}}(t)\}$, define
	\begin{align}
	\label{eq:main_l1_3}
	&\mathbf{\Gamma}_{t}=\mathbf{I},\nonumber\\
	&\mathbf{\Phi}_{t}=\frac{1}{N}\mathbf{h}^{\top}\left(\mathbf{\theta}(t)\right)\mathbf{\Sigma}^{-1},\nonumber\\
	&\mathbf{V}_{t}=\mathbf{y}(t)-\mathbf{h}\left(\mathbf{\theta}_{N}^{*}\right),\nonumber\\
	&\mathbf{T}_{t}=\sqrt{t+1}\left(\mathbf{h}\left(\mathbf{\theta}(t)\right)-\mathbf{h}\left(\mathbf{\theta}_{N}^{*}\right)\right)^{\top}\mathbf{\Sigma}^{-1}\left(\mathbf{h}\left(\mathbf{\theta}(t)\right)-\mathbf{h}\left(\mathbf{\theta}_{N}^{*}\right)\right).
	\end{align}
	
	\noindent From Assumption \ref{as:3}, we have that,
	\begin{align}
	\label{eq:main_l1_4}
	\left\|\mathbf{h}\left(\mathbf{\theta}(t)\right)-\mathbf{h}\left(\mathbf{\theta}_{N}^{*}\right)\right\|\le k_{\mbox{\scriptsize{max}}}\left\|\mathbf{\theta}(t)-\mathbf{\theta}^{*}\right\|,
	\end{align}
	\noindent where $k_{\mbox{\scriptsize{max}}}=\max_{n=1,\cdots,N}k_{n}$, with the $k_{n}$'s defined in Assumption \ref{as:3}.
	\noindent Moreover, from theorem \ref{t1} we have that, with $\tau=1/4$,
	\begin{align}
	\label{eq:main_l1_5}
	\lim_{t\rightarrow\infty}\sqrt{t+1}\left\|\mathbf{\theta}(t)-\mathbf{\theta}_{N}^{*}\right\|^{2}=0~\mbox{a.s.}
	\end{align}
	The above implies that
	\begin{align}
	\label{eq:main_l1_51}
	&\lim_{t\to\infty}\sqrt{t+1}\left(\mathbf{h}\left(\mathbf{\theta}(t)\right)-\mathbf{h}\left(\mathbf{\theta}_{N}^{*}\right)\right)^{\top}\mathbf{\Sigma}^{-1}\left(\mathbf{h}\left(\mathbf{\theta}(t)\right)-\mathbf{h}\left(\mathbf{\theta}_{N}^{*}\right)\right)\nonumber\\
	&\le \lim_{t\to\infty}\sqrt{t+1}\left\|\mathbf{h}\left(\mathbf{\theta}(t)\right)-\mathbf{h}\left(\mathbf{\theta}_{N}^{*}\right)\right\|^{2}\left\|\mathbf{\Sigma}^{-1}\right\|=0.
	\end{align}
	%Finally, we have,
%	\begin{align}
%	\label{eq:main_l1_52}
%	 \lim_{t\to\infty}\sqrt{t+1}\left(\mathbf{h}\left(\mathbf{\theta}(t)\right)-\mathbf{h}\left(\mathbf{\theta}_{N}^{*}\right)\right)^{\top}\mathbf{\Sigma}^{-1}\left(\mathbf{h}\left(\mathbf{\theta}(t)\right)-\mathbf{h}\left(\mathbf{\theta}_{N}^{*}\right)\right)=0.
%	\end{align}

	\noindent From Theorem \ref{t1}, we have $\mathbf{\Phi}_{t}=\frac{1}{N}\mathbf{h}^{\top}\left(\mathbf{\theta}(t)\right)\mathbf{\Sigma}^{-1}\to\frac{1}{N}\mathbf{h}^{\top}\left(\mathbf{\theta}_{N}^{*}\right)\mathbf{\Sigma}^{-1}$ a.s. as $t\to\infty$.

	\noindent Clearly, $\mathbb{E}_{\theta^{*}}\left[\mathbf{V}_{t}|\mathcal{F}_{t}\right]=0$ and $\mathbb{E}_{\theta^{*}}\left[\mathbf{V}_{t}\mathbf{V}_{t}^{\top}|\mathcal{F}_{t}\right]=\mathbf{\Sigma}$. Due to the i.i.d nature of the noise process, the required uniform integrability condition for the process $\{\mathbf{V}_{t}\}$ is also verified. Hence, $\{z_{\mbox{\scriptsize{avg}}}(t)\}$ falls under the purview of Lemma \ref{main_res_l0} and the assertion follows.
\end{IEEEproof}
\begin{IEEEproof}[Proof of Lemma \ref{int_res_2}]
	\noindent Define the process $\{\mathbf{p}(t)\}$ as follows:
	\begin{align}
	\label{eq:int_res_2_pr1}
	\mathbf{p}(t)=\mathbf{z}(t)-\mathbf{1}_{N}\otimes z_{\mbox{\scriptsize{avg}}}(t).
	\end{align}
	\noindent Then $\{\mathbf{p}(t)\}$ evolves as
	\begin{align}
	\label{eq:int_res_2_pr2}
	& \mathbf{p}(t+1)=\frac{t}{t+1}(\mathbf{W}-\mathbf{J})\mathbf{p}(t)+\frac{1}{t+1}\left(\mathbf{h}^{*}\left(\mathbf{\theta}(t)\right)-\frac{\mathbf{1}\mathbf{1}^{\top}}{N}\mathbf{h}^{\top}\left(\mathbf{\theta}(t)\right)\right)J\left(\mathbf{y}(t)\right)\nonumber\\&-\frac{1}{2(t+1)}\left(\mathbf{h}^{*}\left(\mathbf{\theta}(t)\right)\mathbf{\Sigma}^{-1}\mathbf{h}\left(\mathbf{\theta}(t)\right)-\frac{\mathbf{1}_{N}}{N}\otimes\left(\mathbf{h}^{\top}\left(\theta(t)\right)\mathbf{\Sigma}^{-1}\mathbf{h}\left(\theta(t)\right)\right)\right),
	\end{align}
	
	\noindent where $J\left(\mathbf{y}(t)\right)=\mathbf{\Sigma}^{-1}\mathbf{y}(t)$.
	The following lemmas are instrumental for the subsequent analysis. Lemma \ref{int_res_0} is concerned with a stochastic approximation type result which will be used later in the proof, whereas, Lemma \ref{int_res_1} establishes the a.s. boundedness of $J\left(\mathbf{y}(t)\right)$.
	\begin{Lemma}[\hspace{-0.3pt}\cite{kar2010large}]
		\label{int_res_0}
		\noindent Consider the scalar time-varying linear system
		\begin{align}
		\label{eq:int_res_0}
		u(t+1)=(1-r_{1}(t))u(t)+r_{2}(t),
		\end{align}
		\noindent where $\{r_{1}(t)\}$ is a sequence, such that, $0\leq r_{1}(t) \leq 1$ and is given by
		\begin{align}
		\label{eq:int_res_0_1}
		r_{1}(t)=\frac{a_{1}}{(t+1)^{\delta_{1}}}
		\end{align}
		\noindent with $a_{1} >0, 0\leq\delta_{1}\leq 1$, whereas the sequence $\{r_{2}(t)\}$ is given by
		\begin{align}
		\label{eq:int_res_0_2}
		r_{2}(t)=\frac{a_{2}}{(t+1)^{\delta_{2}}}
		\end{align}
		\noindent with $a_{2}>0, \delta_{2}\geq 0$. Then, if $u(0)\geq 0$ and $\delta_{1} < \delta_{2}$, we have
		\begin{align}
		\label{eq:int_res_0_3}
		\lim_{t\to\infty}(t+1)^{\delta_0}u(t)=0,
		\end{align}
		\noindent for all $0\le\delta_{0}<\delta_{2}-\delta_{1}$.
	\end{Lemma}
	
	\begin{IEEEproof}
		\noindent A proof of this Lemma can be found in \cite{kar2010large} in the proof of Lemma 3.3.3 in Chapter 3.
	\end{IEEEproof}

	\begin{Lemma}
		\label{int_res_1}
		Define $J(\mathbf{y}(t))$ as follows:
		\begin{align}
		\label{eq:l1_1}
		J\left(\mathbf{y}(t)\right)=\mathbf{\Sigma}^{-1}\mathbf{y}(t)
		\end{align}
		Then we have
		\begin{align}
		\label{eq:l1_2}
		\mathbb{P}_{\theta^{*}}\left(\lim_{t\to\infty}\frac{1}{(t+1)^{\delta}}||J(\mathbf{y}(t))|| =0\right)=1.
		\end{align}
	\end{Lemma}
	
	\begin{IEEEproof}
		\noindent Consider any $\epsilon_{1} > 0$. By Chebyshev's inequality, we have,
		\begin{align}
		\label{eq:l1proof_1}
		&\mathbb{P}_{\theta^{*}}\left(\frac{1}{(t+1)^{\delta}}\left\|J(\mathbf{y}(t))\right\| > \epsilon_{1}\right)\le \frac{1}{\epsilon_{1}^{1+\frac{1}{\delta}}(t+1)^{1+\delta}}\mathbb{E}_{\theta^{*}}\left[\left\|J(\mathbf{y}(t))\right\|^{1+\frac{1}{\delta}}\right]\nonumber\\
		&=\frac{\mathcal{K}(\theta^{*})}{(t+1)^{1+\frac{1}{\delta}}}
		\end{align}
		
		\noindent where $\mathbb{E}_{\theta^{*}}[\left\|J(\mathbf{y}(t))\right\|^{1+\frac{1}{\delta}}]=\mathcal{K}(\theta^{*}) < \infty$ because the noise in consideration is Gaussian and has finite moments. Moreover, since $\delta > 0$, the sequence $(t+1)^{1+\frac{1}{\delta}}$ is square summable and we obtain
		\begin{align}
		\label{eq:l1proof_2}
		\sum_{t > 0}\mathbb{P}_{\theta^{*}}\left(\frac{1}{(t+1)^{\delta}}\left\|J(\mathbf{y}(t))\right\| > \epsilon_{1}\right) < \infty.
		\end{align}
		
		\noindent Hence, we have from the Borel-Cantelli Lemma, for arbitrary $\epsilon_{1} > 0$,
		\begin{align}
		\label{eq:l1proof_3}
		\mathbb{P}_{\theta^{*}}\left(\frac{1}{(t+1)^{\delta}}\left\|J(\mathbf{y}(t))\right\| > \epsilon_{1}~\mbox{i.o.} \right)=0,
		\end{align}
		
		\noindent where i.o. stands for infinitely often and the claim follows from standard arguments.
	\end{IEEEproof}
	
	\noindent We also have from Lemma \ref{l2} that
	\begin{align}
	\label{eq:iint_res_2_pr11}
	\mathbb{P}\left(\sup_{t\ge 0}\left\|\left(\mathbf{h}^{*}\left(\mathbf{\theta}(t)\right)-\frac{\mathbf{1}\mathbf{1}^{\top}}{N}\mathbf{h}^{\top}\left(\mathbf{\theta}(t)\right)\right)\right\| < \infty\right)=1,
	\end{align}
	\noindent and combining this with lemma \ref{int_res_1}, we have,
	\begin{align}
		\label{eq:iint_res_2_pr12}
		\mathbb{P}\left(\sup_{t\ge 0}\left\|\left(\mathbf{h}^{*}\left(\mathbf{\theta}(t)\right)-\frac{\mathbf{1}\mathbf{1}^{\top}}{N}\mathbf{h}^{\top}\left(\mathbf{\theta}(t)\right)\right)J(\mathbf{y}(t))\right\| < \infty\right)=1.
	\end{align}

	\noindent To prove uniform bounds, we use truncation arguments. For a scalar $d$, let its truncation $(d)^{A_{0}}$ be defined at level $A_{0}$ by
	\begin{align}
	\label{eq:int_res_2_pr3}
	(d)^{A_{0}}=
	\begin{cases}
	\frac{d}{|d|}\min(|d|,A_{0}), & \text{if}~~d \neq 0 \\
	0 ,& \text{if}~~ d=0,
	\end{cases}
	\end{align}	
	\noindent while for a vector, the truncation operator is applied component-wise.
	\noindent To this end, we consider sequences $\{\mathbf{p}_{A_{0}}(t)\}$, which is in turn given by,
	\begin{align}
	\label{eq:int_res_2_pr4}
	& \mathbf{p}_{A_{0}}(t+1)=\frac{t}{t+1}\left(\mathbf{W}-\mathbf{J}\right)\mathbf{p}_{A_{0}}(t)+\frac{1}{t+1}\left(J_{1}\left(\mathbf{y}(t)\right)\right)^{A_{0}(t+1)^{\delta}}\nonumber\\&-\frac{1}{2(t+1)}\left(\left(\mathbf{h}^{*}\left(\mathbf{\theta}(t)\right)\mathbf{\Sigma}^{-1}\mathbf{h}\left(\mathbf{\theta}(t)\right)-\frac{\mathbf{1}_{N}}{N}\otimes\left(\mathbf{h}^{\top}\left(\theta(t)\right)\mathbf{\Sigma}^{-1}\mathbf{h}\left(\theta(t)\right)\right)\right)\right)^{A_{0}},
	\end{align}
	\noindent where $J_{1}\left(\mathbf{y}(t)\right)=\left(\mathbf{h}^{*}\left(\mathbf{\theta}(t)\right)-\frac{\mathbf{1}\mathbf{1}^{\top}}{N}\mathbf{h}^{\top}\left(\mathbf{\theta}(t)\right)\right)J(\mathbf{y}(t))$, $A_{0}>0$ and $\delta > 0$.
	
	\noindent In order to prove the assertion,
	\begin{align}
	\label{eq:int_res_2_pr5}
	\mathbb{P}_{\theta^{*}}\left(\lim_{t\to\infty}(t+1)^{\delta_{0}}\mathbf{p}(t)=\mathbf{0}\right)=1,
	\end{align}
	\noindent it is sufficient to prove that for every $A_{0} > 0$,
	\begin{align}
	\label{eq:int_res_2_pr6}
	\mathbb{P}_{\theta^{*}}\left(\lim_{t\to\infty}(t+1)^{\delta_{0}}\mathbf{p}_{A_{0}}(t)=\mathbf{0}\right)=1,
	\end{align}
	\noindent which is due to the following standard arguments. The pathwise boundedness of the different terms in the recursion for $\mathbf{p}(t)$ as defined in \eqref{eq:int_res_2_pr4}  implies that, for every $\epsilon > 0$, there exists $A_{\epsilon}$ such that
	\begin{align}
	\label{eq:int_res_2_pr7}
	\mathbb{P}_{\theta^{*}}\left(\sup\left\|J_{1}(\mathbf{y}(t))\right\| < A_{\epsilon}(t+1)^{\delta_{0}}\right) > 1-\epsilon,
	\end{align}
and
	
	\begin{align}
	\label{eq:int_res_2_pr8}
	\mathbb{P}_{\theta^{*}}\left(\sup\left\|\left(\mathbf{h}^{*}\left(\mathbf{\theta}(t)\right)\mathbf{\Sigma}^{-1}\mathbf{h}\left(\mathbf{\theta}(t)\right)-\frac{\mathbf{1}_{N}}{N}\otimes\left(\mathbf{h}^{\top}\left(\theta(t)\right)\mathbf{\Sigma}^{-1}\mathbf{h}\left(\theta(t)\right)\right)\right)\right\| < A_{\epsilon}\right) > 1-\epsilon.
	\end{align}
	
	\noindent In particular, \eqref{eq:int_res_2_pr7} follows from the pathwise boundedness of $\{\mathbf{\theta}(t)\}$ proved in Lemma \ref{l2}, whereas, \eqref{eq:int_res_2_pr8} follows from the a.s. convergence in Lemma \ref{int_res_1}. The processes $\{\mathbf{p}(t)\}$ and $\{\mathbf{p}_{A_{\epsilon}}(t)\}$ agree on the set where both of the above mentioned events occur. Hence, it follows that,
	\begin{align}
	\label{eq:int_res_2_pr9}
	\mathbb{P}_{\theta^{*}}\left(\sup\left\|\mathbf{p}(t)-\mathbf{p}_{A_{\epsilon}}(t)\right\|=0\right) > 1-2\epsilon.
	\end{align}
	
	\noindent Invoking the claim in \eqref{eq:int_res_2_pr6}, we have,
	\begin{align}
	\label{eq:int_res_2_pr10}
	\mathbb{P}_{\theta^{*}}\left(\lim_{t\to\infty}(t+1)^{\delta_{0}}\mathbf{p}(t)=\mathbf{0}\right) > 1-2\epsilon.
	\end{align}
	\noindent The assertion then can be proved by taking $\epsilon$ to $0$.
	
	\noindent In order to establish the claim in \eqref{eq:int_res_2_pr6}, for every $A_{0} > 0$, consider the scalar process $\{\hat{p}_{A_{0}}(t)\}_{t\geq 0}$ defined as
	\begin{align}
	\label{eq:int_res_2_pr11}
	\hat{p}_{A_{0}}(t+1)=\left\|\mathbf{I}_{N}-\delta\mathbf{L}-\mathbf{J}\right\|\hat{p}_{A_{0}}(t)+\frac{NA_{0}}{2(t+1)}+\frac{NA_{0}(t+1)^{\delta_{0}}}{t+1},
	\end{align}
	\noindent where $\hat{p}_{A_{0}}(0)$ is initialized as $\hat{p}_{A_{0}}(0)=\left\|\mathbf{p}_{A_{0}}(0)\right\|$ and $\delta$ is as defined in \eqref{eq:dec_upt_l}.
	\noindent From \eqref{eq:int_res_2_pr4}, we have,
	\begin{align}
	\label{eq:int_res_2_pr12}
	& \left\|\mathbf{p}_{A_{0}}(t+1)\right\|\le\frac{t}{t+1}\left\|\left(\mathbf{W}-\mathbf{J}\right)\right\|\left\|\mathbf{p}_{A_{0}}(t)\right\|+\frac{1}{t+1}\left\|\left(J_{1}\left(\mathbf{y}(t)\right)\right)^{A_{0}(t+1)^{\delta}}\right\|\nonumber\\&+\frac{1}{2(t+1)}\left\|\left(\left(\mathbf{h}^{*}\left(\mathbf{\theta}(t)\right)\mathbf{\Sigma}^{-1}\mathbf{h}\left(\mathbf{\theta}(t)\right)-\frac{\mathbf{1}_{N}}{N}\otimes\left(\mathbf{h}^{\top}\left(\theta(t)\right)\mathbf{\Sigma}^{-1}\mathbf{h}\left(\theta(t)\right)\right)\right)\right)^{A_{0}}\right\|\nonumber\\
	&\le \left\|\mathbf{I}_{N}-\delta\mathbf{L}-\mathbf{J}\right\|\left\|\mathbf{p}_{A_{0}}(t)\right\|+\frac{NA_{0}}{2(t+1)}+\frac{NA_{0}(t+1)^{\delta_{0}}}{t+1}.
	\end{align}
	
	\noindent Given the initial condition for $\hat{p}_{A_{0}}(0)$, through an induction argument we have that
	\begin{align}
	\label{eq:int_res_2_pr13}
	\left\|\mathbf{p}_{A_{0}}(t+1)\right\|\le\hat{p}_{A_{0}}(t+1), \forall t.
	\end{align}
	
	\noindent Moreover, we also have that,
	\begin{align}
	\label{eq:int_res_2_pr14}
	\left\|\mathbf{I}_{N}-\delta\mathbf{L}-\mathbf{J}\right\|=\frac{\lambda_{N}\left(\mathbf{L}\right)-\lambda_{2}\left(\mathbf{L}\right)}{\lambda_{N}\left(\mathbf{L}\right)+\lambda_{2}\left(\mathbf{L}\right)}.
	\end{align}
	
	\noindent Using \eqref{eq:int_res_2_pr14} in \eqref{eq:int_res_2_pr11}, we have,
	\begin{align}
	\label{eq:int_res_2_pr15}
	\hat{p}_{A_{0}}(t+1)\le\left(1-\frac{2\lambda_{2}\left(\mathbf{L}\right)}{\lambda_{N}\left(\mathbf{L}\right)+\lambda_{2}\left(\mathbf{L}\right)}\right)\hat{p}_{A_{0}}(t)+\frac{2NA_{0}}{(t+1)^{1-\delta_{0}}},
	\end{align}
	\noindent where $\frac{2\lambda_{2}\left(\mathbf{L}\right)}{\lambda_{N}\left(\mathbf{L}\right)+\lambda_{2}\left(\mathbf{L}\right)} < 1$ and hence the recursion in \eqref{eq:int_res_2_pr15}~comes under the purview of Lemma \ref{int_res_0}. Hence, we have
	\begin{align}
	\label{eq:int_res_2_pr16}
	\mathbb{P}_{\theta^{*}}\left(\lim_{t\to\infty}(t+1)^{\delta_{0}}\hat{p}_{A_{0}}(t)=0\right)=1.
	\end{align}
	
	\noindent Finally, the assertion follows from by invoking \eqref{eq:int_res_2_pr13} and noting that, for arbitrary $A_{0}>0$,
	\begin{align}
	\label{eq:int_res_2_pr17}
	\mathbb{P}_{\theta^{*}}\left(\lim_{t\to\infty}(t+1)^{\delta_{0}}\mathbf{p}_{A_{0}}(t)=0\right)=1.
	\end{align}
\end{IEEEproof}

\section{Proofs of Lemmas in Section \ref{sec:proof_res_cilrt}}
\label{a2}
\noindent\begin{IEEEproof}[Proof of Lemma \ref{le:det1}]
 	
\noindent First, we note that both the matrices $\mathbf{L}\otimes\mathbf{I}_{M}$ and $\mathbf{G}_{H}\mathbf{\Sigma}^{-1}\mathbf{G}_{H}^{\top}$ are symmetric and positive semi-definite. Then the matrix $\mathbf{L}\otimes\mathbf{I}_{M}+\mathbf{G}_{H}\mathbf{\Sigma}^{-1}\mathbf{G}_{H}^{\top}$ is positive semi-definite as it is the sum of two positive semi-definite matrices. To prove that the matrix $\mathbf{L}\otimes\mathbf{I}_{M}+\mathbf{G}_{H}\mathbf{\Sigma}^{-1}\mathbf{G}_{H}^{\top}$ is positive definite, let's assume that it's not positive definite. Hence there exists $\mathbf{x}\in\mathbb{R}^{NM}$, where $\mathbf{x}\neq\mathbf{0}$ such that
\begin{align}
\label{eq:det1_1}
\mathbf{x}^{\top}\left(\mathbf{L}\otimes\mathbf{I}_{M}+\mathbf{G}_{H}\mathbf{\Sigma}^{-1}\mathbf{G}_{H}^{\top}\right)\mathbf{x}=0,
\end{align}
\noindent which further implies that
\begin{align}
\label{eq:det1_2}
\mathbf{x}^{\top}\left(\mathbf{L}\otimes\mathbf{I}_{M}\right)\mathbf{x}=0~~\mbox{and}~~ \mathbf{x}^{\top}\left(\mathbf{G}_{H}\mathbf{\Sigma}^{-1}\mathbf{G}_{H}^{\top}\right)\mathbf{x}=0.
\end{align}

\noindent Moreover, $\mathbf{x}$ can be written as $\mathbf{x}=\left[\mathbf{x}_{1}^{\top},\cdots, \mathbf{x}_{N}^{\top}\right]^{\top}$, with $\mathbf{x}_{n} \in \mathbb{R}^{M}$ for all $n$. Now note that, by the properties of the graph Laplacian \eqref{eq:det1_2} holds if and only if (\emph{iff})
\begin{align}
\label{eq:det1_3}
\mathbf{x}_{n}=\mathbf{g},~\forall n,
\end{align}
\noindent where $\mathbf{g}\in\mathbb{R}^{M}$ and $\mathbf{g}\neq\mathbf{0}$. Hence, from \eqref{eq:det1_2},we have,
\begin{align}
\label{eq:det1_4}
\sum_{n=1}^{N}\mathbf{g}^{\top}\mathbf{H}_{n}^{\top}\mathbf{\Sigma}_{n}^{-1}\mathbf{H}_{n}\mathbf{g}=\mathbf{g}^{\top}\mathbf{G}\mathbf{g}=0,
\end{align}
\noindent which is a contradiction from Assumption \ref{bs:1} as $\mathbf{G}$ is invertible. Hence, we have that $\mathbf{L}\otimes\mathbf{I}_{M}+\mathbf{G}_{H}\mathbf{\Sigma}^{-1}\mathbf{G}_{H}^{\top}$ is positive definite.
\noindent Since $\beta_{t}/\alpha_{t} \to \infty$ as $t\to\infty$, there exists an integer $t_{4}$ (sufficiently large) such that $\forall t\geq t_{4}$ and for all $\mathbf{x}$ with $\left\|\mathbf{x}\right\|=1$,
\begin{align}
\label{eq:det1_5}
&\mathbf{x}^{\top}\left(\beta_{t}\left(\mathbf{L}\otimes\mathbf{I}_{M}\right)+\alpha_{t}\mathbf{G}_{H}\mathbf{\Sigma}^{-1}\mathbf{G}_{H}^{\top}\right)\mathbf{x}\nonumber\\
&=\alpha_{t}\mathbf{x}^{\top}\left(\frac{\beta_{t}}{\alpha_{t}}\left(\mathbf{L}\otimes\mathbf{I}_{M}\right)+\mathbf{G}_{H}\mathbf{\Sigma}^{-1}\mathbf{G}_{H}^{\top}\right)\mathbf{x}\nonumber\\
&\geq\alpha_{t}\mathbf{x}^{\top}\left(\left(\mathbf{L}\otimes\mathbf{I}_{M}\right)+\mathbf{G}_{H}\mathbf{\Sigma}^{-1}\mathbf{G}_{H}^{\top}\right)\mathbf{x}\geq c_{1}\alpha_{t},
\end{align}
\noindent where
\begin{align}
\label{eq:det1_55}
c_{1}=\lambda_{\mbox{\scriptsize{min}}}\left(\left(\mathbf{L}\otimes\mathbf{I}_{M}\right)+\mathbf{G}_{H}\mathbf{\Sigma}^{-1}\mathbf{G}_{H}^{\top}\right).
\end{align}
\noindent We now choose a $t_{3} > t_{4}$ such that $\forall t\geq t_{3}$, $c_{1}\alpha_{t}<1$.

\noindent In order to ensure that all the eigenvalues of $\left(\mathbf{I}_{NM}-\beta_{t}\left(\mathbf{L}\otimes\mathbf{I}_{M}\right)-\alpha_{t}\mathbf{G}_{H}\mathbf{\Sigma}^{-1}\mathbf{G}_{H}^{\top}\right)$ are positive, we choose a $t_{2}$ such that $\forall t\geq t_{2}$,
\begin{align}
\label{eq:det1_7}
\beta_{t}\lambda_{N}(\mathbf{L})+\alpha_{t}\lambda_{\mbox{\scriptsize{max}}}(\mathbf{G}_{H}\mathbf{\Sigma}^{-1}\mathbf{G}_{H}^{\top}) < 1.
\end{align}
It is to be noted that such choices of $t_{3}$ and $t_{2}$ are possible as $\beta_{t},\alpha_{t}\to 0$ as $t\to\infty$. Moreover, the condition in \eqref{eq:det1_7} readily implies that $\lambda_{\mbox{\scriptsize{max}}}\left(\beta_{t}\left(\mathbf{L}\otimes\mathbf{I}_{M}\right)+\alpha_{t}\mathbf{G}_{H}\mathbf{\Sigma}^{-1}\mathbf{G}_{H}^{\top}\right)\le \beta_{t}\lambda_{N}(\mathbf{L})+\alpha_{t}\lambda_{\mbox{\scriptsize{max}}}(\mathbf{G}_{H}\mathbf{\Sigma}^{-1}\mathbf{G}_{H}^{\top}) < 1$ for all $t\geq t_{2}$.
Hence, from \eqref{eq:det1_5}, we have $\forall t\geq t_{1}$, with $t_{1}=\max\{t_{2},t_{3}\}$, and for all $\mathbf{x}$ such that $\left\|\mathbf{x}\right\|=1$,
\begin{align}
\label{eq:det1_8}
\mathbf{x}^{\top}\left(\mathbf{I}_{NM}-\beta_{t}\left(\mathbf{L}\otimes\mathbf{I}_{M}\right)-\alpha_{t}\mathbf{G}_{H}\mathbf{\Sigma}^{-1}\mathbf{G}_{H}^{\top}\right)\mathbf{x}\le 1-c_{1}\alpha_{t},
\end{align} which implies that
\begin{align}
\label{eq:det1_8_1}
 \left\|\mathbf{I}_{NM}-\beta_{t}\left(\mathbf{L}\otimes\mathbf{I}_{M}\right)-\alpha_{t}\mathbf{G}_{H}\mathbf{\Sigma}^{-1}\mathbf{G}_{H}^{\top}\right\|\le 1-c_{1}\alpha_{t},
\end{align}
for all $t\geq t_{1}$.
\end{IEEEproof}
\allowdisplaybreaks[1]
\noindent\begin{IEEEproof}[Proof of Lemma \ref{wish}]
\noindent The following Lemma from \cite{bourin2013decomposition}, will be used in the subsequent analysis.
\begin{Lemma}[\hspace{-0.2pt}\cite{bourin2013decomposition}]
  		\label{nb}
  		Given a positive-semidefinite matrix $\mathbf{P}$ ($Nt\times Nt$), with each of its blocks ($N\times N$) being symmetric, the following result holds for any invariant norm,
  		\begin{align}
  		\label{eq:e6}
  		\left\|\mathbf{P}\right\|\le \left\|\sum_{i=1}^{t}\left[\mathbf{P}\right]_{ii}\right\|.
  		\end{align}
\end{Lemma}
  	\noindent From Lemma \ref{nb}, we have that,
  	\begin{align}
  	\label{eq:e8}
  	\left\|\mathbf{P}_{t}\right\|\le \sum_{i=1}^{t}\left\|\left[\mathbf{P}_{t}\right]_{ii}\right\|.
  	\end{align}
  	\noindent From Lemma \ref{le:det1}. we have that, $\forall t\ge t_{1}$,
  	\begin{align}
  	\label{eq:e9}
  	\left\|\mathbf{A}(u)\right\| \le (1-c_{1}\alpha_{u}),
  	\end{align}which implies
  	\begin{align}
  	\left\|\left[\mathbf{P}_{t}\right]_{ii}\right\|\le \alpha_{i}^{2}\prod_{u=i}^{t-1}(1-c_{1}\alpha_{u})^{2},
  	\end{align}
  	for all $t\geq t_{1}$.
  	\noindent Using \eqref{eq:e9}, the RHS of \eqref{eq:e8} can be rewritten as
  	\begin{align}
  	\label{eq:e10}
  	\sum_{i=1}^{t}\left\|\left[\mathbf{P}_{t}\right]_{ii}\right\|\le c_{3}\prod_{u=t_{1}}^{t-1}(1-c_{1}\alpha_{u})^{2}+\sum_{v=t_{1}}^{t}\alpha_{v}^{2}\prod_{u=v+1}^{t-1}(1-c_{1}\alpha_{u})^{2},
  	\end{align}
  	\noindent where $c_{3}$ is given by
  	\begin{align}
  	\label{eq:e11}
  	c_{3}=\sum_{v=0}^{t_{1}-1}\alpha_{v}^{2}\prod_{u=v+1}^{t_{1}-1}\left\|\mathbf{A}(u)\right\|.
  	\end{align}
  	
  	\noindent	Using the properties of Riemann integration and the inequality $(1-x) \le e^{-x}$, for $x\in (0,1)$, we have,
  	\begin{align}
  	\label{eq:e12}
  	\prod_{u=i}^{t-1}(1-c_{1}\alpha_{u})^{2} \le \left(\frac{i+1}{t}\right)^{2c_{1}\alpha_{0}},
  	\end{align}
  	where, in the derivation, we also use the property that
  	\begin{align}
  	\label{eq:e121}
  	\sum_{u=i+1}^{t}\frac{1}{u} > \ln\left(\frac{t}{i+1}\right).
  	\end{align}
  	\allowdisplaybreaks[1]
  	\noindent On using \eqref{eq:e12}, in \eqref{eq:e10} we have $\forall t \geq t_{1}$,
  	\begin{align}
  	\label{eq:e13}
  	&\sum_{i=1}^{t}\left\|\left[\mathbf{P}_{t}\right]_{ii}\right\|\le c_{3}\prod_{u=t_{1}}^{t-1}(1-c_{1}\alpha_{u})^{2}+\sum_{v=t_{1}}^{t}\alpha_{v}^{2}\prod_{u=v+1}^{t-1}(1-c_{1}\alpha_{u})^{2}\nonumber\\
  	&\le c_{3}\left(\frac{t_{1}+1}{t}\right)^{2c_{1}\alpha_{0}}+\sum_{u=t_{1}+1}^{t-1}\alpha_{u}^{2}\left(\frac{u+1}{t}\right)^{2c_{1}\alpha_{0}}\nonumber\\
  	&= c_{3}\left(\frac{t_{1}+1}{t}\right)^{2c_{1}\alpha_{0}}+\alpha_{0}^{2}\sum_{u=t_{1}+1}^{t-1}\frac{1}{t^{2c_{1}\alpha_{0}}(u+1)^{2-2c_{1}\alpha_{0}}}\nonumber\\
  	&= c_{3}\left(\frac{t_{1}+1}{t}\right)^{2c_{1}\alpha_{0}}+\frac{\alpha_{0}^{2}}{t^{2}}+\alpha_{0}^{2}\sum_{u=t_{1}+1}^{t-2}\frac{1}{t^{2c_{1}\alpha_{0}}(u+1)^{2-2c_{1}\alpha_{0}}}\nonumber\\
  	&\overset{(a)}{\le} c_{3}\left(\frac{t_{1}+1}{t}\right)^{2c_{1}\alpha_{0}}+\frac{\alpha_{0}^{2}}{t^{2}}+\frac{\alpha_{0}^{2}}{t^{2c_{1}\alpha_{0}}}\int_{t_{1}}^{t-1}\frac{1}{(s+1)^{2-2c_{1}\alpha_{0}}}ds \nonumber\\
  	&\le c_{3}\left(\frac{t_{1}+1}{t}\right)^{2c_{1}\alpha_{0}}+\frac{\alpha_{0}^{2}}{t^{2}}+\frac{\alpha_{0}^{2}}{t^{2c_{1}\alpha_{0}}}\left(\frac{t^{2c_{1}\alpha_{0}-1}}{2c_{1}\alpha_{0}-1}\right).
  	\end{align}
  	The above implies that, for all $t\geq t_{1}$,
  	\begin{align}
  	\label{eq:e131}
  	& t\sum_{i=1}^{t}\left\|\left[\mathbf{P}_{t}\right]_{ii}\right\|\nonumber\\
  	&\overset{(b)}{\le} c_{3}\frac{\left(t_{1}+1\right)^{2c_{1}\alpha_{0}}}{t^{2c_{1}\alpha_{0}-1}}+\frac{\alpha_{0}^{2}}{t}+\frac{\alpha_{0}^{2}}{2c_{1}\alpha_{0}-1}.
  	\end{align}
  	where in $(a)$ and $(b)$, we use the fact that $2c_{1}\alpha_0-1 > 1$ by Assumption \ref{bs:6}.
  	\noindent The proof follows by noting that the RHS of \eqref{eq:e131} is a non-increasing function of $t$.
  \end{IEEEproof}

\section{Proof of Theorems in Section \ref{subsec:illu_ex}}
\label{a3}
\noindent\begin{IEEEproof}[Proof of Theorem \ref{th:b3}]

	\noindent The proof for the large deviations upper bound of the probability of false alarm proposed in Theorem \ref{th:b3} exactly follows the derivation of the large deviations upper bound of the probability of false alarm of $\mathcal{CIGLRT-L}$. It follows from \eqref{eq:f1}-\eqref{eq:f4}.
	\noindent The characterization of $\left\|\mathbf{I}_{N}-\beta_{t}\mathbf{L}-\alpha_{t}\mathbf{G}_{H}\mathbf{\Sigma}^{-1}\mathbf{G}_{H}^{\top}\right\|$ exactly follows from \ref{le:det1}. By restricting \ref{le:det1} to the observation model described in \ref{subsec:illu_ex} which satisfies Assumptions \ref{bs:1}-\ref{bs:5} and \ref{bs:7}, we have that on choosing a $t_{2}$ such that $\forall t\geq t_{2}$,
	\begin{align}
	\label{eq:1det1_7}
	\beta_{t}\lambda_{N}(\mathbf{L})+\alpha_{t}\frac{h^{2}}{\sigma^{2}} < 1.
	\end{align}
	\noindent This guarantees that all the eigenvalues of $\mathbf{I}_{N}-\beta_{t}\mathbf{L}-\alpha_{t}\mathbf{G}_{H}\mathbf{\Sigma}^{-1}\mathbf{G}_{H}^{\top}$ are positive.
	\noindent From \ref{le:det1}, we have that there exists $t_{1}$, such that for all $t\geq t_{1}$
	\begin{align}
	\label{eq:1det1_8}
	\left\|\mathbf{I}_{N}-\beta_{t}\mathbf{L}-\alpha_{t}\mathbf{G}_{H}\mathbf{\Sigma}^{-1}\mathbf{G}_{H}^{\top}\right\|\le 1-c_{1}\alpha_{t}.
	\end{align}
	\noindent For notational simplicity we denote $\mathbf{1}_{N}\otimes\mathbf{\theta}^{*}$ as $\mathbf{\theta}_{N}^{*}$. Proceeding as in the proof of Theorem \ref{th:b2}, we have,
	\begin{align}
	\label{eq:1e2}
	\left\|\mathbf{G}_{H}^{\top}\left(\mathbf{\theta}(t)-\tn\right)\right\|\le h\left\|\mathbf{\theta}(t)-\tn\right\|.
	\end{align}
	\noindent
	Recall the representation of $\mathbf{P}_{t}$ and $\gamma_{G}(t)$ as defined in \eqref{eq:e4}-\eqref{eq:e5} and \eqref{eq:e41} in the proof of theorem \ref{th:b2}. Note that, $\mathbf{P}_{t}$ is a block matrix and is symmetric, positive semi definite with each of its individual blocks symmetric as in proof of theorem \ref{th:b2}.
	
	\noindent Proceeding as the proof of Theorem \ref{th:b2} and using Lemma \ref{wish}, we finally have,
	\begin{align}
	\label{eq:1e14}
	t\left\|\mathbf{P}_{t}\right\| \le c_{3}\frac{\left(t_{1}+1\right)^{2c_{1}\alpha_{0}}}{t^{2c_{1}\alpha_{0}-1}}+\frac{\alpha_{0}^{2}}{t}+\frac{\alpha_{0}^{2}}{2c_{1}\alpha_{0}-1}.
	\end{align}
	\noindent For $\mathcal{H}_{1}$, we have,
	\begin{align}
	\label{eq:1g1}
	 z_{n}(kt)=\sum_{j=1}^{N_{1}}\frac{\phi_{n,j}(k-1)}{{(k(t-1)+1)\sigma^{2}}}\sum_{i=0}^{k(t-1)}\mathbf{\theta}_{j}(k(t-1))h\gamma_{j}(i)-\frac{\left(h\left(\mathbf{\theta}_{j}(k(t-1))-\mathbf{\theta}^{*}\right)\right)^{2}}{2}+\frac{h^{2}(\mathbf{\theta}^{*})^{2}}{2}.
	\end{align}
	\noindent For notational simplicity we denote,
	\begin{align}
	\label{eq:1g21} \eta_{2}=\frac{-2N\eta\sigma^{2}+N_{1}h^{2}(\theta^{*})^{2}\left(1-N\sqrt{N}r^{k-1}\right)}{4h^{2}\left(1+N\sqrt{N}r^{k-1}\right)}.
	\end{align}
	
	\noindent Moreover, supposing that the following condition holds
	\begin{align}
	\label{eq:1g22}
	\eta < \frac{N_{1}h^{2}(\theta^{*})^{2}\left(1-N\sqrt{N}r^{k-1}\right)}{2N\sigma^{2}},
	\end{align}
	
	\noindent we have on proceeding as in \eqref{eq:g2} that the probability of miss can be characterized as follows
	\begin{align}
		\label{eq:1g2}
		\mathbb{P}_{1,\theta^{*}}\left(z_{n}(kt) < \eta\right)\le a1+a2+a3,
		\end{align}
	\noindent where $(a1)$, $(a2)$, $(a3)$ are given by
		\begin{align}
		\label{eq:1g23}
		& a1=\mathbb{P}_{1,\theta^{*}}\left(h^{2}\frac{\left\|\mathbf{\theta}(k(t-1))-\tn\right\|^{2}}{2\sigma^{2}}> \frac{-2N\eta\sigma^{2}+N_{1}h^{2}(\theta^{*})^{2}\left(1-N\sqrt{N}r^{k-1}\right)}{4\sigma^{2}\left(1+N\sqrt{N}r^{k-1}\right)}\right)\nonumber\\
		& a2=\mathbb{P}_{1,\theta^{*}}\left(\sum_{j=1}^{N_{1}}\frac{\phi_{n,j}(k-1)}{(k(t-1)+1)\sigma^{2}}\sum_{i=0}^{k(t-1)}\left(\mathbf{\theta}_{j}(k(t-1))-\mathbf{\theta}^{*}\right)h\gamma_{j}(i)< \frac{\eta}{4}-\frac{N_{1}h^{2}(\mathbf{\theta}^{*})^{2}\left(\frac{1}{N}-\sqrt{N}r^{k-1}\right)}{8\sigma^{2}}\right)\nonumber\\
		& a3=\mathbb{P}_{1,\theta^{*}}\left(\sum_{j=1}^{N_{1}}\frac{\phi_{n,j}(k-1)}{(k(t-1)+1)\sigma^{2}}\sum_{i=0}^{k(t-1)}\left(\mathbf{\theta}^{*}\right)h\gamma_{j}(i)< \frac{\eta}{4}-\frac{N_{1}h^{2}(\mathbf{\theta}^{*})^{2}\left(\frac{1}{N}-\sqrt{N}r^{k-1}\right)}{8\sigma^{2}}\right)
		\end{align}
	
	\noindent First, we characterize $(a1)$. Following as in \eqref{eq:g2}-\eqref{eq:g4} in the proof of theorem \ref{th:b2}, we have that if
	$\lambda < c_{4}$, where $c_{4}$ is given by
	\begin{align}
	\label{eq:1gg1}
	c_{4}= \frac{\sigma^{2}}{h^{2}\left(c_{3}\frac{\left(t_{1}+1\right)^{2c_{1}\alpha_{0}}}{kt_{1}^{2c_{1}\alpha_{0}-1}}+\frac{\alpha_{0}^{2}}{kt_{1}}+\frac{\alpha_{0}^{2}}{2c_{1}\alpha_{0}-1}\right)},
	\end{align}
	\begin{align}
	\label{eq:1gg2}
	\det\left(\mathbf{I}_{Nkt}-kt\lambda\mathbf{P}_{kt}\left(\mathbf{I}_{kt}\otimes\mathbf{G}_{H}\mathbf{\Sigma}^{-1}\mathbf{G}_{H}^{\top}\right)\right) \ge \left(1-\frac{kth^{2}\lambda\left\|\mathbf{P}_{kt}\right\|}{\sigma^{2}}\right)^{Nkt}.
	\end{align}
	
	We also have from \eqref{eq:g5},
	\begin{align}
	\label{eq:1gg3}
	\limsup_{t\to\infty}kt\left\|\mathbf{P}_{kt}\right\|\le\frac{\alpha_{0}^{2}}{2c_{1}\alpha_{0}-1}.
	\end{align}
	
	Now, on specializing the expressions in the proof of theorem \ref{th:b2} by using specifics of the scalar observation model in \ref{subsec:illu_ex}, we have,
	\begin{align}
		\label{eq:1g6}
		&\mathbb{P}_{1,\theta^{*}}\left(\left\|\mathbf{\theta}(k(t-1))-\tn\right\|^{2}> \eta_{2}\right)\nonumber\\
		&\le e^{-\lambda\eta_{2}kt}\times \left(\det\left(\mathbf{I}_{NKt}-kt\lambda\mathbf{P}_{kt}\left(\mathbf{I}_{kt}\otimes\mathbf{G}_{H}\mathbf{\Sigma}^{-1}\mathbf{G}_{H}^{\top}\right)\right)\right)^{-1/2}\nonumber\\
		&\le e^{-\lambda\eta_{2}kt}\times \left(1-kt\lambda\left\|\mathbf{P}_{kt}\right\|h^{2}/\sigma^{2}\right)^{-Nkt/2}\nonumber\\
		&\Rightarrow \frac{1}{kt}\log\left(\mathbb{P}_{1,\theta^{*}}\left(\left\|\mathbf{\theta}(k(t-1))-\tn\right\|^{2}> \eta_{2}\right)\right)\nonumber\\
		&\le -\lambda\eta_{2}-\frac{N}{2}\log\left(1-kt\lambda\left\|\mathbf{P}_{kt}\right\|h^{2}/\sigma^{2}\right)\nonumber\\
		&\Rightarrow \limsup_{t\to\infty}\frac{1}{kt}\log\left(\mathbb{P}_{1,\theta^{*}}\left(\left\|\mathbf{\theta}(k(t-1))-\tn\right\|^{2}> \eta_{2}\right)\right)\nonumber\\
		&\le -\lambda\eta_{2}-\frac{N}{2}\log\left(1-\frac{\lambda\alpha_{0}^{2}h^{2}}{\sigma^{2}(2c_{1}\alpha_{0}-1)}\right).
		\end{align}	
	\noindent Let $LD(\lambda)=\lambda\eta_{2}+\frac{N}{2}\log\left(1-\frac{\lambda\alpha_{0}^{2}h^{2}}{\sigma^{2}(2c_{1}\alpha_{0}-1)}\right)$.	
	\noindent We first note that $LD(0)=0$. So, in order to have a positive exponent, the function $LD(.)$ needs to be strictly increasing in an interval of the form, $\left[0,c_{5}\right]$, where $0 < c_{4} \le c_{5}$, with $c_{4}$ as defined in \eqref{eq:1gg1} which is formalized as follows:
	\begin{align}
	\label{eq:1g61}
	\lambda < \frac{(2c_{1}\alpha_{0}-1)\sigma^{2}}{\alpha_{0}^{2}h^{2}}-\frac{N}{2\eta_{2}}=c_{4}^{*},
	\end{align}
	\noindent with $\eta_{2}$ as defined in \eqref{eq:1g21}.	
	\noindent In order to have a positive exponent, the RHS of \eqref{eq:1g61} needs to be positive and hence, we require,
	\begin{align}
	\label{eq:1g62}
	&\frac{(2c_{1}\alpha_{0}-1)\sigma^{2}}{\alpha_{0}^{2}h^{2}}-\frac{N}{2\eta_{2}} > 0\nonumber\\
	&\Rightarrow \eta <\frac{N_{1}h^{2}(\theta^{*})^{2}\left(1-N\sqrt{N}r^{k-1}\right)}{2N\sigma^{2}}-\frac{\alpha_{0}^{2}h^{4}\left(1+N\sqrt{N}r^{k-1}\right)}{(2c_{1}\alpha_{0}-1)\sigma^{4}}.
	\end{align} 	
	\noindent We note that the condition derived in \eqref{eq:1g62} is tighter than \eqref{eq:1g22}. Now, combining the threshold condition derived above in \eqref{eq:1g62} and the one derived in \eqref{eq:f5}, we have the following condition on the parameter $\theta^{*}$
	\begin{align}
	\label{eq:1g63}
	\frac{N_{1}h^{2}(\theta^{*})^{2}\left(1-N\sqrt{N}r^{k-1}\right)}{2N\sigma^{2}} > \frac{\alpha_{0}^{2}h^{4}\left(1+N\sqrt{N}r^{k-1}\right)}{(2c_{1}\alpha_{0}-1)\sigma^{4}}+\frac{\left(\frac{1}{N}+\sqrt{N}r^{k-1}\right)N_{1}}{2}
	\end{align}
	\noindent which ensures that $(a1)$ decays exponentially.
	\noindent Now, when we analyze $(a2)$ and $(a3)$ in \eqref{eq:1g2}, we note that $a2$ involves an additional time-decaying term, i.e., $\theta_{j}(k(t-1))
	-\theta^{*}$ which contributes to the large deviations exponent as well. Hence, the exponent which will dominate among $(a2)$ and $(a3)$, would be the exponent of their sum.
	\noindent Using the condition derived in \eqref{eq:1g22} and the union bound on $a3$, we have,
	\begin{align}
		\label{eq:1g7}
		&\mathbb{P}_{1,\theta^{*}}\left(\sum_{j=1}^{N_{1}}\frac{\phi_{n,j}(k-1)}{(k(t-1)+1)\sigma^{2}}\sum_{i=0}^{k(t-1)}\theta^{*}h\gamma_{j}(i).< \frac{\eta}{4}-\frac{N_{1}h^{2}(\mathbf{\theta}^{*})^{2}\left(\frac{1}{N}-\sqrt{N}r^{k-1}\right)}{8\sigma^{2}}\right)\nonumber\\
%		&\le \sum_{j=1}^{N_{1}}\mathbb{P}_{1,\theta^{*}}\left(\frac{\phi_{n,j}(k-1)}{(k(t-1)+1)\sigma^{2}}\sum_{i=0}^{k(t-1)}-\theta^{*}h\gamma_{j}(i) > -\frac{\eta}{4N_{1}}+\frac{h^{2}(\mathbf{\theta}^{*})^{2}\left(\frac{1}{N}-\sqrt{N}r^{k-1}\right)}{8\sigma^{2}}\right)\nonumber\\
%		&\le\sum_{j=1}^{N_{1}}\mathbb{Q}\left(\frac{-\frac{\eta\sigma\sqrt{k(t-1)+1}}{4N_{1}}+\frac{\sigma\sqrt{k(t-1)+1}h^{2}(\mathbf{\theta}^{*})^{2}\left(\frac{1}{N}-\sqrt{N}r^{k-1}\right)}{8\sigma^{2}}}{\phi_{n,j}(k-1)h\theta^{*}}\right)\nonumber\\
		&\le\mathbb{Q}\left(\frac{-\frac{\eta\sigma\sqrt{k(t-1)+1}}{4}+\frac{\sigma\sqrt{k(t-1)+1}N_{1}h^{2}(\mathbf{\theta}^{*})^{2}\left(\frac{1}{N}-\sqrt{N}r^{k-1}\right)}{8\sigma^{2}}}{N_{1}h\theta^{*}\left(\frac{1}{N}+\sqrt{N}r^{k-1}\right)}\right)\nonumber\\
		&\Rightarrow \limsup_{t\to\infty}\frac{1}{kt}\log\left(\mathbb{P}_{1,\theta^{*}}\left(\sum_{j=1}^{N_{1}}\frac{\phi_{n,j}(k-1)}{(k(t-1)+1)\sigma^{2}}\sum_{i=0}^{k(t-1)}\theta^{*}h\gamma_{j}(i)< \frac{\eta}{4}-\frac{N_{1}h^{2}(\mathbf{\theta}^{*})^{2}\left(\frac{1}{N}-\sqrt{N}r^{k-1}\right)}{8\sigma^{2}}\right)\right)\nonumber\\
		&\le -\left(\frac{\left(-\frac{\eta}{4}+\frac{N_{1}h^{2}(\mathbf{\theta}^{*})^{2}\left(\frac{1}{N}-\sqrt{N}r^{k-1}\right)}{8\sigma^{2}}\right)^{2}}{2N_{1}h^{2}(\theta^{*})^{2}\left(\frac{1}{N}+\sqrt{N}r^{k-1}\right)^{2}/\sigma^{2}}\right).
		\end{align}
	
	\noindent Combining \eqref{eq:1g7} and \eqref{eq:1g6}, we have,
	\begin{align}
	\label{eq:1g8}
	&\limsup_{t\to\infty}\frac{1}{kt}\log\left(\mathbb{P}_{1,\theta^{*}}\left(z_{n}(kt) < \eta\right)\right)\nonumber\\
	&\le \max\left\{-\left(\frac{\left(-\frac{\eta}{4}+\frac{N_{1}h^{2}(\mathbf{\theta}^{*})^{2}\left(\frac{1}{N}-\sqrt{N}r^{k-1}\right)}{8\sigma^{2}}\right)^{2}}{2N_{1}h^{2}(\theta^{*})^{2}\left(\frac{1}{N}+\sqrt{N}r^{k-1}\right)^{2}/\sigma^{2}}\right), -LD\left(\min\left\{c_{4},c_{4}^{*}\right\}\right)\right\}=LD_{1}\left(\eta\right),
	\end{align}
	\noindent We specifically focused on the sub-sequence $\{z_{n}(kt)\}$ for the derivation of large deviations\footnote{By large deviations exponent, we mean the exponent associated with our large deviations upper bound.} exponent in this proof. It can be readily seen that other time-shifted sub-sequences (with constant time-shifts upto $k$ units) also inherit a similar large deviations upper bound as by construction, (see \eqref{eq:dec_stat_lin_12} for example), the decision statistic $z_{n}(kt)$ stays constant on the time interval $\left[kt, kt+k-1\right]$. Hence, the large deviations upper bound can be extended as a large deviations upper bound for the sequence $\{z_{n}(t)\}$.
	
	\end{IEEEproof}

\bibliographystyle{IEEEtran}
\bibliography{CentralBib,glrt,dsprt}

\end{document}